%% file: main.tex
\documentclass[fleqn,usenatbib]{mnras}

\usepackage{newtxtext,newtxmath}

\usepackage[T1]{fontenc}

\DeclareRobustCommand{\VAN}[3]{#2}
\let\VANthebibliography\thebibliography
\def\thebibliography{\DeclareRobustCommand{\VAN}[3]{##3}\VANthebibliography}

\usepackage{graphicx}	

\usepackage{float} 
\usepackage{flafter} 

\usepackage{natbib}
\usepackage{graphicx}%
\usepackage{multirow}%

\usepackage{amssymb,amsfonts}%

\usepackage{amsmath}

\usepackage{amsthm}%
\usepackage{mathrsfs}%
\usepackage[title]{appendix}%
\usepackage{xcolor}%
\usepackage{textcomp}%
\usepackage{manyfoot}%
\usepackage{booktabs}%
\usepackage{algorithm}%
\usepackage{algorithmicx}%

\usepackage{listings}%

\usepackage[utf8]{inputenc}

\usepackage{booktabs}       
\usepackage{nicefrac}       

\usepackage{xcolor}         
\usepackage{wrapfig}        

\usepackage{comment}

\usepackage{blindtext}
\usepackage{booktabs}
\usepackage{varwidth}
\usepackage{caption}

\usepackage{graphicx}
\usepackage{}

\usepackage{paralist}
\usepackage{mathrsfs}
\usepackage{upgreek}
\usepackage{cleveref}
\usepackage{multirow} 
\usepackage{enumitem}

\usepackage{soul,xcolor} 

\usepackage{algorithm}
\usepackage[noend]{algpseudocode}

\usepackage{caption}
\usepackage{subcaption}

\usepackage{tikz}
\usetikzlibrary{shapes, arrows}
\usepackage{capt-of}

\usetikzlibrary{arrows.meta}
\tikzset{>={Latex[width=2mm,length=2mm]}}

\usepackage{url} 

\usepackage{mathtools}
\usepackage{adjustbox}

\defcitealias{rosenbaum1983central}{RR83}
\defcitealias{wright2020photometric}{W20}

\usepackage{etoolbox}   
\makeatletter
\patchcmd{\@maketitle}{\artauthors}{{\artauthors}}{}{}
\makeatother

\usepackage{filecontents}
\usepackage{multibib}
\newcites{supp}{Supplementary References}

\usepackage{hyperref}       

\let\cline\cmidrule

\title[Photo-$z$ calibration via StratLearn-Bayes]{Improved Weak Lensing Photometric Redshift Calibration via StratLearn and Hierarchical Modeling}

\author[M. Autenrieth et al.]{
Maximilian Autenrieth,$^{1}$\thanks{E-mail: m.autenrieth19@imperial.ac.uk}
Angus H. Wright,$^{2}$
Roberto Trotta,$^{3,4,5,6}$
David A. van Dyk,$^{1}$
\newauthor{
David C. Stenning,$^{7}$
Benjamin Joachimi,$^{8}$
}
\\
$^{1}$Department of Mathematics, Imperial College London, 180 Queen’s Gate, London SW7 2AZ, UK\\
$^{2}$Ruhr University Bochum, Faculty of Physics and Astronomy, Astronomical Institute (AIRUB), \\ \ \ German Centre for Cosmological Lensing, 44780 Bochum, Germany\\
$^{3}$SISSA -- International School for Advanced Studies, Via Bonomea 265, 34136 Trieste, Italy\\
$^{4}$Department of Physics, Imperial College London, Blackett Laboratory, Prince Consort Rd, SW72AZ, London, UK \\
$^{5}$Italian Research Center on High Performance Computing, Big Data and Quantum Computing\\
$^{6}$INFN -- National Institute for Nuclear Physics, Via Valerio 2, 34127 Trieste, Italy\\
$^{7}$Department of Statistics and Actuarial Science, Simon Fraser University\\
$^{8}$Department of Physics and Astronomy, University College London, Gower Street, London WC1E 6BT, United Kingdom\\
}

\pubyear{2024}

\begin{document}
\label{firstpage}
\pagerange{\pageref{firstpage}--\pageref{lastpage}}
\maketitle

\begin{abstract}
Discrepancies between cosmological parameter estimates from cosmic shear surveys and from recent Planck cosmic microwave background measurements challenge the ability of the highly successful $\Lambda$CDM model to describe the nature of the Universe. To rule out systematic biases in cosmic shear survey analyses, accurate redshift calibration within tomographic bins is key. In this paper, we improve photo-$z$ calibration via Bayesian hierarchical modeling of full 
galaxy photo-$z$ conditional densities, 
by employing \textit{StratLearn}, a recently developed statistical methodology, which accounts for systematic differences in the distribution of the spectroscopic training/source set and the photometric target set.
Using realistic simulations that were designed to resemble the KiDS+VIKING-450 dataset, we show that \textit{StratLearn}-estimated conditional densities improve the galaxy tomographic bin assignment, and that our \textit{StratLearn}-Bayesian framework leads to nearly unbiased estimates of the target population means. This leads to a factor of $\sim 2$ improvement upon the previously best photo-$z$ calibration method. Our approach delivers a maximum bias per tomographic bin of $\Updelta \langle z \rangle =  0.0095 \pm 0.0089$, with an average absolute bias of $0.0052 \pm 0.0067$ across the five tomographic bins. 
\
\end{abstract}

\begin{keywords}
cosmology: observations -- large-scale structure of Universe -- methods: statistical -- galaxies: distances and redshifts
\end{keywords}



\section{Introduction}\label{sec:introduction}
Cosmological parameter estimation from the cosmic microwave background (CMB; \citealt{2020A&A...641A...1P}) 
and from tomographic cosmic shear measurements
\citep[e.g., ][]{asgari+2021,Abbott+2022,sugiyama+2023} 
lead to discrepancies 
in the estimated clustering strength of dark matter (see \citealt{2022JHEAp..34...49A} for a recent review on cosmic tensions). 
Such systematic discrepancies could challenge the highly successful dark energy and cold dark matter paradigm ($\Lambda$CDM) in describing the true nature of the Universe. Of course, such a claim needs critical and detailed consideration of the surveys and analysis steps performed by the various collaborations to rule out systematic biases in the different procedures, which might explain said discrepancies.

Extensive explorations of various survey, model, and analysis modifications have been performed in recent cosmological interpretations of the current generation of cosmic shear surveys, the Dark Energy Survey (DES), the Hyper-Suprime Camera (HSC) survey, and the Kilo-Degree Survey (KiDS). These highlight that inaccuracies and statistical uncertainty in the line-of-sight distribution of galaxies as determined by photometric redshifts can limit, and potentially bias, constraints on cosmological parameters \citep[cf.][]{2023OJAp....6E..36D,2023MNRAS.524.5109R}. 

In cosmic shear tomography, galaxies are assigned to (pre-defined) tomographic redshift bins \citep{1999ApJ...522L..21H} based on an estimate of their photometric redshift (photo-$z$). For recent reviews on photometric redshift estimation and its application in large galaxy surveys see \citet{2019NatAs...3..212S} and \citet{2022ARA&A..60..363N}, respectively. If the 
estimated population redshift distribution (in a tomographic bin) differs systematically from the (non-observable) true redshift distribution, the parameter estimates from cosmic shear tomography might be systematically biased.

For instance, if the estimated redshift distribution is systematically lower than it is in reality, then observed gravitational distortions are attributed to  an overly dense and too highly clustered matter distribution. It is thus essential to obtain accurate redshift distribution estimates. In particular, it is crucial to obtain an unbiased estimate of the first moment (mean) of the true underlying redshift population distribution (per tomographic bin), in 
order to avoid such systematic biases in the analysis \citep{amara2008systematic, Reischke2023}. 
This is because the accuracy of cosmic shear cosmological measurements is highly dependent on the accuracy of the first moment of the binned 
redshift population distributions, but much less sensitive to the higher-order moments: \cite{Reischke2023}, for example, demonstrate that a one-sigma shift 
in the desired cosmological model parameters (for a {\em Euclid}-like survey) is induced when the first moment of the redshift distributions  
is mis-specified at the level of $<1\%$. Conversely, a similar bias is only introduced for the second-order moment with a $\sim 10\%$ mis-specification, 
and all higher-order moments can be essentially ignored \cite[Figure 2]{Reischke2023}.

Several redshift calibration methods have been investigated to improve cosmic shear tomography. \cite{wright2019kids+} group these approaches into three categories: 
(i) cross-correlation with reference galaxy samples that have precise and accurate redshifts 
\citep{schneider2006using,newman2008calibrating,mcquinn2013using,morrison2017wizz}; (ii) stacking of individual galaxy redshift distributions \citep{hildebrandt2012cfhtlens,hoyle2018dark,tanaka2018photometric,malzhogg2022}; 
and (iii) direct redshift calibration \citep{lima2008estimating,hildebrandt2016rcslens,hildebrandt2020kids+,buchs2019phenotypic,wright2020photometric}. The idea of direct redshift calibration is to reweight the distribution of spectroscopic redshift, obtained only for a small and non-representative subsample of the data (source/training data), to match the distribution of the photometric target data.

In recent work, \cite{masters+2015,buchs2019phenotypic}; and  
\citet[][hereafter \citetalias{wright2020photometric}]{wright2020photometric}
develop direct redshift calibration methods based on self organising maps (SOM; \citealt{kohonen1982self}). In \citetalias{wright2020photometric}, their implementation of SOM-based direct calibration is shown to outperform previously proposed methods on comprehensive simulations designed to realistically resemble the KIDS+VIKING-450 dataset 
(\citealp{wright2019kids+}; \citetalias{wright2020photometric}; \citealp{hildebrandt2020kids+}). 
Their (and previous) methods obtain a tomographic bin assignment via a Bayesian-Photometric-Redshift estimate \citep{benitez2000bayesian}, 
further denoted as $z_B$, calculated for each galaxy. 
While improving on other direct redshift calibration methods, the SOM method proposed in \citetalias{wright2020photometric} still leads to potentially concerning bias in some tomographic bins, and mitigates these biases by introducing additional systematic selections to the data. Such selections lead to fewer sources available for science, and therefore constitute a source of increased statistical uncertainty in down-stream cosmological analyses.

In this paper, we propose a different strategy to improve redshift calibration, based on galaxy (object level) conditional photo-$z$ density estimates. More precisely, we employ a recently proposed statistical method, \textit{StratLearn} \citep{autenrieth2021stratifiedASA}, that allows principled photo-$z$ conditional density estimation under non-representative source/training data.
\textit{StratLearn} alleviates (or bypasses) the problem of non-representative source/training data (identified as covariate shift),  
by subgrouping the data into strata based on estimated propensity scores, a pivotal methodology in causal inference \citep{rosenbaum1983central}. In our context, the propensity score is the probability of a galaxy being assigned to the spectroscopic training/source set given the observed covariates (i.e., photometric magnitudes/colors). 
\cite{autenrieth2021stratifiedASA} demonstrate that fitting conditional density estimators within strata, constructed by partitioning the data based on the estimated propensity scores, 
improves 
full conditional photo-$z$ density estimates under-non representative source/training data. 
Here, we show that the \textit{StratLearn} conditional densities\footnote{We note that \textit{StratLearn} is a general-purpose statistical method for learning under covariate shift. While \cite{autenrieth2021stratifiedASA} show the effectiveness of conditional density estimation within the \textit{StratLearn} framework, the conditional density estimators 
themselves are not part of the \textit{StratLearn} methodology, and have been proposed elsewhere \citep{izbicki2017photo,izbicki2016nonparametric}. For conciseness, we loosely refer to the conditional density estimates as `\textit{StratLearn} conditional densities'.} can be used directly to improve the tomographic bin assignment, by assigning each galaxy to the tomographic bin with its highest conditional probability. 
In a second step, we construct a Bayesian hierarchical framework to model summaries of each galaxy's conditional density (within tomographic bins), leading to nearly unbiased estimates of the mean redshift of each tomographic bin. We evaluate our novel \textit{StratLearn}-Bayes approach on comprehensive simulations constructed by \citetalias{wright2020photometric}, demonstrating a substantial reduction of bias compared to the previously proposed SOM calibration method. 

While the primary sensitivity of cosmic shear is to the first moment of the redshift distribution, other cosmological probes, which also require redshift distribution estimation and calibration, are more sensitive to the accurate recovery of higher-order redshift distribution moments. \cite{Reischke2023} show that higher-order moments have much more influence on bias in an analysis of photometric galaxy clustering. Additionally, cosmic shear surveys will become increasingly sensitive to higher-order moments with increasing statistical power. As such, it is sensible to consider how we can best estimate the full redshift distribution. 
While our \textit{StratLearn}-Bayes method is specifically designed to obtain accurate estimates of the first moments of the redshift distributions, we demonstrate how estimated propensity scores can be used in a direct redshift calibration scheme to obtain accurate estimates of the redshift population distribution shapes. 

The remainder of the paper is structured as follows.
In Section~\ref{sec:notation}, we specify notation and we 
formally introduce the underlying covariate shift scenario, arising through the non-representativeness of the training/source data. In Section~\ref{sec:related_literature}, we summarize the direct redshift calibration method. We then briefly introduce the supervised learning task with a focus on conditional density estimation under the covariate shift scenario. In Section~\ref{sec:methodology}, we formally introduce our approach. In Section~\ref{sec:StratLearn}, we provide a detailed description of how we estimate conditional densities under covariate shift within \textit{StratLearn}. We then specify how these conditional densities can be used to improve galaxy tomographic bin assignment (Section~\ref{sec:StratLearn_bin_assignment}). In Section~\ref{sec:Bayesian_model}, we demonstrate how summaries of the estimated conditional densities can be employed in a Bayesian hierarchical framework to accurately estimate the redshift population means (within a tomographic bin). 
In Section~\ref{sec:inverse-PS}, we demonstrate  
how inverse propensity score weighting (inverse-PS) can be employed to estimate the redshift population shapes for each tomographic bin. 
In Section~\ref{sec:numerical_demonstrations}, we present numerical evaluation of our method. We first introduce the simulation setting in Section~\ref{sec:data}. 
We then evaluate our new bin assignment with a comparison to previously used $z_B$ bin assignment 
(Section~\ref{sec:results_bin_assignment}). 
We present our redshift calibration results in Section~\ref{sec:results_bias}, and illustrate 
the inverse-PS population distribution estimates in Section~\ref{section:population_shapes_results}. 
Finally, in Section~\ref{sec:discussion}, we conclude with a discussion of our findings and implications for future weak lensing survey analyses.


\section{Addressing Non-Representative Training Data}\label{sec:notation&background}

\subsection{Non-representative Spectroscopic Data and Covariate Shift}\label{sec:notation}
Let $z_i$ be the true spectroscopic redshift of galaxy $i$, and $x_i$ be the vector of its observed photometric magnitudes/colors (the exact choice of covariates is described in Section~\ref{sec:technical_details}). 
In a cosmic shear analysis, we have access to a relatively small set of galaxies with measured spectroscopic redshift, since obtaining spectroscopy for millions of objects is observationally expensive (over the magnitude range in question). 
For our purposes, spectroscopically-measured redshifts can be considered equal to the true redshift. We denote this spectroscopic set as source (or training) data $D_S= \{(x_S^{(i)}, z_S^{(i)} )\}_{i = 1}^{n_s}$, with $n_s$ galaxies sampled at random 
from the joint distribution $p_S(x,z)$. 
The so-called ``photo-$z$ estimation problem'' \citep{hildebrandt2010phat,izbicki2017photo,freeman2017unified,dey2023conditionally} 
is to find a redshift estimate that can be deployed on 
a much larger set of galaxies, for which only the photometric data $x_T$ are available, but not spectroscopically measured redshifts, $z_T$. We denote this photometric set as our target data $D_T= \{x_T^{(i)}\}_{i = 1}^{n_t},$ with $n_t$ unlabelled samples from the joint distribution $p_T(x,z)$ (with $n_T \gg n_S$). 
The problem is compounded by the fact that $p_S(x,z) \neq p_T(x,z) $, i.e., the spectroscopic source and photometric target distributions differ systematically due to selection effects in the acquisition of spectroscopy based on characteristics of the photometric magnitudes, leading to $p_S(x) \neq p_T(x)$. However, the conditional distribution of redshift $z$ given the magnitudes $x$ are the same in spectroscopic source and photometric target data, i.e., $p_S(z|x) = p_T(z|x)$. This situation, in which $p_S(z|x) = p_T(z|x)$ but $p_S(x) \neq p_T(x)$, is called ``covariate shift'' in the statistical learning literature \citep{moreno2012unifying}. 
If such covariate shift is not accounted for, machine learning or other statistical methods that aim to learn the relationship between the covariates and redshift can perform poorly; the training set is not representative of the target/test, meaning that patterns learned from the training set are not generalizable.

\subsection{Direct Redshift Calibration}\label{sec:related_literature}

Since in the covariate shift scenario $p_S(z|x) = p_T(z|x)$ but $p_S(x) \neq p_T(x)$, it generally follows that the redshift distribution of the spectroscopic set differs from that of the target,  $p_S(z) \neq p_T(z)$.
Direct redshift calibration methods reweight the spectroscopic redshift sample to match the photometric redshift distribution \citep{lima2008estimating}. 

More precisely, 
under the covariate shift scenario, it holds that 
\begin{align}\label{form:joint_weighted_source}
    p_T(z,x) &= p_T(z|x) p_T(x) \\
                 &= p_S(z|x) p_T(x)   \\
                 &= p_S(z,x) \frac{p_T(x)}{p_S(x)}   \label{form:joint_weighted_source3}    
\end{align} 
That is, one can express the joint target distribution by reweighting the joint source distribution. Precisely, $p_T(z,x) = \omega(x) p_S(z,x),$ with weights $\omega(x) = p_T(x)/p_S(x)$. In practice, one can reweight galaxies in the spectroscopic source set (with weights $\omega(x)$) to obtain a representative sample of the joint target distribution. 
In principle, looking at the marginal sample of $z$ in the weighted joint distribution thus provides us a consistent estimate of the target redshift distribution $p_T(z)$.

Accurate estimation of the weights $\omega(x)$ is key for direct redshift calibration methods.
\cite{lima2008estimating} and \cite{hildebrandt2020kids+} implement a k-nearest-neighbor (kNN) method for weight estimation. \citetalias{wright2020photometric} demonstrate improvement over the kNN method by computing the weights via an SOM method, a form of unsupervised neural network which can map a high-dimensional covariate space 
to a lower-dimensional grid.

Unfortunately, weighting methods typically entail high variance, particularly if there is a small number of objects with very large weights.
In addition, finding a suitable set of weights $\omega$ is not trivial, but key for direct calibration methods. Noisy, inaccurate weights might lead to potentially severe bias and strongly increased variance.

\subsection{Photometric Redshift Regression}

Instead of reweighting, our approach uses the labelled spectroscopic source data as a training set to fit supervised full conditional density models. Our trained models then deliver a non-parametric estimate of the full conditional redshift (photo-$z$) distribution for each galaxy in the photometric target data (conditional on its observed covariates), $\hat{f}(z|x)$.  
If source and target data follow the same distribution, conditional density estimators aim to minimize the {\it generalized} risk under the $L^2-$loss (generalized in that the underlying loss can be negative), given by:
\begin{align} \label{formula:fzxrisk_source}
    \hat{R}_S(\hat{f}) = & \frac{1}{n_S} \sum_{k=1}^{n_S} \int \hat{f}^2 (z|x_S^{(k)}) dz  -  2 \frac{1}{n_S} \sum_{k=1}^{n_S}  \hat{f} (z_S^{(k)}|x_S^{(k)}).
\end{align} 
(see Section~\ref{sec_supp:conditional_densities} for the derivations of  (\ref{formula:fzxrisk_source}) and \cite{izbicki2017photo} for further details). 
To provide intuition for the {\it generalized} risk in (\ref{formula:fzxrisk_source}), note that the second term 
of (\ref{formula:fzxrisk_source}) averages the values of the conditional density estimates at the true spectroscopic redshift (known for the source data); this is optimized if the true redshift is at (or close to) the mode of the conditional density estimate $\hat{f}(z_i|x_i)$, with $\hat{f}(z_i|x_i)$ being very tall and narrow 
(the Dirac delta distribution at the true redshift value is the optimal limiting case). In contrast, the first term 
of (\ref{formula:fzxrisk_source}), which integrates the squared conditional density estimates over the redshift range (without information of the true redshift), is minimized for wide and (nearly) uniform conditional density estimates, thus penalizing highly localized predictions. Thus, overall, estimates that are very certain (i.e., low variance), but fail to cover the truth lead to a high risk.

In the presence of covariate shift, however, obtaining accurate target estimates requires minimization of the target risk $\hat{R}_T(\hat{f})$, which is obtained by replacing all source samples in (\ref{formula:fzxrisk_source}) with target samples, which typically means 
$ \hat{R}_S(\hat{f}) \neq \hat{R}_T(\hat{f})$. The challenge is to minimize $\hat{R}_T(\hat{f})$ without access to the target true redshift $z_T$. In the next section, we provide a summary of our approach, called \textit{StratLearn} \citep{autenrieth2021stratifiedASA}, which allows minimization of $\hat{R}_T(\hat{f})$ under the covariate shift scenario.

\section{Bayesian Photometric Redshift Calibration via StratLearn}\label{sec:methodology}

\subsection{Photo-z Conditional Densities within \textit{StratLearn}}\label{sec:StratLearn}

\textit{StratLearn} allows target risk minimization by subgrouping the source and target data into strata based on estimated propensity scores. Within strata, the joint distribution of target data and source data are approximately the same, 
and target risk can thus be minimized via source risk minimization. In the following, we provide a detailed description of the procedure.

Let $S$ be a binary indicator variable, with $s_i = 1$ indicating the assignment of galaxy $i$ to the spectroscopic source set ($s_i = 0$ indicates assignment to the photometric target set). In the context of this paper, the propensity score is the probability of a galaxy $i $ being in the spectroscopic source data, given its observed covariates (photometry) $x_i$, i.e., 
\begin{align}\label{form:PS_covariate_shift}
e(x_i) := P(s_i = 1|x_i), \text{ with } 0 < e(x_i) < 1.
\end{align} 
In practice, we obtain an estimate $\hat{e}(x_i)$ of (\ref{form:PS_covariate_shift}) via binary, probabilistic classification of source and target data using a logistic regression model with all the photometric magnitudes/colors as independent predictor variables (main effects) and the source/target set assignment variable $S$ 
as the binary dependent variable.
We then subgroup (stratify) the source and target data into $k=5$ strata 
based on the quintiles of the estimated propensity score distribution $\hat{e}(x)$. The use of five strata is 
suggested by \cite{autenrieth2021stratifiedASA}, based on numerical evidence provided by \cite{cochran1968effectiveness} that subgrouping into $k=5$ strata removes at least $90\%$ of the bias for many continuous distributions.

By Proposition 1 of \cite{autenrieth2021stratifiedASA}, within strata, 
\begin{equation}\label{form:stratified_joint_dist}
  p_{T_j}(z,x)  \approx  p_{S_j}(z,x), \text{ for } j \in 1,\dots , k,
\end{equation}
where $S_j$ indicates conditioning on assignment to the $j$'th source stratum (analogously for target $T_j$). It directly follows that $\hat{R}_{Tj}(\hat{f}) \approx \hat{R}_{Sj}(\hat{f})$ within strata $j \in 1,\dots , k$. 
Thus, we can minimize the target risk $\hat{R}_{Tj}(\hat{f})$ within strata by minimizing the source risk $\hat{R}_{Sj}(\hat{f})$ within strata. See \cite{autenrieth2021stratifiedASA} for details.

Given the strata conditional on the estimated propensity score, we can now fit any supervised model on the spectroscopic source data within each stratum and predict on its respective photometric target stratum. As suggested in \cite{autenrieth2021stratifiedASA}, within each strata, we employ a weighted average (convex combination) of two conditional density estimators: 
{\it ker-NN} \citep{izbicki2017photo}   and {\it Series} \citep{izbicki2016nonparametric}.
The kernel nearest neighbor estimator (ker-NN) computes the conditional density of an object via a kernel smoothed histogram of the redshift of its $k$ nearest neighbors in the respective source stratum. 
The spectral series estimator (Series) 
adapts a lower-dimensional subspace of the $x$-space as the intrinsic dimension of the data, based on data-dependent eigenfunctions of a kernel-based operator
\citep{izbicki2016nonparametric}. 
Details can be found in \cite{izbicki2017photo} and \cite{izbicki2016nonparametric}. 
Previous studies \citep{izbicki2017photo,autenrieth2021stratifiedASA} indicate that each estimator appears to perform better in a different data regime,  
and combining them leads to a more robust estimator.\footnote{We refer the interested reader to the ensemble learning literature for further background \citep{wolpert1992stacked,van2007super,naimi2018stacked}.}

We individually optimize the conditional density estimators (ker-NN, Series) by minimizing (\ref{formula:fzxrisk_source}) in each source stratum separately. 
The final \textit{StratLearn} conditional density estimate is obtained by combining the ker-NN and Series conditional density estimates $\hat{f}_{\text{ker-NN}}(z|x) $ and $\hat{f}_{\text{Series}}(z|x)$ by optimizing 
\begin{align} \label{formula:comb}
        \hat{f}(z|x) =  (1 -\alpha) \hat{f}_{\text{Series}}(z|x) + \alpha \hat{f}_{\text{ker-NN}}(z|x)  ,  
\end{align}
with $0 \leq \alpha \leq 1$.
The parameter $\alpha $ is optimized 
to minimize the generalized risk
\begin{align} \label{formula:fzxrisk_source_and_target}
    \hat{R}_{S2}(\hat{f}) = & \frac{1}{n_T} \sum_{k=1}^{n_T} \int \hat{f}^2 (z|x_T^{(k)}) dz  -  2 \frac{1}{n_S} \sum_{k=1}^{n_S}  \hat{f} (z_S^{(k)}|x_S^{(k)}) 
\end{align} 
within each strata. 
We note that (\ref{formula:fzxrisk_source_and_target}) only differs from (\ref{formula:fzxrisk_source}) in that the first term is averaged over the photometric sample, $x_T$, rather than the spectroscopic sample, $x_S$, 
which does not require any target redshift $z_T$. Finally, (\ref{formula:comb}) provides a galaxy-by-galaxy full conditional density redshift estimate $\hat{f}(z_i|x_i)$.\footnote{Note that for better readability in (\ref{formula:fzxrisk_source}) and (\ref{formula:fzxrisk_source_and_target}) we use superscripts $(k)$ to enumerate objects, elsewhere we use subscripts $i$.}  
Some illustrative examples of the resulting galaxy conditional density estimates are shown in Figure~\ref{figure:fzx_examples}. Section~\ref{sec:technical_details} 
provides additional details on computation and parameter optimization of the conditional density estimators.


\subsection{Tomographic Bin Assignment}\label{sec:StratLearn_bin_assignment}

For cosmic shear analysis, the photometric galaxies are assigned to groups along the line-of-sight called tomographic bins. These bins are constructed with the best available proxy for the true line-of-sight distance of the galaxies. For wide-field photometric surveys, bins are typically constructed based on a redshift estimate  determined from broad-band photometry (called the photometric redshift estimate, or photo-$z$ estimate). The KiDS survey employs photometric redshifts estimated using the Bayesian-Photometric-Redshift code (BPZ; \citealt{benitez2000bayesian}), which constructs a posterior probability distribution of redshift given a source's observed photometry. This code produces a posterior mode point-estimate of photometric redshift, $z_B$, which is subsequently used for tomographic binning. \citetalias{wright2020photometric} 
assign the photometric galaxies to five non-overlapping top-hat photometric redshift bins: $(0.1,0.3], (0.3,0.5], (0.5,0.7], (0.7,0.9], (0.9,1.2]; $ we denote these ranges as bins 1 through 5, respectively. 
Galaxies with $z_B$ estimates outside of the five bin ranges $(z_B \leq 0.1$ and $z_B > 1.2)$ are discarded.

\begin{figure*}
\centering
\begin{minipage}[t]{.85\textwidth}
  \centering
    \includegraphics[width=0.32\columnwidth]{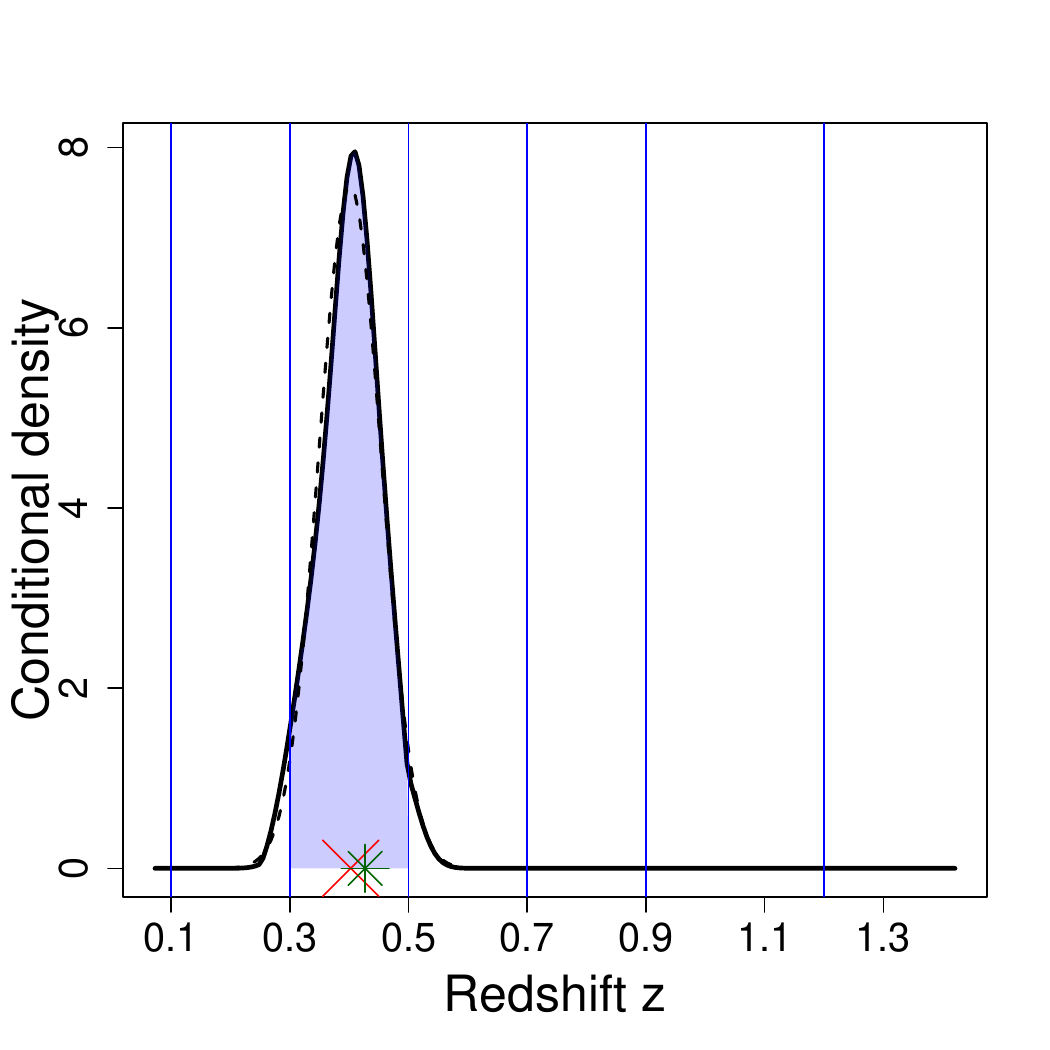}
    \includegraphics[width=0.32\columnwidth]{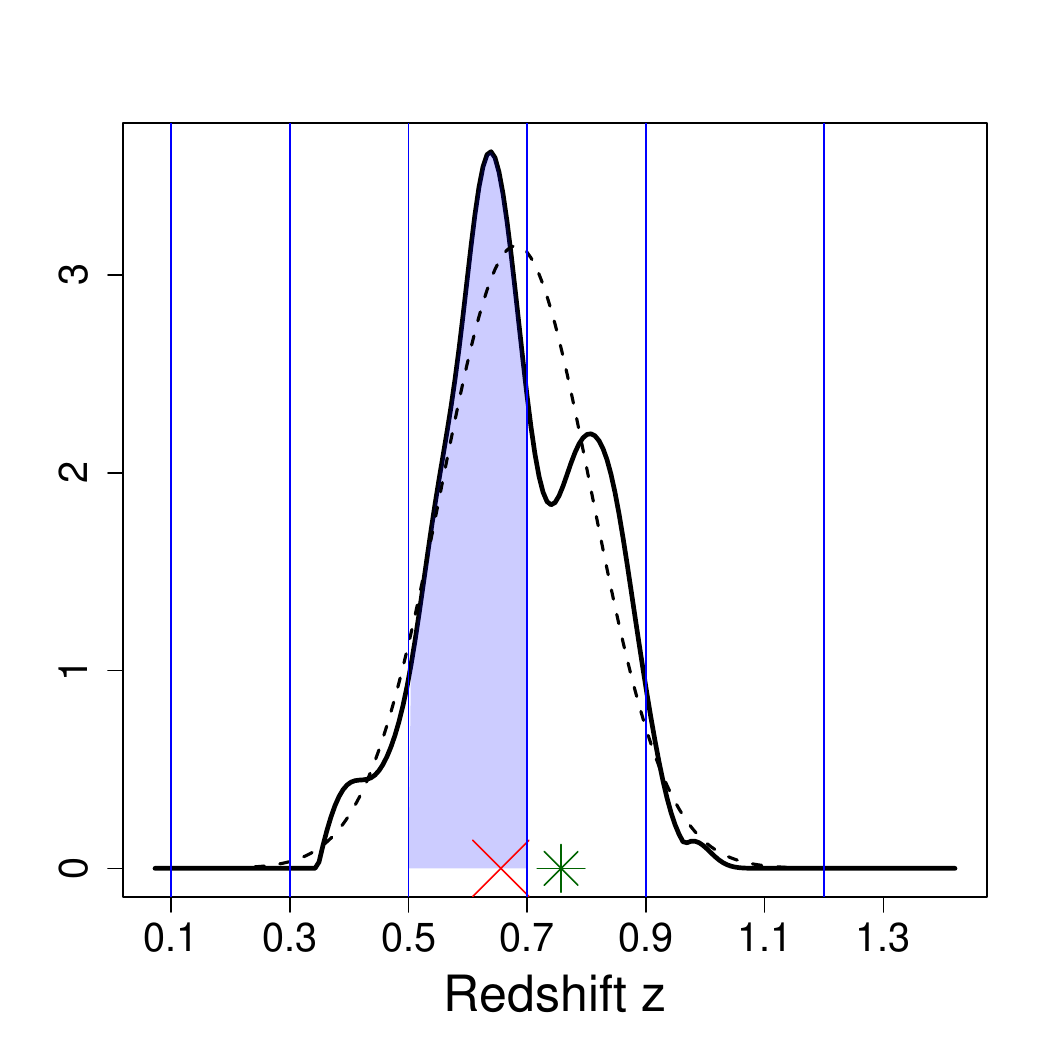}
    \includegraphics[width=0.32\columnwidth]{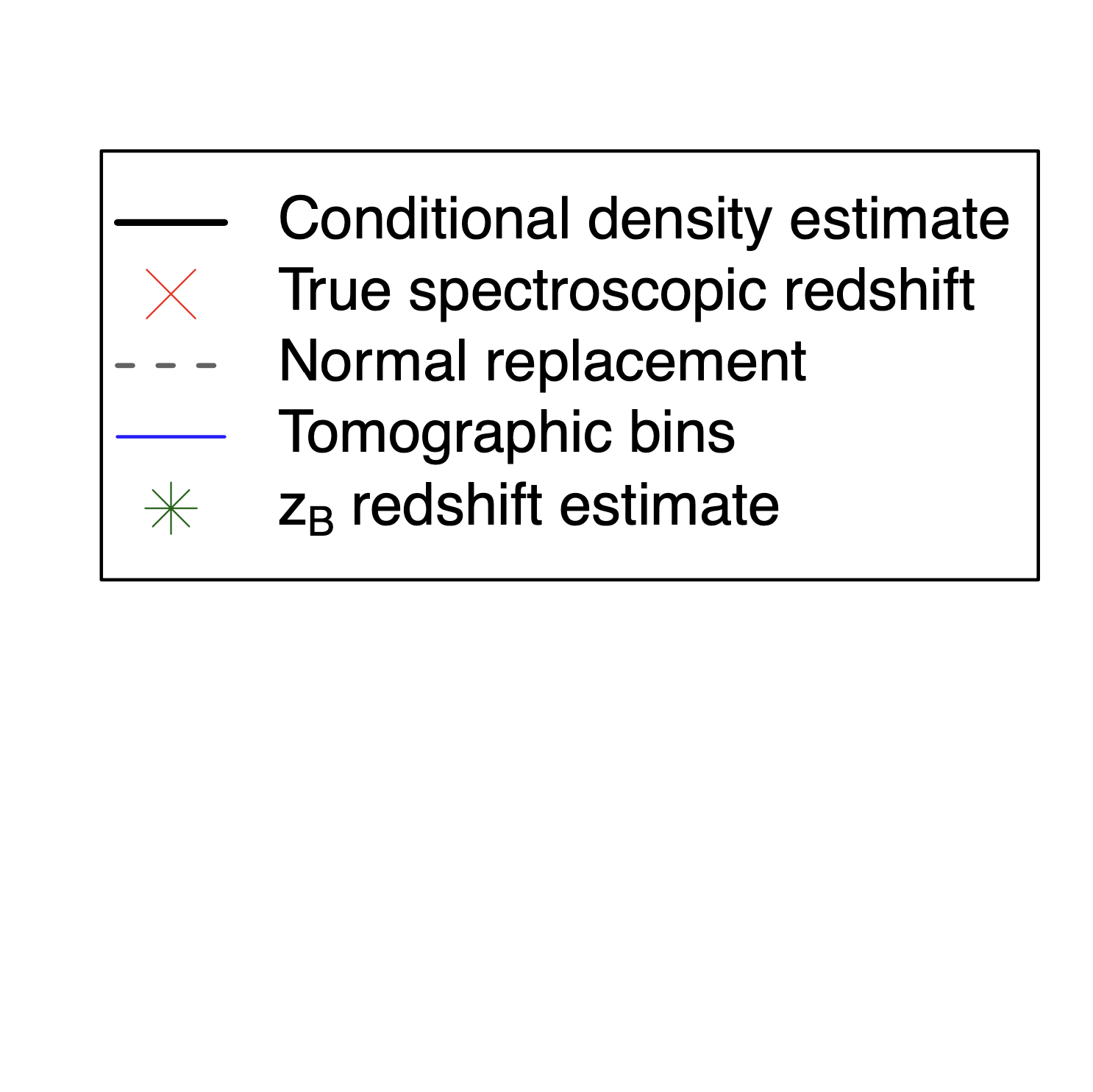}
        \includegraphics[width=0.32\columnwidth]{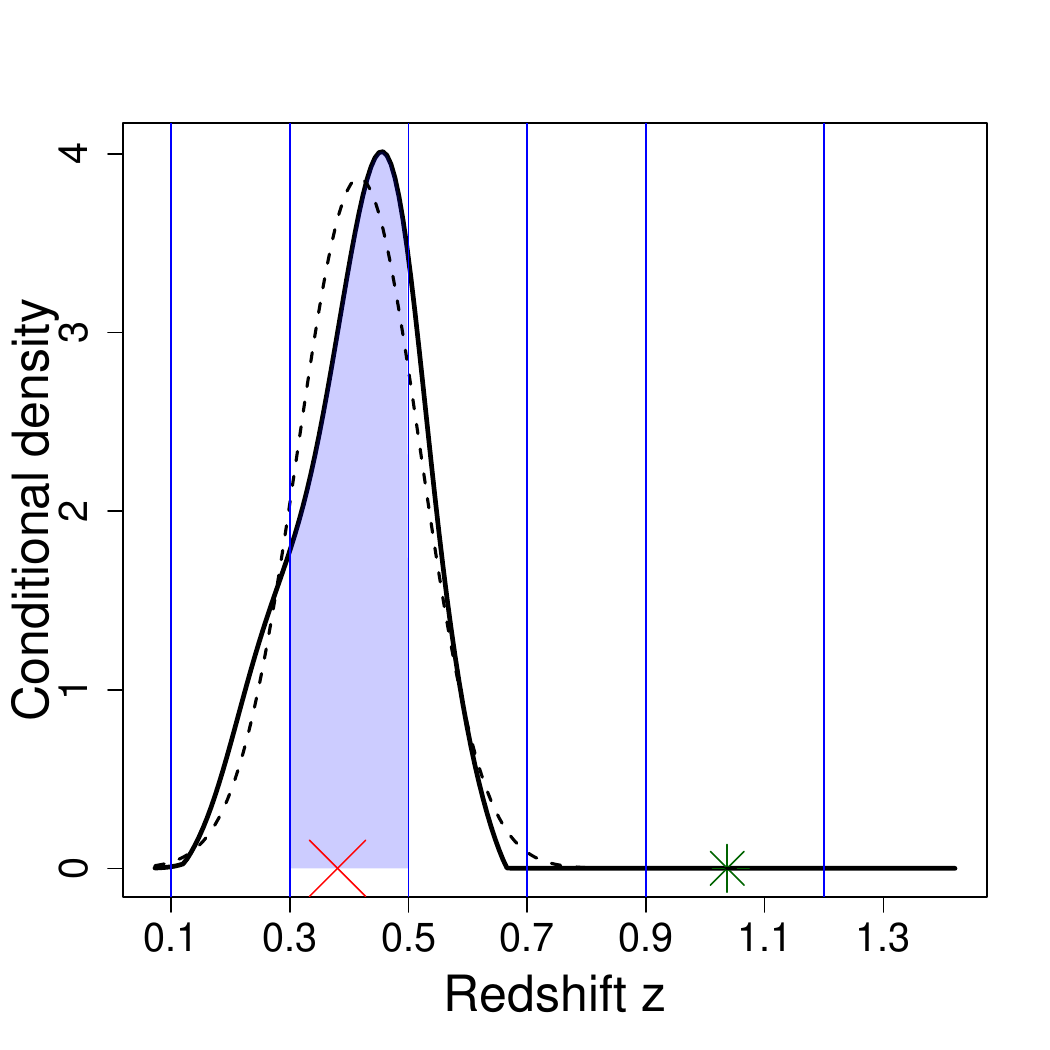}
        \includegraphics[width=0.32\columnwidth]{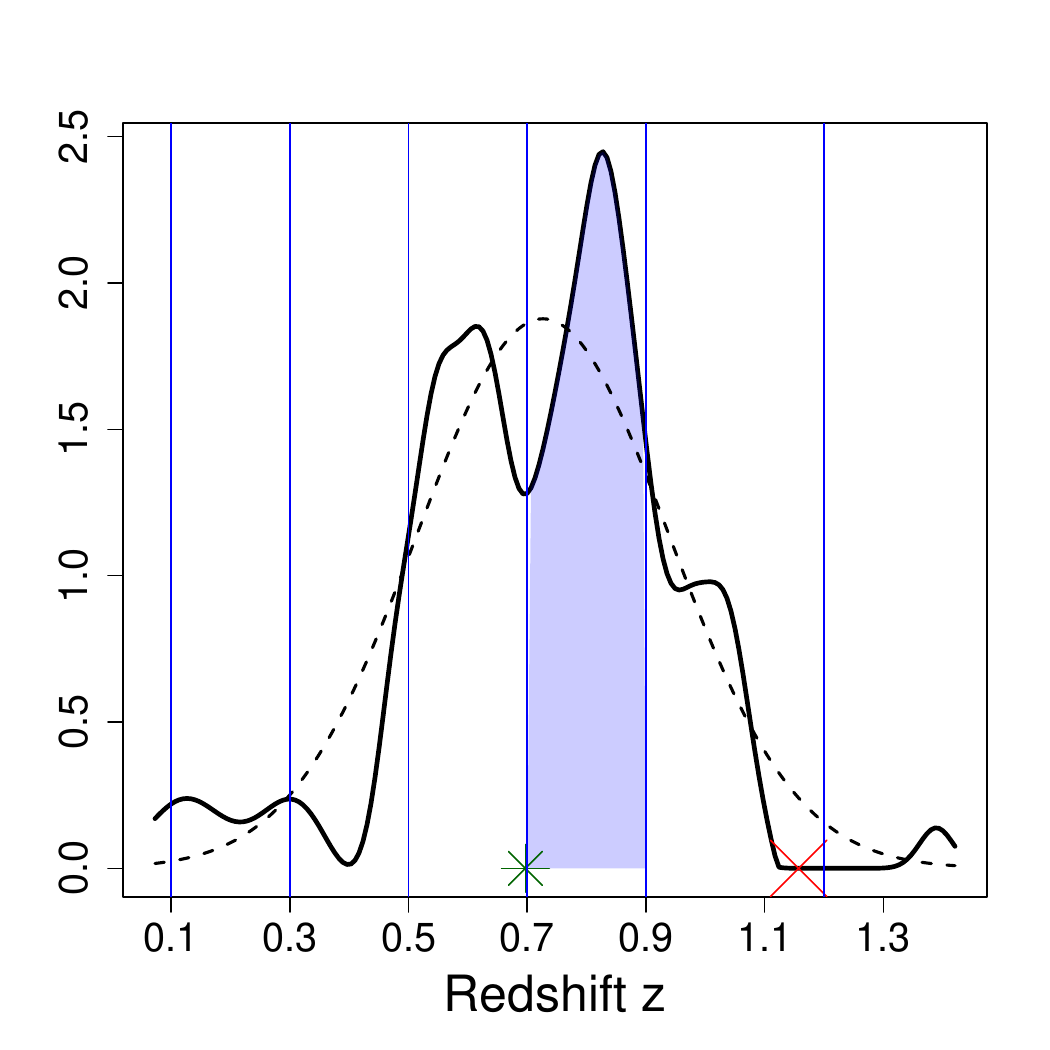}
        \includegraphics[width=0.32\columnwidth]{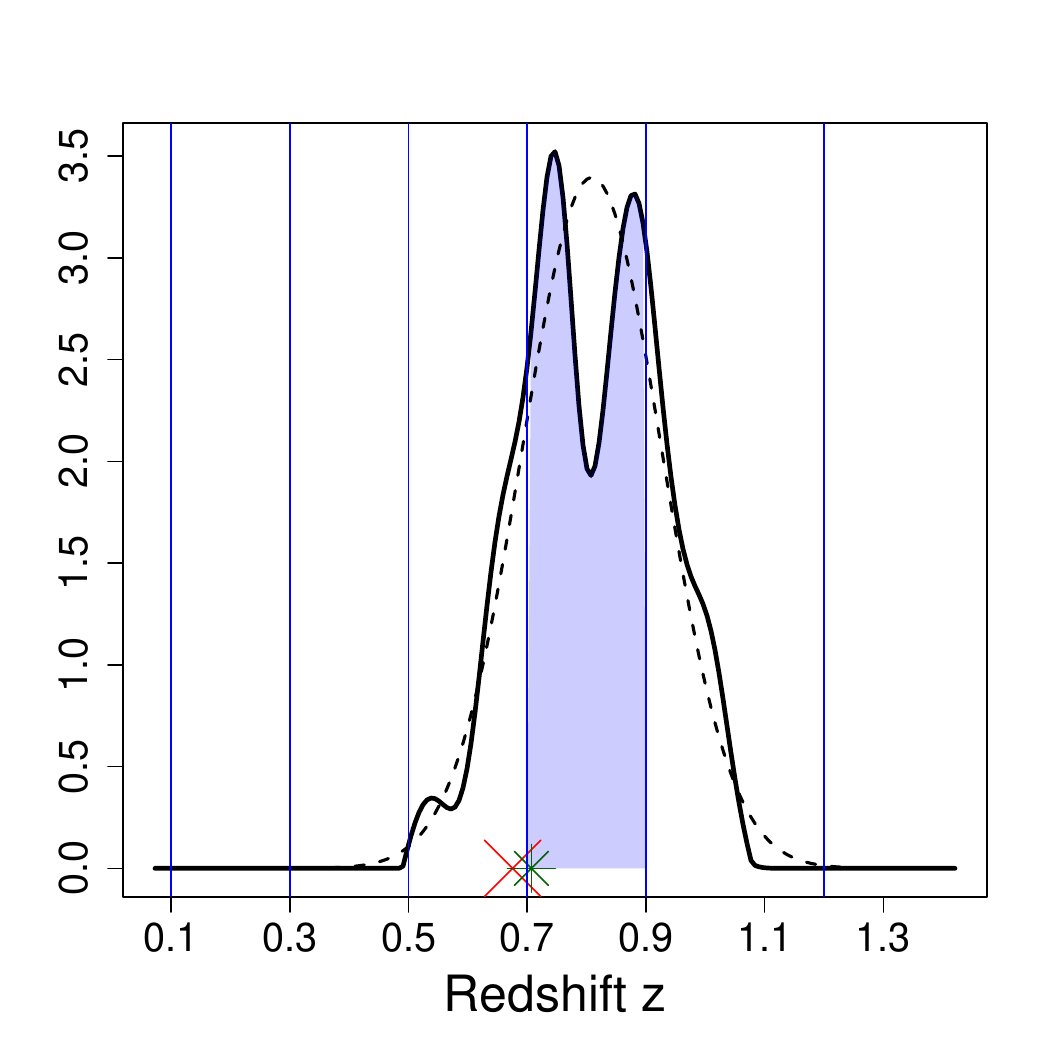}
        \includegraphics[width=0.32\columnwidth]{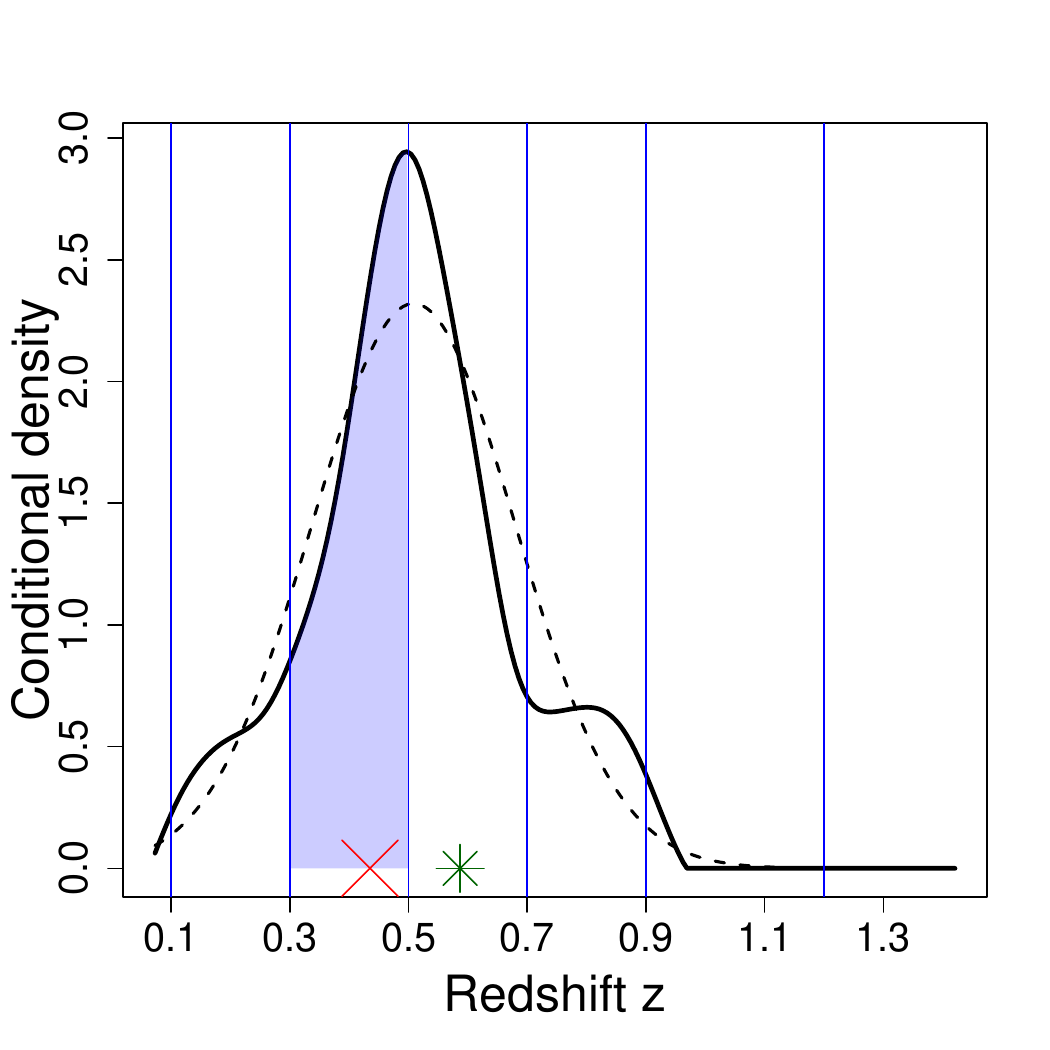}
        \includegraphics[width=0.32\columnwidth]{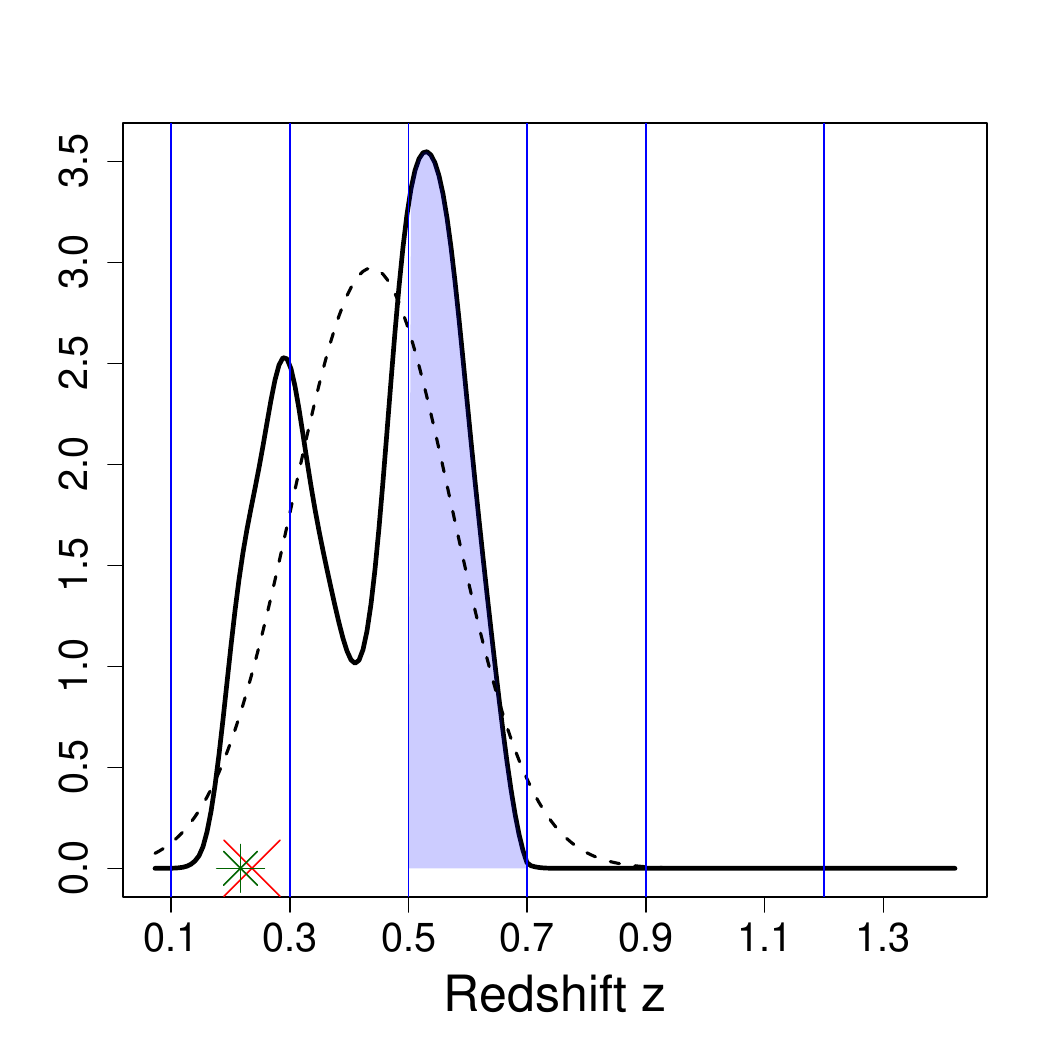}
        \includegraphics[width=0.32\columnwidth]{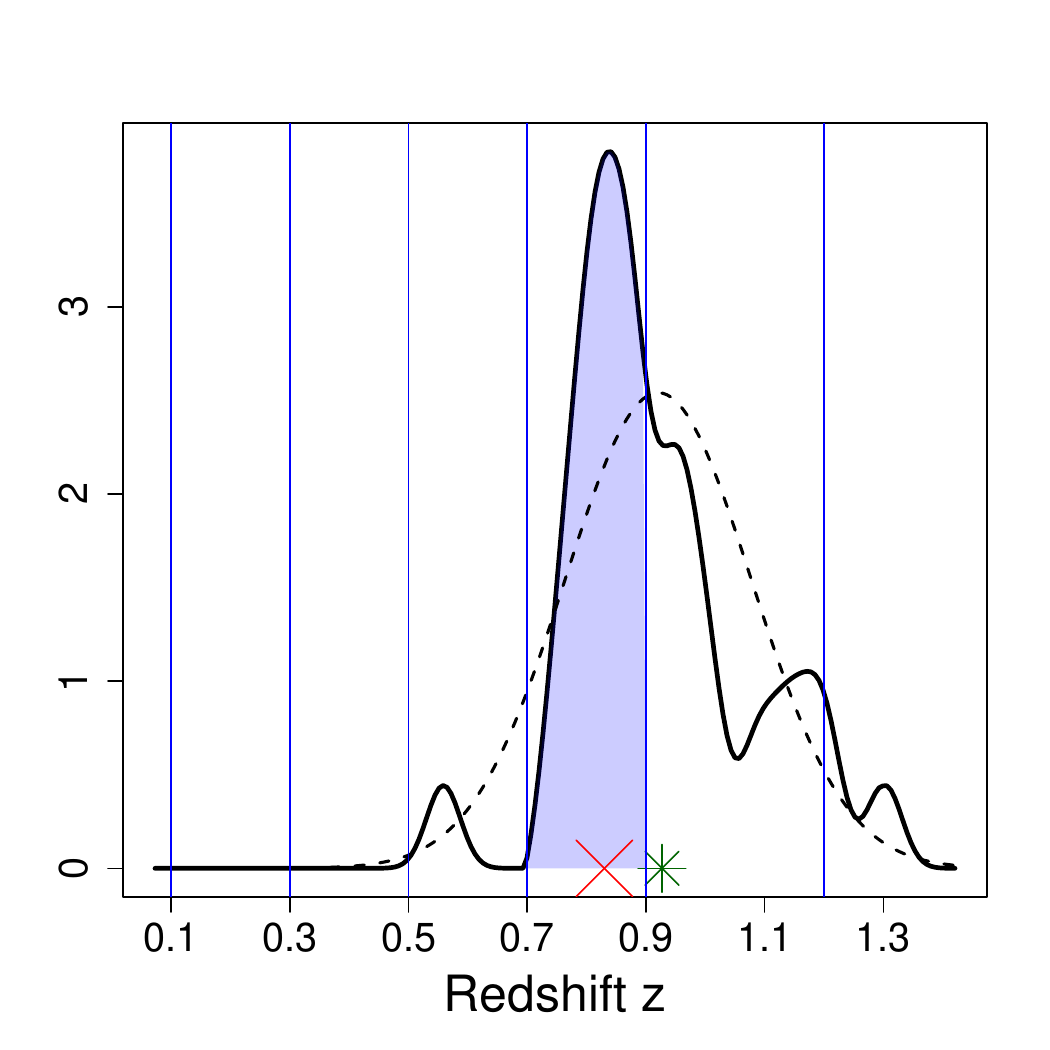}
    \caption{\baselineskip=15pt Examples of the  conditional density estimates $\hat{f}(z_i|x_i)$, for galaxies in the photometric target samples, illustrated on the tomographic bin grid. The \textit{StratLearn} assigned bin (the one containing the highest conditional probability) is shaded in blue. The true spectroscopic redshift is shown by the red cross. The $z_B$ estimate is shown by the green star.
    A fraction of $30 \%$ of the conditional density estimates appear to  be roughly bell-shaped 
    like in the top left example, but many conditional densities can be skewed and multimodal. Normal distributions with the same means and variances as the conditional density examples $\hat{f}(z_i|x_i)$ are added as dashed lines (as discussed in Section~\ref{sec:Bayesian_model}.)  \label{figure:fzx_examples}} 
\end{minipage}
\end{figure*}

A central step of our proposed photo-$z$ calibration method is the computation of galaxy-by-galaxy full conditional density redshift estimates $\hat{f}(z_i|x_i)$. 
Instead of relying on $z_B$, we can therefore use the full conditional density estimates $\hat{f}(z_i|x_i)$ to provide an alternative tomographic bin assignment for each galaxy $i$. A natural choice is to assign each galaxy to the bin which contains the highest conditional probability: let $b(i)$ be the bin assignment of galaxy $i$, then
\begin{align}
    b(i) = \text{argmax}_m \int_{B(m)} \hat{f}(z_i |x_i) dz, \quad m= 1, \dots, 5,l,r,  
\end{align}
with $B(i)$, $i=1,\dots,5$ specifying the five tomographic bin redshift ranges. $B(l):=\{z|z \leq 0.1\}$ and $B(r):=\{z|z>1.2\}$ specify two end bins for galaxies outside of the bin ranges; galaxies assigned to the end bins are not used in the analysis. Figure~\ref{figure:fzx_examples} shows examples of conditional density estimates, plotted on the tomographic redshift bin grid. The assigned bins $b(i)$ with highest conditional probabilities are shaded.

\subsection{Bayesian Hierarchical Modelling of Conditional Densities} \label{sec:Bayesian_model}

In this section, we detail our Bayesian hierarchical framework for accurate estimation of the redshift population mean within each tomographic bin, given the object-level (galaxy) conditional density estimates. 
Employing a hierarchical Bayesian framework allows us to model the conditional density estimates in a statistically principled framework,  with optimal shrinkage on the object-level 
photo-$z$ estimates, allowing more precise population mean estimates. 
Figure \ref{figure:BHM_DAG} provides an overview of our hierarchical Bayesian framework, with details described hereafter.

\newcommand\Square[1]{+(-#1,-#1) rectangle +(#1,#1)}
\tikzstyle{square} = [Square, draw, text centered, minimum height=2em, text width=20em]
\tikzstyle{terminator} = [rectangle, draw, text centered, minimum height=2em, text width=20em]
\tikzstyle{block} = [rectangle, draw, text centered, text width=20em, minimum height=2em]
\tikzstyle{block_m} = [rectangle, draw, text centered, text width=20em, minimum height=2em] 

\tikzstyle{connector} = [draw, -latex']
\tikzstyle{decision} = [diamond, draw, text badly centered, minimum height=2.2em, inner sep=0pt, text width=4.3em]
\tikzstyle{descr} = [fill=black, inner sep=-0pt]
\tikzstyle{latent} = [circle, draw, text centered, text width=2em, minimum height=2.5em, line width=0.35mm]
\tikzstyle{data} = [rectangle, fill=gray!40, draw, text centered, text width=2em, minimum height=2em, line width=0.35mm]
\tikzstyle{data_wide} = [rectangle, fill=gray!40, draw, text centered, text width=4em, minimum height=2em, line width=0.35mm]
\tikzstyle{outer_box} = [rectangle,  draw, text centered, text width=15em, minimum height=16em, thick] 
\tikzstyle{outer_box2} = [rectangle,  draw, text centered, text width=15em, minimum height=15em, thick] 

\begin{figure}
\begin{center}
\begin{tikzpicture}

    \node [latent] at (-3,0) (pop_mean) {$\mu_b$};
    \node [latent] at (-1,0) (pop_sd) {$\sigma_b^2$};
    \node [data] at (0.5,0) (pop_sd_est) {$\hat{\sigma}_b^2$};
    
    \node [latent] at (-2,-1.5) (latent_z) {$z_i$};
    \node [data_wide] at (-2,-3) (fzx) {$\hat{f}(z_i|x_i)$};
    \node [data] at (-0.25,-3) (x_bands) {$x_i$};
    \node [data] at (-3,-4.5) (obj_mean) {$\hat{\xi}_i$};
    \node [data] at (-1,-4.5) (obj_sd) {$\hat{\tau}_i^2$};
    \node [outer_box2] at (-1.65,-2.95) (for_box) { };
    \node[draw=none] at (-0.55, -1.15) (for_loop) {$i= 1,\dots,n_{Tb}$ };

    \draw[->, thick] (pop_sd) node[]{} -- (pop_sd_est) node[]{};
    \draw[->, thick] (pop_mean) node[]{} -- (latent_z) node[]{};
    \draw[->, thick] (pop_sd) node[]{} -- (latent_z) node[]{};
    \draw[->, thick] (latent_z) node[]{} -- (fzx) node[]{};
    \draw[->, thick] (x_bands) node[]{} -- (fzx) node[]{};
    \draw[->, thick] (fzx) node[]{} -- (obj_mean) node[]{};
    \draw[->, thick] (fzx) node[]{} -- (obj_sd) node[]{};
\end{tikzpicture}%
\end{center}
\captionof{figure}{Graphical representation of our Gaussian hierarchical Bayesian model for the estimation of the redshift population  mean in each tomographic bin, based on (summaries) of the photo-$z$ conditional density estimates $\hat{f}(z_i|x_i)$. Observed quantities are illustrated in shaded squares. Unobserved parameters are illustrated via circles. 
\label{figure:BHM_DAG}
}
\end{figure}
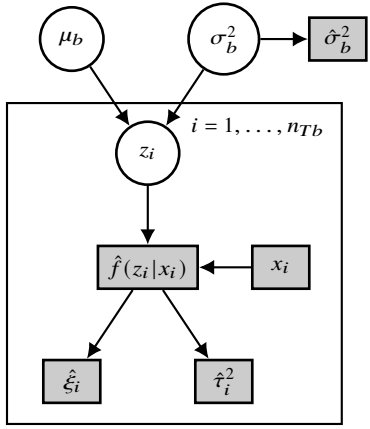

On the object (galaxy) level,
$\hat{f}(z_i|x_i)$ is an estimate of the conditional density $p(z_i|x_i)$. 
Via Bayes theorem, the conditional density $p(z_i|x_i)$ can be expressed as
\begin{align}\label{form:object_Bayes_uniform}
 p(z_i|x_i) \propto p(x_i |z_i) p(z_i).
\end{align}
The estimation of the conditional densities $\hat{f}(z_i|x_i)$ is performed before and outside of the hierarchical Bayesian model fit (as described in Section~\ref{sec:StratLearn}) and without incorporation of prior information on the redshift distributions. 
By assuming a flat prior on $z_i$ (e.g., a wide uniform prior that covers the expected photometric 
redshift range),\footnote{We note that in future work a more informative prior on the object level redshift distributions could principally be included via our hierarchical Bayesian model in (\ref{form:joint_posterior2}). } 
we have $p(z_i) \propto 1$. Then,  (\ref{form:object_Bayes_uniform}) simplifies to 
\begin{align}\label{form:switch_condiionals}
    p(z_i|x_i) \propto p(x_i |z_i).
\end{align}

On the population level, 
recall that we aim to accurately estimate the population mean $\mu_b$ of $z_i$ ($i =1,\dots,n_{Tb}$, with $n_{Tb}$ being the number of galaxies within tomographic bin $b$). Accurate estimation of $\mu_b$ is crucial to avoid systematic biases in the downstream cosmic shear analysis \citep{amara2008systematic,Reischke2023}. 
Thus, we model the redshift population within each bin with a normal distribution -- a convenient choice that facilitates the introduction of a hierarchical Bayesian framework 
and a reasonable simplification given that we are primarily interested in the population mean.  
Specifically, at the redshift population level within bin $b$, we model
\begin{align}
    \text{Population Level:} \quad z_i| \mu_b, \sigma_b \overset{\text{indep.}}{\sim} N(\mu_b, \sigma_b^2), \label{form:pop_distribution}
\end{align}
with $\sigma_b^2$ being the redshift population variance.

Thus, we formulate the joint posterior distribution $p(z_1, \dots, z_{n_{Tb}}, \mu_b, \sigma_b | X_{\bold{{n_{Tb}}}})$, with $X_{\bold{{n_{Tb}}}}:=\{x_i\}_{i=1}^{n_{Tb}}$, 
via 
\begin{align}
    p(z_1, &\dots, z_{n_{Tb}},  \mu_b, \sigma_b | X_{\bold{{n_{Tb}}}}) \nonumber\\
    \propto &p(X_{\bold{{n_{Tb}}}}|z_1, \dots, z_{n_{Tb}}, \mu_b, \sigma_b ) 
    p(z_1, \dots, z_{n_{Tb}} | \mu_b, \sigma_b ) p(\mu_b, \sigma_b ) \label{form:joint_posterior} \\
     = &p(\mu_b, \sigma_b ) \prod_i p(x_i|z_i) p(z_i| \mu_b, \sigma_b )\label{form:joint_posterior2}\\
     \propto &p(\mu_b, \sigma_b ) \prod_i p(z_i|x_i) p(z_i| \mu_b, \sigma_b ), \label{form:joint_posterior3}
\end{align} 
where (\ref{form:joint_posterior}) to (\ref{form:joint_posterior2}) holds due to the independence in (\ref{form:pop_distribution}) and the conditional independence $x_i \Perp  (\{z_j\}_{j \neq i}, \mu_b, \sigma_b) | z_i$. That is, given the redshift $z_i$ for an object $i$, the distribution of its photometry $x_i$ does not depend on other observed redshifts, nor on the parameters describing the population of redshift. (\ref{form:joint_posterior2}) to (\ref{form:joint_posterior3}) follows from (\ref{form:switch_condiionals}).
Since we are not targeting the object-level redshifts $z_i$ themselves, but rather an accurate estimate of the population-level mean, $\mu_b$, we integrate over the individual galaxies' redshifts to obtain the marginal posterior distribution
\begin{align} \label{form:marginal_posterior} 
    p(\mu_b, \sigma_b | X_{\bold{{n_{Tb}}}})  & \propto p(\mu_b, \sigma_b ) \prod_i \int p(z_i|x_i) p(z_i| \mu_b, \sigma_b ) dz_i. 
\end{align}

\paragraph*{Replacing $p(z_i|x_i)$ with a Gaussian approximation:}
Although we could substitute the estimates, $\hat{f}(z_i|x_i)$, of $p(z_i|x_i)$ directly into (\ref{form:marginal_posterior}), the required integrals would be computationally expensive. 
Instead, 
we simplify the problem by modelling each $p(z_i|x_i)$ with a normal distribution:
\begin{align}\label{form:object_level_normal}
    z_i |x_i \overset{\text{indep.}}{\sim} N(\hat{\zeta}_i, \hat{\tau}_i^2),
\end{align}
where the estimate of the object-level mean $\hat{\zeta}_i$ is simply the mean of the conditional density estimate, $\hat{f}(z_i|x_i)$, while the object-level Gaussian variance $\hat{\tau}_i^2$ is obtained by computing the variance of $\hat{f}(z_i|x_i)$.  
More precisely, by treating $\hat{f}(z_i|x_i)$ as a histogram evaluated on $K$ bins, we have
\begin{align}
    \hat{\tau}_i^2 = \frac{1}{\sum_k{\hat{f}_{m_k}(z_i|x_i)}} \sum_k \hat{f}_{m_k}(z_i|x_i)(m_k - \hat{\zeta}_i)^2,
\end{align}
where each $k = 1, \dots, K$ specifies a histogram bin with location $m_k$, and $\hat{f}_{m_k}$ is the value of the conditional density (histogram) at location $m_k$.\footnote{We note that by bins $k$ (with locations $m(k)$), we refer to the density (histogram) bins and not the tomographic redshift bins.}
The summary statistics $\hat{\zeta}_i$ and $\hat{\tau}_i^2$ are observed quantities 
summarized by $\hat{X}_{\bold{{n_{Tb}}}}:=\{\hat{\zeta}_i, \hat{\tau}_i \}_{i=1}^{n_{Tb}}$.

There are two reasons behind our replacement of the conditional density estimates $\hat{f}(z_i|x_i)$ by normal distributions.  First, by modeling both the population- and object-level distributions as Gaussians, the Bayesian posterior distribution for the population mean can be calculated analytically. This allows us to scale our model to the large photometric data set at hand. 
Second, and more importantly, modelling the conditional densities as Gaussians leads to almost unbiased estimates of the population means in all tomographic bins, as demonstrated in Section~\ref{sec:results_bias}.\footnote{In our simulations, we also investigated a hierarchical model based on the conditional densities $\hat{f}(z_i|x_i)$ directly (without Normal replacement). In this case, we obtained the posterior distributions via MCMC sampling using only a small fraction of the target data due to computational limitations. Surprisingly, (given the results from the subset) the Normal-Normal model led to better estimates of the population means than using the conditional density estimates directly. This effect was consistently observed with many simulation settings, e.g., using various subsets of the (non-weighted) full photometric sample (from the blue dashed distribution in Figure~\ref{fig:resample_vs_reweight}), and for different subsets of the weighted full photometric sample (from the green dashed distribution in Figure~\ref{fig:resample_vs_reweight}), demonstrating some robustness w.r.t. the underlying simulation.}

With this approximation, 
(\ref{form:marginal_posterior}) 
can be written as
\begin{align}\label{form:joint_posterior_gaussian}
    p(\mu_b, \sigma_b | \hat{X}_{\bold{{n_{Tb}}}})  & \propto p(\mu_b, \sigma_b ) \prod_i \int 
    N(z_i| \hat{\zeta}_i, \hat{\tau}_i^2)  N(z_i | \mu_b, \sigma_b^2)
    dz_i, 
\end{align}
where $N(t| \theta, \phi^2)$ is the probability density function (pdf) of a normal distribution with mean $\theta$ and variance $\phi^2,$ evaluated at  $t$.

The integral in (\ref{form:joint_posterior_gaussian}) can be solved analytically (see Section~\ref{supp_sec:add_model_details} in the Appendix for theoretical justifications) to obtain the (joint) marginal posterior density 
\begin{align}\label{form:joint_marginal}
    p(\mu_b, \sigma_b | \hat{X}_{\bold{{n_{Tb}}}})  & \propto p(\mu_b, \sigma_b ) \prod_i
    N(\hat{\zeta}_i | \mu_b, \hat{\tau}_i^2 + \sigma_b^2).
\end{align}

Writing $p(\mu_b, \sigma_b ) = p(\mu_b|\sigma_b)p(\sigma_b)$ and adopting a uniform conditional prior density $p(\mu_b|\sigma_b) \propto 1$, yields the conditional posterior distribution of $\mu_b$ given $\sigma_b$:
\begin{align}\label{form:post_mu_cond_sigma}
    \mu_b | \sigma_b, \hat{X}_{\bold{{n_{Tb}}}} \sim N(\tilde{\mu}_b, V_{\mu_b}), 
\end{align}
with
\begin{align}\label{form:post_mu_cond_sigma_params}
\tilde{\mu}_b = \frac{\sum_i \frac{1}{\hat{\tau}_i^2 + \sigma_b^2} \hat{\zeta}_i }{\sum_i \frac{1}{\hat{\tau}_i^2 + \sigma_b^2} }  
\quad \text{and } V_{\mu_b}^{-1} = \sum_i \frac{1}{\hat{\tau}_i^2 + \sigma_b^2},
\end{align}
with $ V_{\mu_b}^{-1}$ being the total precision. 

Since we are not interested in the posterior uncertainty of $\sigma_b$, we choose an empirical Bayesian approach by setting $\sigma_b$ to a fixed value estimated from the data, i.e., by choosing $p(\sigma_b) = \delta(\sigma_b - \hat{\sigma}_b)$. An obvious choice for the estimate $\hat{\sigma}_b$ is the MAP of the marginal posterior $p(\sigma_b| \hat{X}_{\bold{{n_{Tb}}}})$ (shown in the Appendix (\ref{supp_form:posterior_sigma})). However, in our simulations, we found that the MAP estimate strongly and consistently underestimates $\sigma_b$. We thus choose a different strategy to estimate $\sigma_b$ from the data as detailed below. 

Finally, given an estimate of $\sigma_b$, an estimate of $\mu_b$ can be obtained analytically via (\ref{form:post_mu_cond_sigma_params}), as $\tilde{\mu}_b$, the MAP estimate of $\mu_b$.

\paragraph*{Population variance estimation via stacking of conditional densities:}

Given the poor performance of the MAP estimate for $\sigma_b$, we instead estimate the population variance $\sigma_b^2$ via a ‘‘stacked estimate" of the marginal redshift population distribution $p_b(z)$ of galaxies within tomographic bin $b$. More precisely, we obtain an estimate $\hat{p}^{\text{stack}}_b(z)$ of $p_b(z)$ by averaging (stacking) the conditional densities within bin, that is, 
\begin{align}\label{form:stack_marginal_est}
    \hat{p}^{\text{stack}}_b(z) = \frac{1}{n_{Tb}} \sum_j \hat{f}(z_j |x_j).
\end{align}
with $x_j, j = 1,\dots,n_{Tb}$, being the photometric magnitudes of the observed galaxies within tomographic bin $b$. While quite intuitive, the form of (\ref{form:stack_marginal_est}) is justified more formally in Section~\ref{section:append_stacking_justification}. 

An estimate for the redshift population variance $\sigma_b^2$ can then be obtained by calculating the variance of $\hat{p}^{\text{stack}}_b(z)$, via
\begin{align}\label{form_stacked_variance_estimate}
    \hat{\sigma}_b^2 = \frac{1}{\sum_k{\hat{p}^{\text{stack}}_{b,m(k)}(z)}} \sum_k\hat{p}^{\text{stack}}_{b,m(k)}(z)(m(k) - \hat{\mu}_b^{\text{stack}})^2,
\end{align} 
where $k = 1, \dots, K$ specifies the (density/histogram) bin with location $m_k$, $\hat{p}^{\text{stack}}_{b,m(k)}$ is the value of the stacked density (histogram) $\hat{p}^{\text{stack}}_{b}$ of bin $b$ at location $m(k)$, and $\hat{\mu_b}^{\text{stack}}$ is the mean of $\hat{p}_b^{\text{stack}}(z)$. 
An estimate of the population standard deviation $\sigma_b$ is then obtained by simply taking the square-root of (\ref{form_stacked_variance_estimate}). 

We compare two versions of (\ref{form_stacked_variance_estimate}). First, we compute (\ref{form_stacked_variance_estimate}) via stacking over $\hat{f}(z_i|x_i)$, the galaxy conditional density estimates obtained via \textit{StratLearn}. We denote this method as option \textit{StratLearn}-Bayes (A). Second, we substitute the conditional density estimates $\hat{f}(z_i|x_i)$ by their normal replacements 
described in (\ref{form:object_level_normal}). We denote this option as \textit{StratLearn}-Bayes (B).

\section{Estimating the Population Distribution via Inverse-Propensity Score Weighting}\label{sec:inverse-PS}

As we demonstrate below, our hierarchical Bayesian framework delivers highly accurate and precise estimates of the redshift means within each tomographic bin, the quantity of main interest. Its use of Gaussian distributions for the redshift populations, however, precludes realistic distribution shapes. Here we propose a different approach for estimation of the redshift population distributions, where we use propensity scores for direct redshift calibration, thereby yielding an estimate of the full tomographic redshift distribution.

As described in Section~\ref{sec:related_literature}, direct redshift calibration methods depend on the estimation of weights $\omega(x) = p_T(x)/p_S(x)$, used to reweight a spetroscopic sample (with known true redshifts) to obtain an estimate of the redshift distribution of the photometric sample (per tomographic bin).
The weights $\omega(x)$ can also be expressed via 
\begin{equation} \label{form:weights_PS_spectro}
 \omega(x) = \frac{p_T(x)}{p_S(x)} = \frac{p(s = 1)}{p(s = 0)} \frac{p(s = 0 | x) }{p(s = 1 | x) } \propto \left( \frac{1 }{p(s = 1 | x) } -1  \right).
\end{equation}
We can thus obtain an estimate of the weights $w(x)$ by employing the inverse of the propensity score (inverse-PS), via the right-hand side of (\ref{form:weights_PS_spectro}). 
To estimate the tomographic binned redshift distributions, we first 
obtain a \textit{StratLearn} conditional density estimate $\hat{f}(z_i|x_i)$ as described in Section~\ref{sec:StratLearn} for each galaxy in both the photometric and the spectroscopic set. Based on these estimates, each galaxy (in both sets) is assigned to its respective tomographic bin, following the \textit{StratLearn} binning strategy described in
Section~\ref{sec:StratLearn_bin_assignment}.\footnote{We note that the binning of the spectroscopic set (as previously performed by \citetalias{wright2020photometric}) is needed when performing direct redshift calibration of the binned photometric set, since the photometric bin assignment is based on (summaries) of the conditional density estimates (as described in Section~\ref{sec:StratLearn_bin_assignment}). 
The conditional density estimates implicitly incorporate information of source redshift (through the fitting process described in Section~\ref{sec:StratLearn}). To prevent unmeasured confounding (information encoded in the spectroscopic redshift, but not in the magnitudes/colors) the same selection function is applied for source and target data. }

For each tomographic bin, following (\ref{form:joint_weighted_source3}), we obtain an estimate of the binned joint target distribution $p_{Tb}(z,x)$ via 
\begin{align}\label{form:binned_direct_calibration}
p_{Tb}(z,x) = \omega_b(x) p_{Sb}(z,x),    
\end{align}
with  $ p_{sb}(z,x)$ being the binned joined source distribution of tomographic bin $b$, and with weights $\omega_b(x)$ computed via inverse-PS following (\ref{form:weights_PS_spectro}) for each bin $b$.
We estimate the propensity scores (employing logistic regression as detailed in Section~\ref{sec:data})  
based on the covariates of the spectroscopic source galaxies and photometric target galaxies in the respective tomographic bin.   
In practice, we employ the relation in (\ref{form:binned_direct_calibration}) by reweighting galaxies in the binned spectroscopic source data using the respective estimated inverse-PS weights (obtained via (\ref{form:weights_PS_spectro})). 
We then obtain an estimate of the photometric redshift distribution $\hat{p}_b(z)$ (for each tomographic bin $b$) 
by looking at the marginal sample of $z$ in the weighted joint distribution. 
This method is numerically demonstrated in Section~\ref{section:population_shapes_results}.

\section{Numerical Demonstrations}\label{sec:numerical_demonstrations}

\subsection{Simulation Study}
\label{sec:data}
We explore the performance of our framework using the comprehensive set of realistic simulations introduced in \citetalias{wright2020photometric}. 
The simulations aim to mimic the KiDS+VIKING-450 dataset, presented in 
\cite{wright2019kids+} and \cite{hildebrandt2020kids+}, starting from the MICE2 simulation \citep{fosalba2015micea,crocce2015mice,carretero2015algorithm,hoffmann2015measuring} and based on a framework provided in \cite{van2020testing}.
In the following, we provide a summary of the simulated data employed in our study; see \citetalias{wright2020photometric} for a full description of the construction and validation of the simulations.

\paragraph*{Photometric Survey:} 
The simulations are designed to mimic the wide-field, multi-band photometric dataset of KiDS+VIKING-450. 
The KiDS+VIKING-450 dataset consists of imaging in nine photometric bands ($ugriZYJHK_{\rm s}$): the four optical bands (${ugri}$) are observed as part of the KiDS survey \citep{kuijken+2019} using the VLT Survey Telescope \citep[VST;][]{capaccioli+2005} located at the European Southern Observatory's Cerro Paranal observatory in Chile, and the five near-infrared filters ($ZYJHK_{\rm s}$) are observed as part of the VIKING survey \citep{edge+2013} using the Visible and Infrared Survey Telescope for Astronomy \citep[VISTA;][also located at Cerro Paranal]{emerson+2006,dalton+2006}. 
The first 450 square degrees of joint imaging from the two surveys forms the KiDS+VIKING-450 cosmic shear survey (referred to simply as `KiDS' hereafter).  
The simulated photometric data $D^{u}_T$ (where the superscript $u$ refers to ``unweighted''; below we describe a pre-processing step to produce a shear-measurement weighted photometric sample as employed in the downstream scientific analysis) 
consists of ${\sim}21 \times 10^6$ galaxies, for each of which a simulated measurement of its position, lensing convergence, morphological information, 
and model magnitudes in the $\textit{ugriZYJHK}_s$-bands is provided. Magnitudes include photometric noise, realistic to KiDS survey data \citepalias[][Section \ref{sec:data}]{wright2020photometric}.  
A large proportion of galaxies $({\sim} 17\%)$ have a flux error greater than or equal to the flux measurement in at least one band, and are therefore flagged as `non-detections' in the KiDS photometric processing pipeline.  
Figure~\ref{figure:missing_magnitudes_and_colors} in the Appendix illustrates the full pattern of such cases.
The flux measurement of such non-detections was removed prior to our analysis and only placeholder/indicator values were available to indicate these non-detection cases. 
We thus treat these cases as ``missing data" (details on processing of these cases appear at the end of this section). 
While the spectroscopic redshift $z$ is  unavailable for galaxies in the photometric set, a Bayesian redshift estimate $z_B$ is also provided (see Sect. \ref{sec:StratLearn_bin_assignment}), which was used in previous works to assign galaxies to the tomographic bins. 

Finally, a cosmic shear-measurement weight $\hat{w}_i$ is provided for each galaxy, which relates to the quality of its shear measurement, and which filters through to the cosmological analysis for cosmological parameter estimation. 
As a preprocessing step, 
we resample the data $D^{u}_{Ti}$ proportionally to the cosmic shear-measurement weights $\hat{w}_i$. In this way, we obtain the target sample $D_T$, with  $|D_T|=12.48 \times 10^6$ galaxies. 
Figure~\ref{fig:resample_vs_reweight} in the Appendix demonstrates that there is a negligible difference in targeting the shear-measurement resampled distribution (obtained via the above preprocessing step) and targeting the shear-measurement weighted distribution (as done in \citetalias{wright2020photometric}). Our results on the resampled data thus hold without loss of generality.

\paragraph*{Spectroscopic Survey:} 

\begin{figure}
\centering
\begin{minipage}{.95\columnwidth}
  \centering
    \includegraphics[width=0.99\columnwidth]{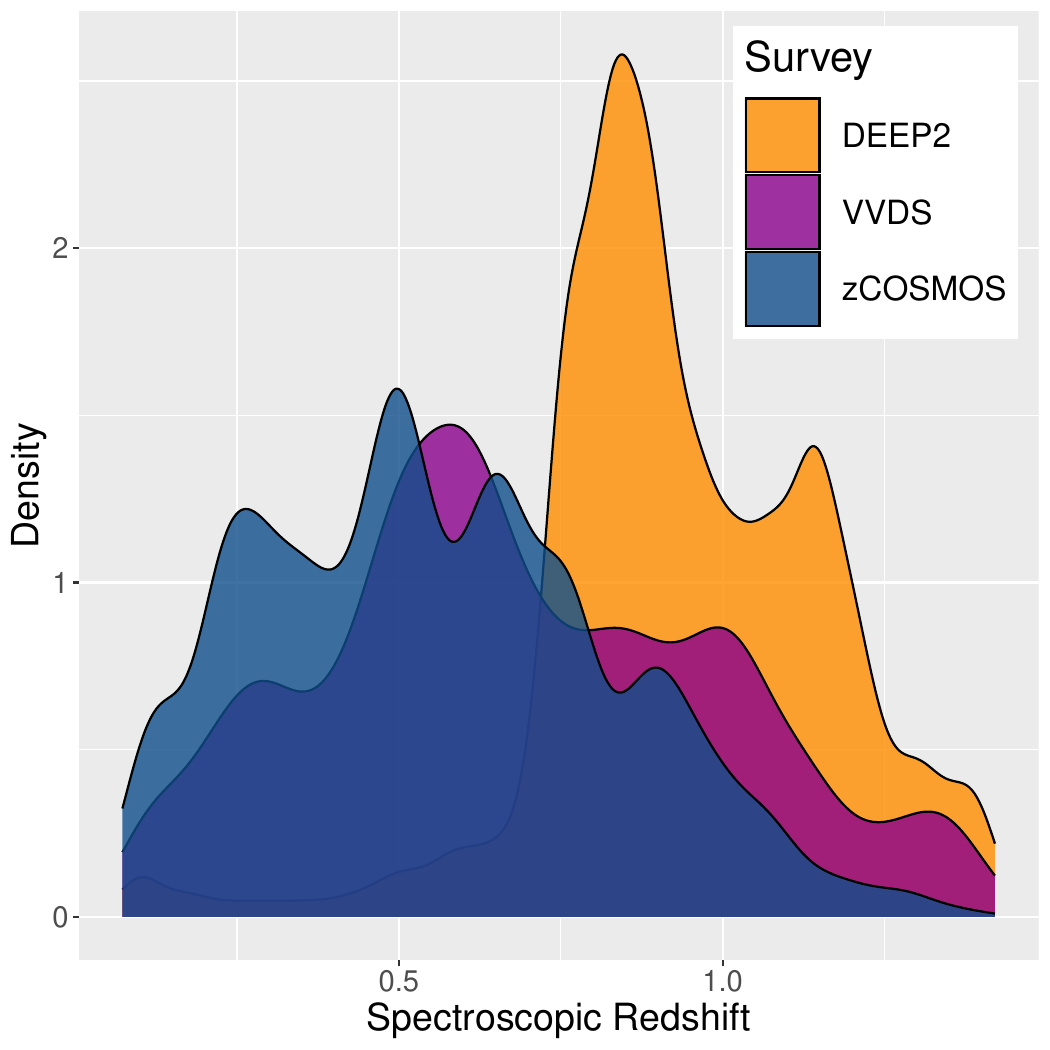}
  \caption{\baselineskip=15pt 
  Spectroscopic redshift distributions of the three spectroscopic surveys used as source data. 
  \label{figure:redshift_dist_spectro_stacked}}
\end{minipage}%
\hspace{0.3cm}
\begin{minipage}{.95\columnwidth}
  \centering
  \vspace{1.05cm}
  \includegraphics[width=0.99\columnwidth]{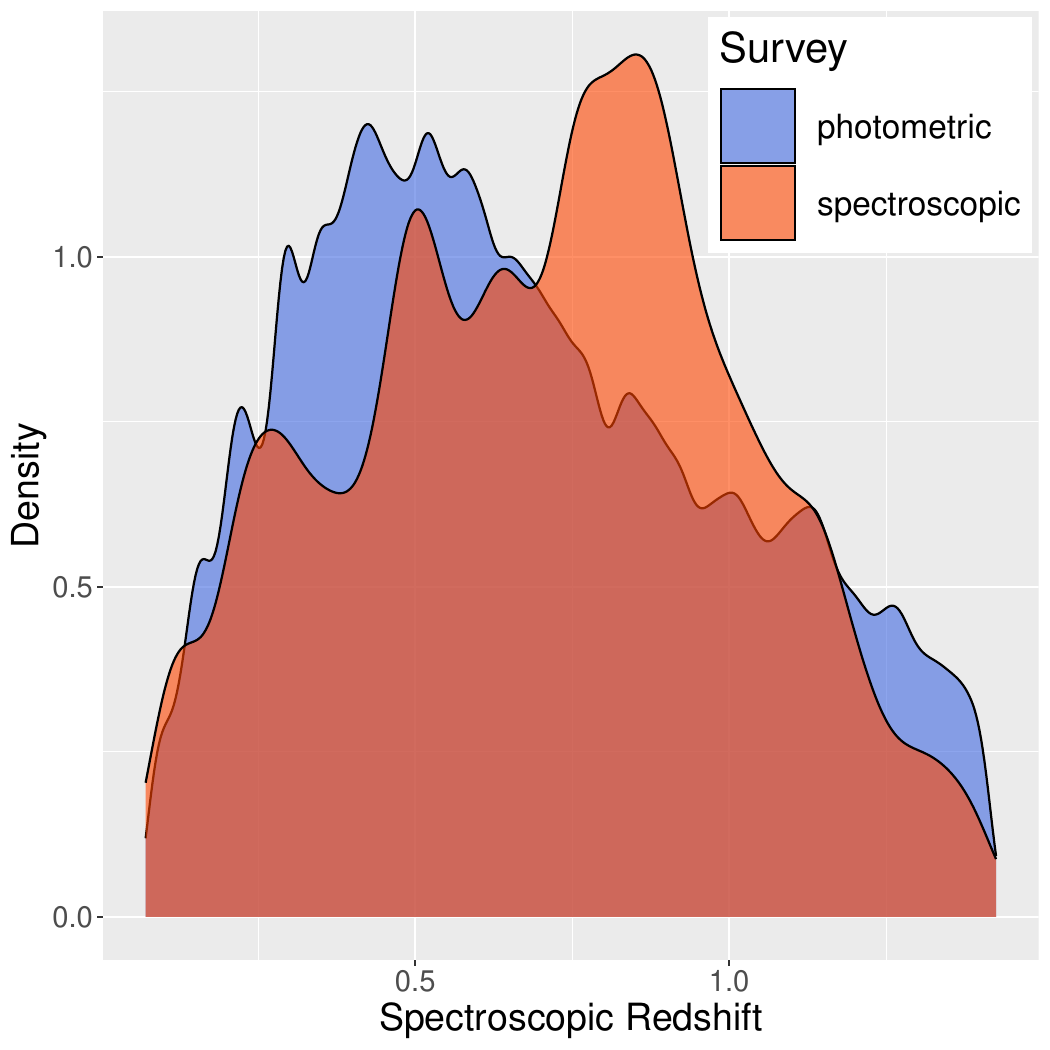}
 \caption{\baselineskip=15pt Spectroscopic (true) redshift distributions of the photometric simulated target data $D_T$ (blue, not available in practice) compared with the spectroscopic source data (red). \label{figure:redshift_dist_spectro_photo}}
\end{minipage}
\end{figure}

A much smaller spectroscopic source data set $D_S$, with $|D_S| = 21,537$, galaxies is provided, composed of simulated data 
mimicking three surveys: zCOSMOS ($9930$ galaxies); DEEP2 ($6919$ galaxies); and VVDS ($4688$ galaxies), altogether spanning a redshift range of $ 0.07 \leq z \leq 1.43$. The spectroscopic redshift distributions of the three surveys are illustrated in Figure~\ref{figure:redshift_dist_spectro_stacked}. The spectroscopic source set is not a representative sample of the photometric target distribution (selection effects are described in \citetalias{wright2020photometric}). Figure~\ref{figure:redshift_dist_spectro_photo} illustrates the density of the spectroscopic source redshift distribution  (red), and the (true) redshift distribution of the photometric simulated target data (blue; not available in practice). While both distributions cover the same redshift range, the difference in densities is immediately apparent, an effect of the underlying covariate shift. For each galaxy in the spectroscopic set, the same set of covariates as provided for galaxies in the photometric set is available. In addition, an accurate spectroscopic measurement of the true redshift $z$ is available for each galaxy, with measurement error that is negligible for our purposes. 

To account for sampling variance, 100 independent spectroscopic catalogues are provided, each with above described specifications. These correspond to 100 independent fields (lines-of-sights, abbreviated as LoS). The fields of the three spectroscopic surveys are independent of each other across the 100 LoS.

\paragraph*{Choice of Covariates and Handling of Missing Data:}\label{sec:technical_details}
To obtain the conditional density estimates for all objects in the photometric target data, we choose as covariates the $r$-band magnitude and the 8 colors: ($u-g$, $g-r$, $r-i$, $i-Z$, $Z-Y$, $Y-J$, $J-H$, $H-Ks$), a setup previously adopted (e.g., \citetalias{wright2020photometric}; \citealt{izbicki2017photo}; \citealt{autenrieth2021stratifiedASA}). We note that using colors instead of magnitudes does not worsen the ``missing data" pattern, as illustrated in Figure~\ref{figure:missing_magnitudes_and_colors}. 
As a preprocessing step, all covariates are scaled to have mean zero and standard deviation one. In the \textit{StratLearn} framework, the missing data pattern has to be taken into account in the propensity score estimation step, and in the computation of the conditional density estimators within strata. For propensity score estimation, we use mean imputation of the 9 covariates to fill the missing values; we also add 9 binary indicator variables as dependent variables (main effects) to the logistic regression propensity score model, which describe the missingness of each covariate. Figure~\ref{figure:PS_distributions} in the Appendix illustrates the distributions of the estimated propensity scores for source and target data. The support of the target propensity score distribution is well covered by the support of the source propensity score distribution, demonstrating the availability of source galaxies that match the covariate space of the target galaxies. 

The computation of the conditional density estimators ({\it ker-NN} and {\it Series}) requires the calculation of Euclidean distances between the covariate vectors of each galaxy. In the missing data cases, we compute the pairwise distances of two galaxies using only the covariate values with measurements (no missingness) for both galaxies. 
The large size of the photometric target set causes additional computational challenges: for prediction of the conditional densities on the target data, distance matrices between photometric target set and spectroscopic source set are required, which is not computationally feasible for the entire target set at once. We thus process the prediction on the photometric target set in batches of 60,000 target samples. 
Table~\ref{table:strata_composition_60K} shows the strata composition of spectroscopic source and photometric target data for one random batch, illustrating that there is enough spectroscopic source/training data in each stratum to fit the conditional density estimators within strata separately. While there is a slight discrepancy between the average redshift in source (0.71) and target data (0.68) overall, most strata have well-balanced redshift means between source and target, an indicator of reduced covariate shift after the propensity score stratification \citep{autenrieth2021stratifiedASA}. Other batches demonstrate a similar pattern. To reduce the computational burden, for each LoS,  we use a fixed 
set of hyperparameters for prediction of the conditional density estimators on all target batches.
For each LoS, we obtained the fixed hyperparameter set by optimizing (\ref{formula:fzxrisk_source}) and (\ref{formula:fzxrisk_source_and_target}), separately for each stratum, using one initial strata composition, with a randomly selected target batch.\footnote{Optimization of (\ref{formula:fzxrisk_source}) was performed by splitting the source data within each strata in a training and validation set (one half each). The parameters which led to the best predictive performance on the source validation sets (in each stratum) were then selected for each conditional density estimator (ker-NN and Series) separately. The final optimization in (\ref{formula:fzxrisk_source_and_target}) was then performed on the same source strata validation sets, using the optimized ker-NN and Series source validation set predictions.
The hyperparameters for the five strata and for all 100 LoS are illustrated in the Appendix, Figures~\ref{figure:hyperparam_Series}, \ref{figure:hyperparam_KerNN} and \ref{figure:hyperparam_Comb}.} 
Using batches of photometric target data has the advantage that distance matrices can be stored in memory, and predictions can be processed in parallel on several batches.\footnote{We performed all computations on a CPU cluster employing up to $\sim150$ CPU simultaneously. }

\begin{table}
\centering
\caption{Composition of the five \textit{StratLearn} strata. The number of galaxies and  the average spectroscopic (true) redshift is presented in each source and target stratum. (Composition of one random batch of 60,000 photometric samples is shown for illustration).
\label{table:strata_composition_60K}}
\begin{tabular}{ l l r r  } 
\hline
Stratum & Set & \#galaxies  & Mean $z$ \\ 
\hline
\hline
\multirow{1}{*}{1} 
&  Source &  6091 & 0.74 \\ 
&  Target & 10217 & 0.74 \\ 
\hline
\multirow{1}{*}{2} 
&  Source &  5036 & 0.77 \\ 
&  Target & 11271 & 0.74 \\ 
\hline
\multirow{1}{*}{3} 
&  Source & 4351 & 0.72 \\ 
&  Target &  11957 & 0.72 \\ 
\hline
\multirow{1}{*}{4} 
&  Source & 3668 & 0.65 \\ 
&  Target & 12639 & 0.66 \\ 
\hline
\multirow{1}{*}{5} 
&  Source & 2391 & 0.58 \\ 
&  Target &  13916 & 0.57 \\
\hline
\hline
\multirow{1}{*}{All} 
&  Source & 21537 & 0.71 \\ 
&  Target &  60000 & 0.68 \\ 
\hline
\end{tabular}
\end{table}

\subsection{Improved Bin Assignment Accuracy}\label{sec:results_bin_assignment}

In this section, we evaluate the accuracy of the new tomographic bin assignment, obtained via \textit{StratLearn}-based conditional density estimates, as described in Section~\ref{sec:StratLearn_bin_assignment}, and illustrated in Figure~\ref{figure:fzx_examples}.  

In Table~\ref{table:accuracy_bin_assignment}, we compare our bin assignment with the standard practice of using $z_B$ for the assignment, across five different classification performance metrics, demonstrating improvement in all of them. On average across the 100 LoS, the  \textit{StratLearn} binning assigns the photometric target galaxies to the correct tomographic bin in $62.2\%$ of the cases (considering the five tomographic bins, and both end bins separately). 
This is a substantial improvement over the $z_B$ binning,  with an accuracy of $52.5\%$. \textit{StratLearn} improves both the sensitivity (true positive rate) and the specificity (true negative rate) of tomographic bin assignment compared to $z_B$, thus also leading to an improvement of the balanced accuracy and Cohen's kappa, which take into account the imbalance of class (bin) proportions.\footnote{The balanced accuracy is defined as (specificity + sensitivity)/2. The Cohen's Kappa measures the relative performance of the classifier with the performance of a random guess (based on the class frequency). Both metrics take on values between 0 and 1 (with 1 being a perfect classifier).}
We note that the standard deviations of all performance measures across the 100 LoS is relatively low (Table~\ref{table:accuracy_bin_assignment}), which demonstrates that the improvement is consistent throughout the 100 LoS 
(the $z_B$ assignment is the same for all 100 LoS).

\begin{table}
\centering
\caption{Tomographic bin assignment performance evaluated over 100 LoS, comparing the \textit{StratLearn} bin assignment (following Section~\ref{sec:StratLearn_bin_assignment}) and the $z_B$ bin assignment. The average (sd) of the performance metrics computed for each of the 100 LoS is reported for \textit{StratLearn}. Using $z_B$, the bin assignment is consistently the same for all 100 LoS. For all metrics, higher values indicate better performance.} \label{table:accuracy_bin_assignment}
\begin{tabular}{lcc}
  \hline 
\multirow{2}{*}{} & \textit{StratLearn} & $z_B$ \\ 
\cline{2-3}
Performance metric & mean (sd) & mean (sd) \\
\hline
  \hline
Accuracy & 0.622  (0.003) & 0.526 (-) \\ 
  Balanced Accuracy & 0.718  (0.003) & 0.706 (-) \\ 
  Sensitivity & 0.502  (0.006) & 0.493 (-) \\ 
  Specificity & 0.934  (0.001) & 0.918 (-) \\ 
  Cohen's Kappa & 0.439  (0.006) & 0.415 (-)\\ 
   \hline
\end{tabular}
\end{table}

\begin{figure}
\centering
\begin{subfigure}[b]{0.95\columnwidth}
   \includegraphics[width=1\columnwidth]{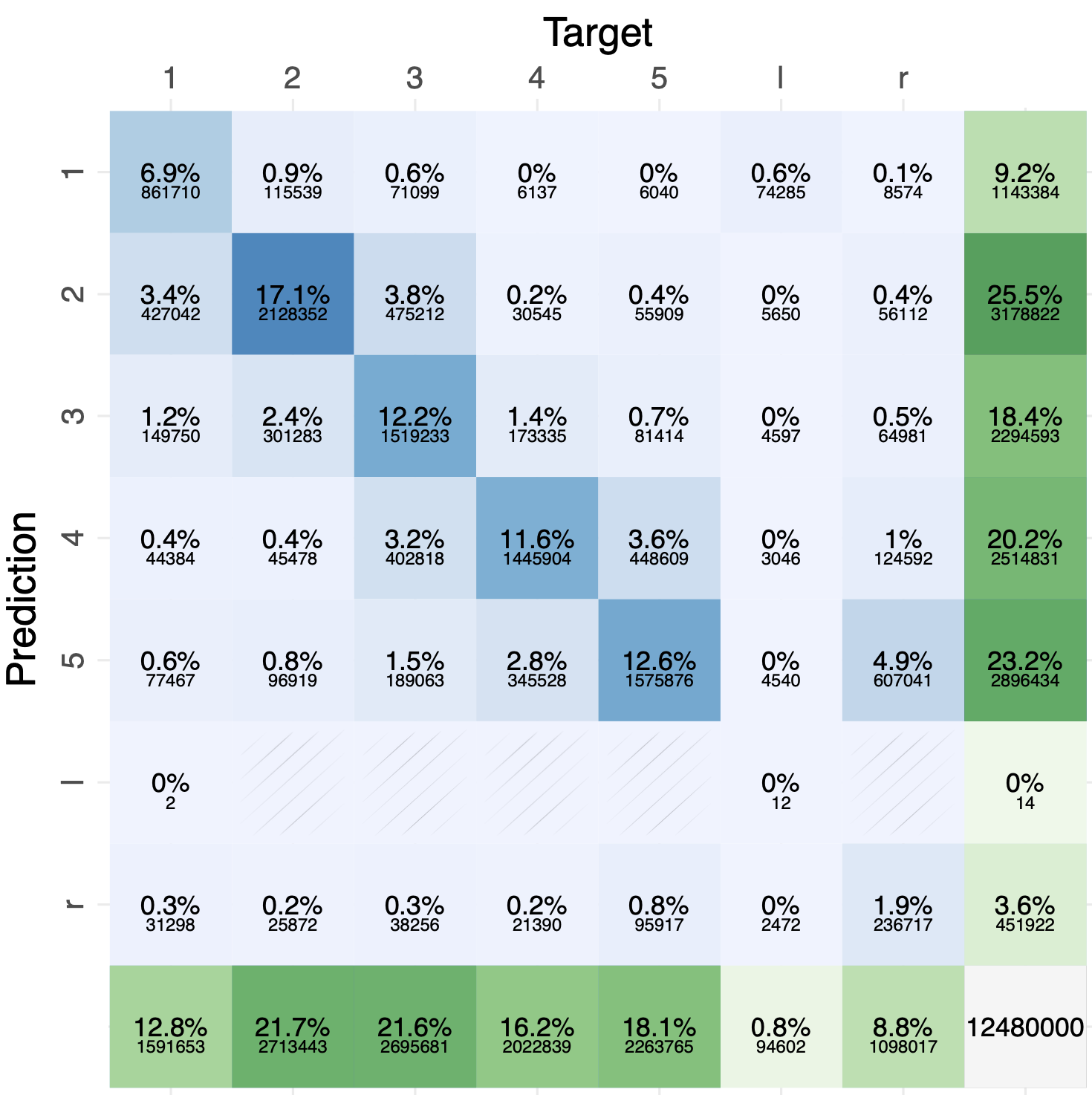}
   \caption{StratLearn tomographic bin assignment.}
   \label{figure:Conf_matrix_StratLearn} 
\end{subfigure}
\begin{subfigure}[b]{0.95\columnwidth}
    \vspace{0.3cm}
   \includegraphics[width=1\columnwidth] {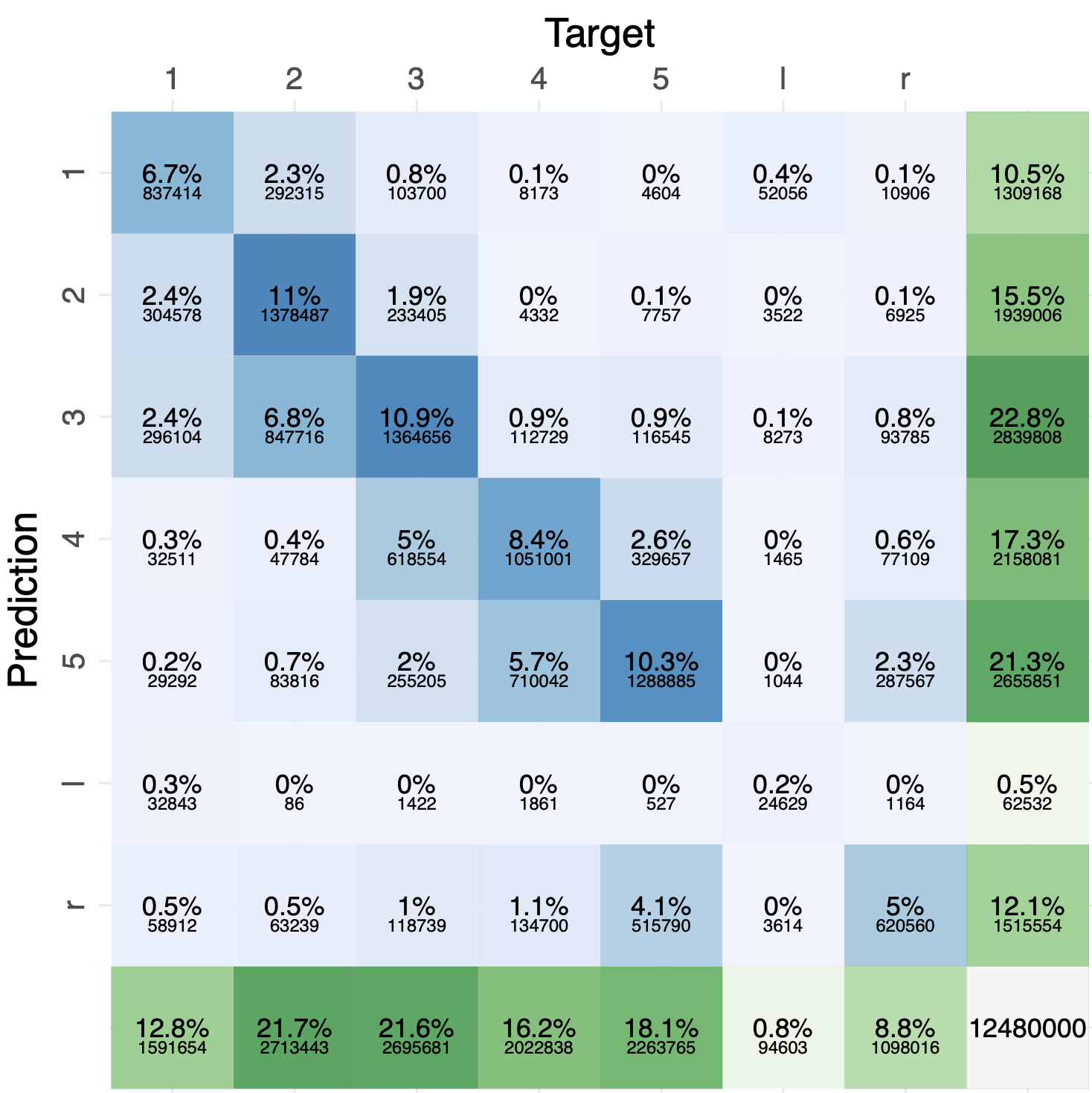}
   \caption{$z_B$ tomographic bin assignment.}
   \label{figure:Conf_matrix_ZB}
\end{subfigure}
\caption{(a) Confusion matrix of the \textit{StratLearn} tomographic bin assignment (averaged over 100 LoS).
(b) Confusion matrix of the $z_B$ tomographic bin assignment. 
The labels `l' and `r' refer to the left and right end bins respectively (for galaxies outside the tomographic bin ranges).
}
\end{figure}

Figures~\ref{figure:Conf_matrix_StratLearn} and \ref{figure:Conf_matrix_ZB} show the confusion matrices of tomographic bin assignment using \textit{StratLearn} and $z_B$ (averaged over the 100 LoS). The confusion matrices demonstrate that \textit{StratLearn} improves the bin assignment across all five tomographic bins (top five diagonal values), and most substantially in the second bin (with $z \in  (0.3,0.5]$), the one with the largest fraction of galaxies ($21.7\%$). In this bin, \textit{StratLearn} improves over the $z_B$ bin assignment by more than $55\%$.

\begin{figure}
\centering
\begin{minipage}{1\columnwidth}
  \centering      
  \hspace*{-1.1cm}  
      \includegraphics[width=1.3\columnwidth]{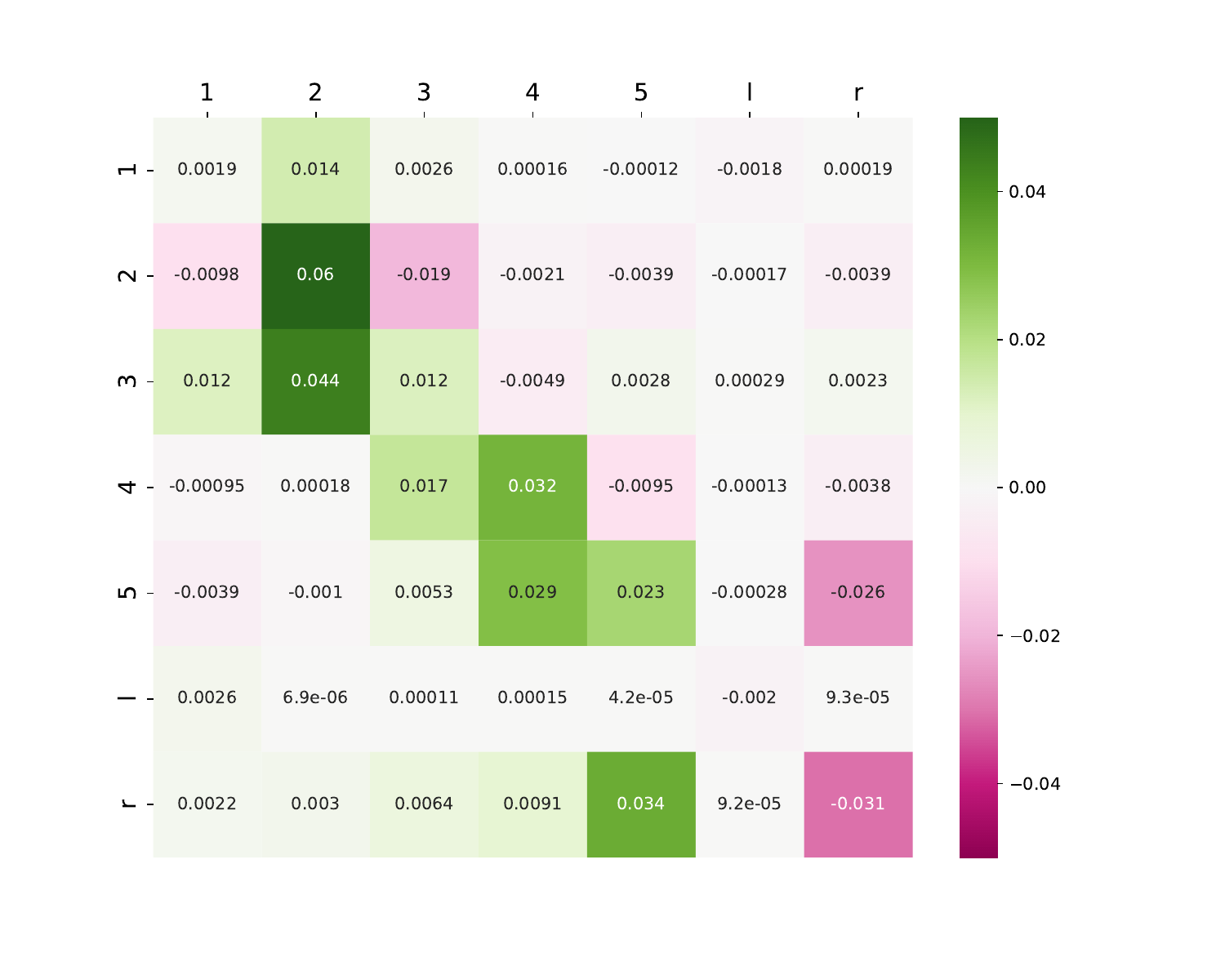}
      \vspace{-1cm}
 \caption{\baselineskip=1pt Heatmap of confusion matrix (accuracy) differences between \textit{StratLearn} and $z_B$. On the diagonal, the difference of \textit{StratLearn} - $z_B$ accuracy's is shown. Off-axis, the difference of $z_B$ - \textit{StratLearn} is shown. Thus higher values (in green) illustrate that \textit{StratLearn} performs better than the $z_B$ estimate. \label{figure:Heatmap_StratLearn-ZB}}
 \end{minipage}
\end{figure}

The heatmap in Figure~\ref{figure:Heatmap_StratLearn-ZB} provides a visual comparison of the confusion matrices in Figures~\ref{figure:Conf_matrix_StratLearn} and \ref{figure:Conf_matrix_ZB}. 
Green squares in Figure~\ref{figure:Heatmap_StratLearn-ZB} correspond to an improvement of $\textit{StratLearn}$ over $z_B$, while pink squares correspond to better performance of the $z_B$ assignment. The heatmap is computed by subtracting the diagonal values of Figure~\ref{figure:Conf_matrix_ZB}, the bin assignment accuracy of $z_B$, from the diagonal values of Figure~\ref{figure:Conf_matrix_StratLearn}, the bin assignment accuracy of \textit{StratLearn}; on the off-diagonal, the sign is reversed, so that green (positive) values denote {\it StratLearn} improvement everywhere. The improvement of \textit{StratLearn} is particularly strong in the top five diagonal squares, the five tomographic bins, which are of highest interest for the scientific analysis.

Figure~\ref{figure:conf_matrix_StratLearn_vs_ZB} in the Appendix illustrates the changes in bin assignment of \textit{StratLearn} vs. $z_B$, showing a moderate to strong disagreement of \textit{StratLearn} and $z_B$ in most of the bins. The reassignment of galaxies, and especially the improved bin assignment accuracy of \textit{StratLearn}, might thus substantially improve cosmological results -- this will be subject of a future, dedicated study.


\subsection{Improved Population Mean Estimates Accuracy}\label{sec:results_bias}

The main purpose of this study is to obtain accurate estimates of the (true) redshift population means within tomographic bins. The foremost criteria to evaluate redshift calibration methods \citep{2022ARA&A..60..363N} is the mean discrepancy 
\begin{align}\label{form:mean_discrepancy}
    \mathbb{E}[ \hat{\mu_b} - \mu_b^{\text{true}} ] \simeq \frac{1}{L} \sum_{l = 1}^L (\hat{\mu}_{b,l} - \mu_{b,l}^{\text{true}} ),  \quad (= \widehat{\text{bias}}_b)
\end{align}
where $L = 100$ is the number of LoS, $\hat{\mu}_{b, l}$ the estimated mean redshift and $ \mu_{b,l}^{\text{true}}$ the true redshift mean for LoS $l$ for galaxies assigned to tomographic bin $b$. 
We note that (\ref{form:mean_discrepancy}) is not a bias in a strict statistical sense, since the true redshift mean (within tomographic bin) varies across lines of sights. However, being consistent with the notation of previous studies, we will loosely refer to (\ref{form:mean_discrepancy}) as ``bias". 
In addition to (\ref{form:mean_discrepancy}), we are interested in the standard deviation, SD, of the mean differences across the 100 LoS: 
\begin{align}
\text{SD}(\hat{\mu}_b -  \mu_b^{\text{true}} ) = \sqrt{\frac{\sum_{l= 1 }^L (\mu_{b,l}^{\text{diff}} -  \widehat{\text{bias}}_b)^2 }{L - 1} }
\end{align}
with $\mu_{b,l}^{\text{diff}} = \hat{\mu}_{b,l} -  \mu_{b,l}^{\text{true}}$, for $l = 1, \dots, L$ and $b=1, \dots, 5$.

\begin{table*}
  \centering 
    \caption{Mean discrepancy (bias) computed over 100 lines of sight for different calibration methods, and different bin assignment strategies. We abbreviate the \textit{StratLearn} bin assignment as \textit{SL}. The add-on (gold) 
    denotes quality cuts applied to the data according to \protect\citetalias{wright2020photometric}. 
    The Galaxies column shows the total number of galaxies (in millions) available in the five tomographic bins. \label{table:bias_results}}
\begin{tabular}{lllrrrrrr}
  \hline
 & Binning & Galaxies [M]& Bin 1 & Bin 2 & Bin 3 & Bin 4 & Bin 5 & $|$Average$|$ \\ 
    \hline
     \textit{StratLearn}-Bayes (A)  &  \textit{SL}& 12.02 & 0.0123 & -0.0076 & 0.0053 & -0.0010 & 0.0001 & 0.0053 \\ 
    \textit{StratLearn}-Bayes (B) & \textit{SL}& 12.02 & 0.0095 & -0.0092 & 0.0047 & -0.0013 & 0.0012 & 0.0052 \\
   SOM & \textit{SL} (gold)&11.48 & -0.0084 & 0.0022 & 0.0156 & 0.0117 & 0.0148 & 0.0105\\
  \hline
  \textit{StratLearn}-Bayes (A) & $z_B$& 10.90 & 0.0259 & 0.0127 & 0.0084 & 0.0003 & -0.0231 & 0.0141 \\ 
   \textit{StratLearn}-Bayes (B) & $z_B$& 10.90 & 0.0228 & 0.0117 & 0.0071 & -0.0002 & -0.0236 & 0.0131 \\ 
   SOM & $z_B$ (gold)&10.17 & -0.0005 & 0.0036 & 0.0135 & 0.0147 & -0.0102 & 0.0085 \\
   \hline
        \end{tabular} 
\end{table*}

Table~\ref{table:bias_results} presents the bias results obtained for our novel \textit{StratLearn}-Bayes method, with a comparison to the previously best performing method, the SOM. 
Based on the \textit{StratLearn} binning, our method options \textit{StratLearn}-Bayes (A) and (B) 
lead to an average absolute bias of 0.0053 and 0.0052 across the five tomographic bins, an improvement of $\sim 40\%$ w.r.t\ the SOM method with 
$z_B$ binning, which leads to an average absolute bias of 0.0085. 
We further note that the SOM (with $z_B$ binning) method requires systematic quality cuts, which reduce the data size for the scientific analysis (we return to this point below).

We also apply the SOM calibration method using the new \textit{StratLearn} binning, applying quality cuts as described in \citetalias{wright2020photometric}, which leads to an increase of bias (0.105 absolute average bias) compared to SOM with the $z_B$ binning (0.0085 absolute average bias). Using the \textit{StratLearn}-Bayes model on the $z_B$ bin assignment also leads to an increase in bias to 0.0141 and 0.0131 (on absolute average across the five bins). Such a reduction in performance could in fact be expected: the \textit{StratLearn}-Bayes model is based on the modelling of the \textit{StratLearn} object level (galaxy) conditional density estimates, but
by applying a different binning (e.g., via $z_B$) 
additional (external) errors 
are introduced in the assignment of galaxies per tomographic bin. The \textit{StratLearn}-Bayes framework is not designed for correction of such external errors (biases), which lead to systematic shifts of the population mean estimates. For instance, if the $z_B$ galaxy bin assignment is correlated with the variance of \textit{StratLearn} galaxy conditional density estimates, then the (tomographic bin) population mean estimate in (\ref{form:post_mu_cond_sigma_params}) can be systematically shifted. We thus advise against the combination of \textit{StratLearn}-Bayes based on $z_B$ binning, and advocate for the use of \textit{StratLearn}-Bayes via the \textit{StratLearn}-based binning, which leads to the best performance.

Table~\ref{table:sd_results} shows the standard deviation (SD) population scatters from the 100 LoS. \textit{StratLearn}-Bayes with \textit{StratLearn} binning leads to slightly increased standard deviations of 0.0066 (option (A)) and 0.0067 (option (B)) on average across the five tomographic bins, compared to the SOM method based on $z_B$ binning with an average of 0.0051. The results in Table~\ref{table:sd_results} indicate that the standard deviation results are related to the binning strategy, rather than to the calibration method. Using SOM calibration on the \textit{StratLearn} binning (with gold quality cuts) leads to comparable increase in SD of 0.0066 on average across the five bins. On the other hand, using the \textit{StratLearn}-Bayes model applied on the $z_B$ binning leads to a decrease in SD to an average of 0.0048 and 0.0047, even lower than applying SOM on the $z_B$ binning.

In general, we note that the similarity in results of the \textit{StratLearn}-Bayes options (A) and (B) demonstrate robustness with respect to the computation of the population variance (last paragraph of Section~\ref{sec:Bayesian_model}). Both methods outperform the previously best method (SOM on $z_B$ binning).  Given a slight improvement of bias reduction, we propose the application of \textit{StratLearn}-Bayes (B) as our best method.

Finally, our proposed method, \textit{StratLearn}-Bayes (B), leads to a maximum bias within tomographic bins of $\Updelta \langle z \rangle =  0.0095 \pm 0.0089$ (in bin 1).  
In contrast, using the previously best calibration method, SOM based on $z_B$ binning, leads to maximum biases of $0.0135 \pm 0.0052$ and $0.0147 \pm 0.0040$ (in bin 3 and 4). 
In addition, SOM based on $z_B$ binning requires systematic quality cuts (gold selection), which are not necessary for our methodology. 

The improved accuracy that we see with our proposed method brings the biases down to $ \Updelta \langle z \rangle < 0.01 $ in all bins. This threshold has been chosen 
in previous work as delineating `negibile' and `non-negligible' biases \citepalias{wright2020photometric}. 
Moreover, our method produces biases that are consistent with zero within $1.5\sigma$ in all bins, whereas the SOM method produces biases that are inconsistent with zero at the level of $\sim 3.7\sigma$ in the fourth bin. 
As such, our method is intrinsically less biased given the same calibrating data and target wide-field population, while retaining a greater number of sources for scientific analysis.

\begin{table*}
\centering
\caption{
As in Table~\ref{table:bias_results}, but showing standard deviation (SD) computed over 100 lines of sight for the different calibration methods, and different bin assignment strategies. \label{table:sd_results}}
\begin{tabular}{llrrrrrr}
  \hline
 & Binning & Bin 1 & Bin 2 & Bin 3 & Bin 4 & Bin 5 & $|$Average$|$ \\ 
    \hline
   \textit{StratLearn}-Bayes (A)  & \textit{SL} & 0.0087 & 0.0065 & 0.0046 & 0.0052 & 0.0082 & 0.0066 \\ 
    \textit{StratLearn}-Bayes (B)  & \textit{SL} & 0.0089 & 0.0065 & 0.0045 & 0.0052 & 0.0082 & 0.0067 \\ 
   SOM & \textit{SL} (gold) & 0.0085 & 0.0064 & 0.0059 & 0.0058 & 0.0066 & 0.0066 \\
  \hline
    \textit{StratLearn}-Bayes (A) & $z_B$ & 0.0055 & 0.0048 & 0.0049 & 0.0037 & 0.0051 & 0.0048 \\   
  \textit{StratLearn}-Bayes (B) & $z_B$ & 0.0052 & 0.0048 & 0.0047 & 0.0036 & 0.0051 & 0.0047 \\ 
    SOM & $z_B$ (gold) &  0.0055 & 0.0061 & 0.0052 & 0.0040 & 0.0049 & 0.0051 \\
   \hline
\end{tabular}
\end{table*}

\subsection{Larger Sample Size for Weak Lensing Analysis}
In Table~\ref{table:sample_sizes_tomography}, we show the absolute numbers of galaxies obtained via the different tomographic bin assignment strategies and quality cuts. 
In bins 2, 4, and 5, the number of galaxies is higher when using \textit{StratLearn} binning compared to $z_B$. In bins 1 and 3, the number of galaxies is slightly higher using the $z_B$ binning. Overall, due to the improved binning accuracy of \textit{StratLearn}, there is an approximately $10\%$ increase in the number of available galaxies for science (summing over all bins) when using \textit{StratLearn} for tomography instead of $z_B$. 
We note that \textit{StratLearn} assigns substantially less galaxies to the right end bin than $z_B$ (see Figure~\ref{figure:Conf_matrix_StratLearn},\ref{figure:Conf_matrix_ZB}), leading to a lower proportion of galaxies that are falsely removed from the analysis (the five tomographic bins), but also to a higher proportion of galaxies that actually are in the right end bin (having redshift greater than 1.2), but are assigned to one of the five tomographic bins (mostly to bin 4 and 5). Given the small biases of \textit{StratLearn}-Bayes in bins 4 and 5 (Table~\ref{table:bias_results}), the inclusion of such high redshift galaxies does not seem to have a negative impact on the calibration, but the positive effect of increasing the available data size within tomographic bins.

Table~\ref{table:sample_sizes_tomography} also provides the number of galaxies within bin after applying the gold selection, as introduced by \citetalias{wright2020photometric}. 
We note that the gold selection cut is not needed when applying the \textit{StratLearn}-Bayes approach, while it is a necessary step to obtain the SOM results. Thus, compared to the previously best combination of bin assignment and calibration method in \citetalias{wright2020photometric}, the \textit{StratLearn}-Bayes approach leads to an increase of galaxies available for science of ${\sim} 18\%$. 

For weak lensing analyses, the relevant statistic  is the increase in the effective number of sources incorporating the shape measurement weight. \cite{heymans+2012} derive the metric for effective number density of weak lensing sources as 
\begin{equation}
    n_{\rm eff}=\frac{1}{A}\frac{(\sum_N w)^2}{\sum_N (w^2)},
\end{equation}
where $w$ is the shape-measurement weight for each source $i\in N$, and $A$ is the survey area in square-arcmin. 
The change in the $n_{\rm eff}$ due to the SOM gold selection and quality control is described as $\Delta n_{\rm eff}=n_{\rm eff}^{\rm gold} / n_{\rm eff}^{\rm all}$. \citetalias{wright2020photometric} quote this metric for SOM calibration with quality control in their Table 2, finding values of $\sim 0.8$ in all bins. This suggests that, for a reanalysis of cosmic shear with  our \textit{StratLearn}-Bayes approach, we would increase the available lensing sample statistical power by a similar $\sim20\%$ in each tomographic bin.

\begin{table*}
\centering
\caption{Sample sizes (in millions) within tomographic bins obtained via different bin assignment strategies and quality cuts (mean and standard deviation computed over 100 LoS).
With (gold), we refer to the gold selection quality cuts described in 
\protect\citetalias{wright2020photometric}. 
\label{table:sample_sizes_tomography}
}
\begin{tabular}{lccccccc}
  \hline
& & Bin 1 & Bin 2 & Bin 3 & Bin 4 & Bin 5 & Total \\ 
& & $(0.1,0.3]$ & $ (0.3,0.5]$& $(0.5,0.7]$ &$ (0.7,0.9] $ &$ (0.9,1.2] $ &$(0.1,1.2] $ \\
  \hline
  \hline
\textit{StratLearn} & mean &  1.14 &  3.18 &  2.29 &  2.51 &  2.90 & 12.03 \\ 
            & (sd) & (0.112) & (0.207) & (0.189) & (0.204) & (0.220) &  \\ 
  \hline
  \textit{StratLearn} (gold) & mean  &  1.06 &  2.94 &  2.21 &  2.51 &  2.75 & 11.48 \\ 
            & (sd) & (0.089) & (0.158) & (0.171) & (0.205) & (0.223) &  \\ 
    \hline
    \hline
  $z_B$ & mean & 1.31 &  1.94 &  2.84 &  2.16 &  2.66 & 10.90 \\ 
            & (sd) & (-) & (-) & (-) & (-)& (-) &  \\ 
    \hline
  $z_B$ (gold) & mean &  1.15 &  1.91 &  2.36 &  2.10 &  2.65 & 10.17 \\ 
            & (sd) & (0.034) & (0.007) & (0.071) & (0.047) & (0.003) &  \\ 
   \hline
\end{tabular}
\end{table*}

\subsection{Population Distribution Estimates}\label{section:population_shapes_results}

In the previous sections, we demonstrate the ability of the \textit{StratLearn}-Bayes method to accurately and precisely estimate the redshift population means, which is most crucial for photo-$z$ calibration in the weak lensing analysis. 
Since realistic estimates of the population distribution shapes will become more influential in cosmic shear analysis and for photometric galaxy clustering (as discussed in Section~\ref{sec:introduction}), 
here we numerically demonstrate how propensity scores can be employed via inverse-PS weighting (as introduced in Section~\ref{sec:inverse-PS}) to improve estimation of the whole shape of the distribution.

\begin{figure*}
\centering
\begin{minipage}{.95\textwidth}
  \centering
    \includegraphics[width=0.32\columnwidth]{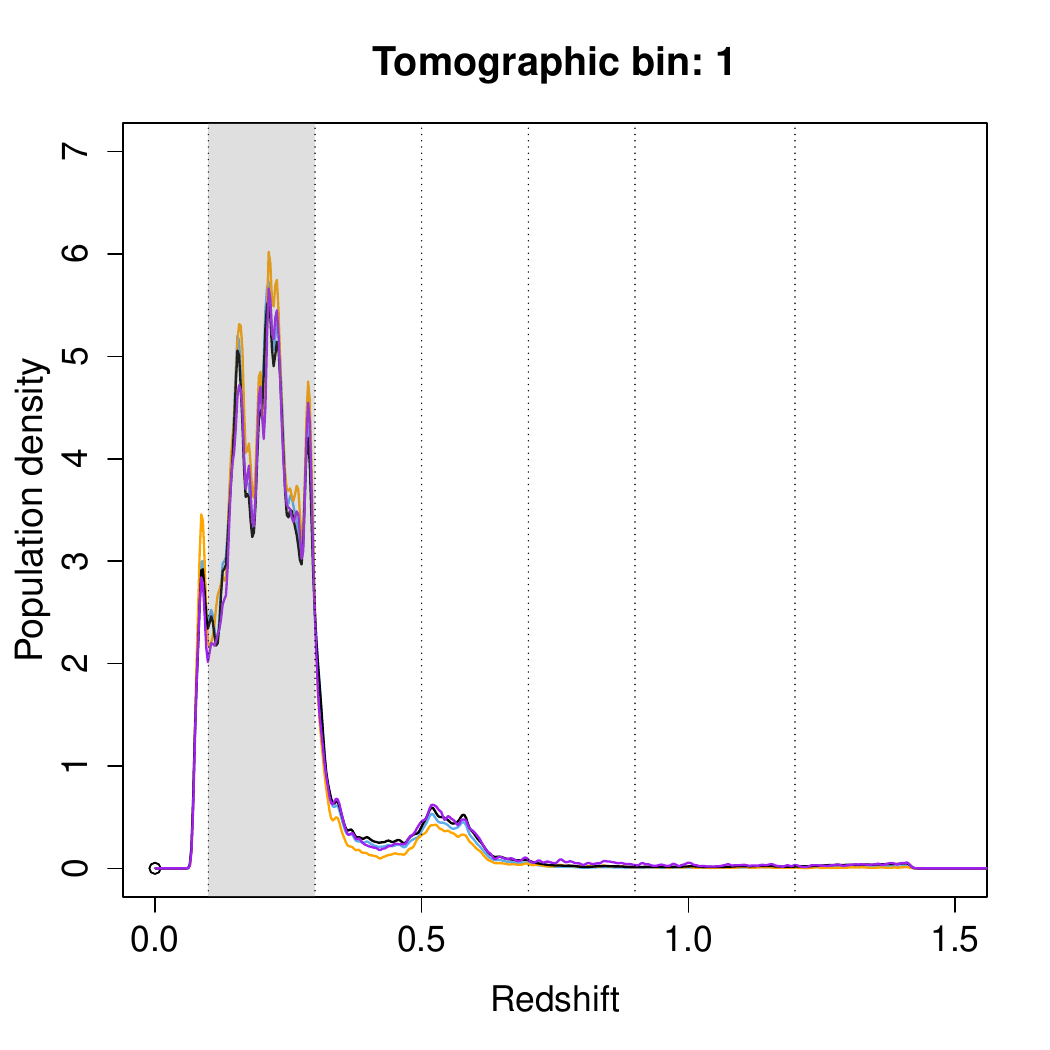}
    \includegraphics[width=0.32\columnwidth]{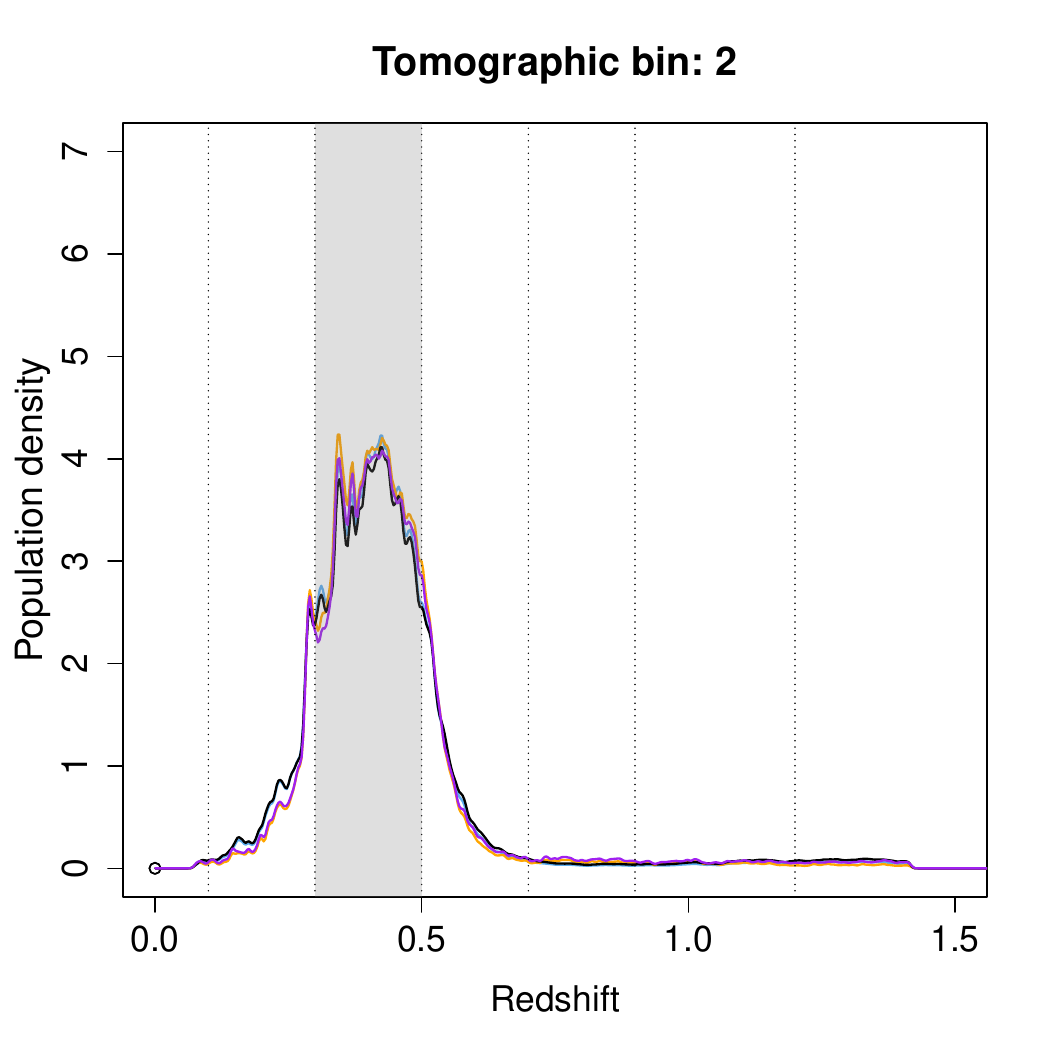}
    \includegraphics[width=0.32\columnwidth]{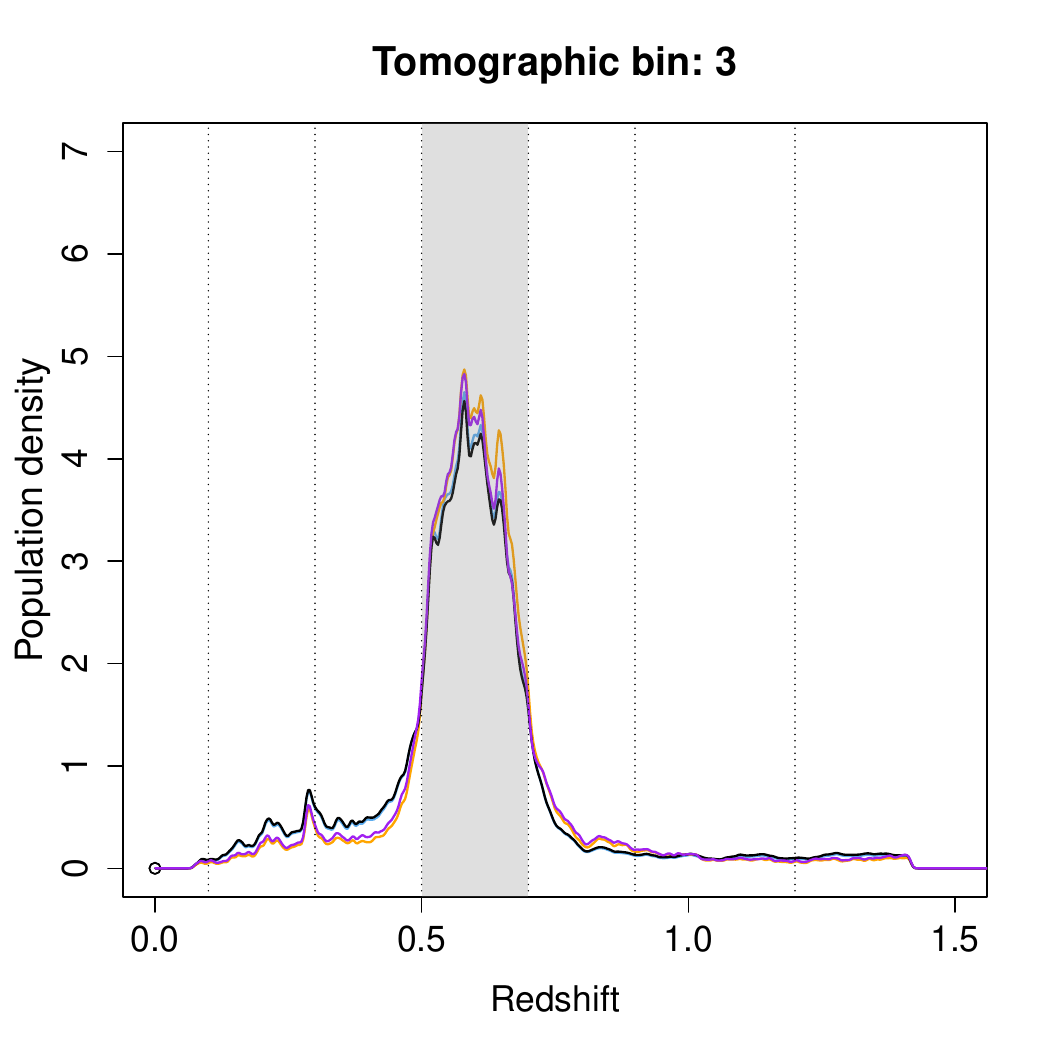}
        \includegraphics[width=0.32\columnwidth]{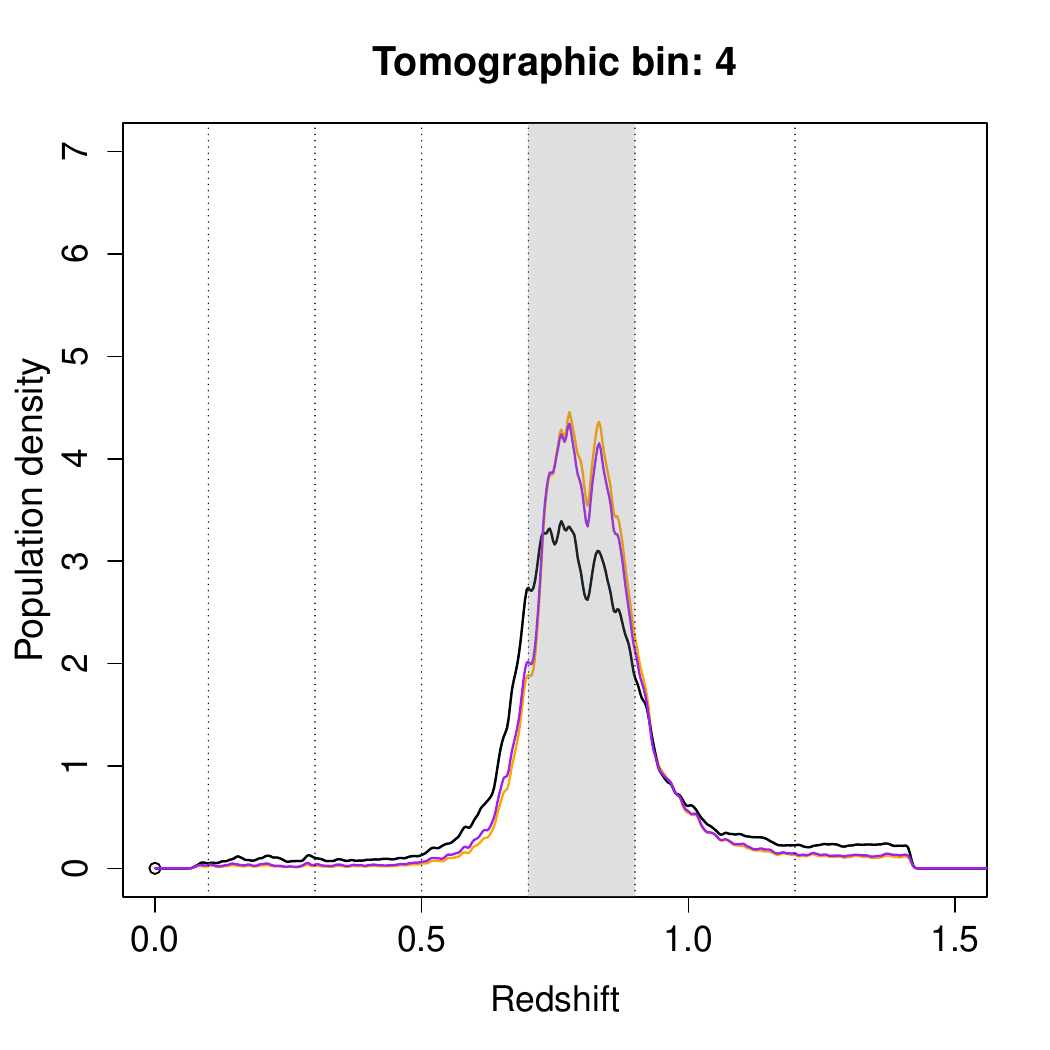}
        \includegraphics[width=0.32\columnwidth]{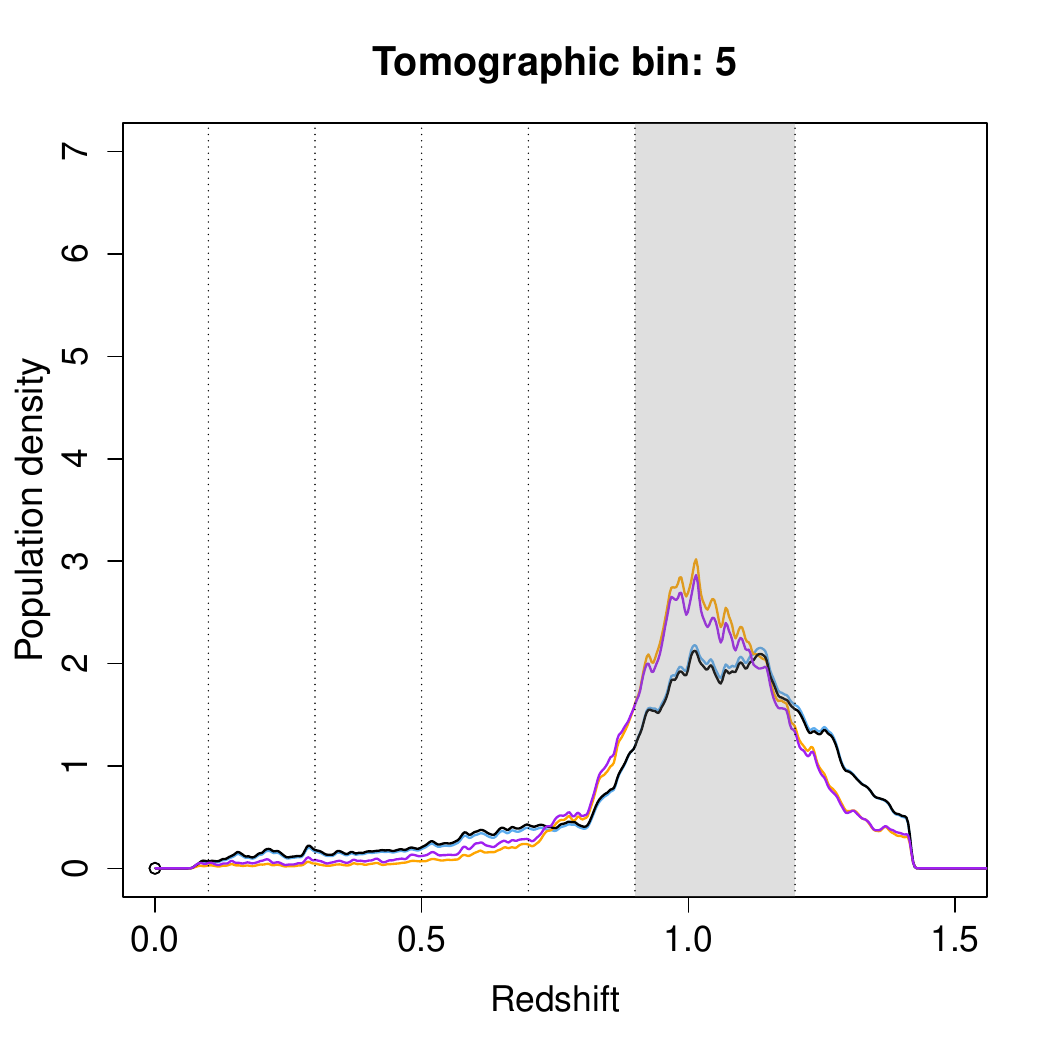}
        \includegraphics[width=0.32\columnwidth]{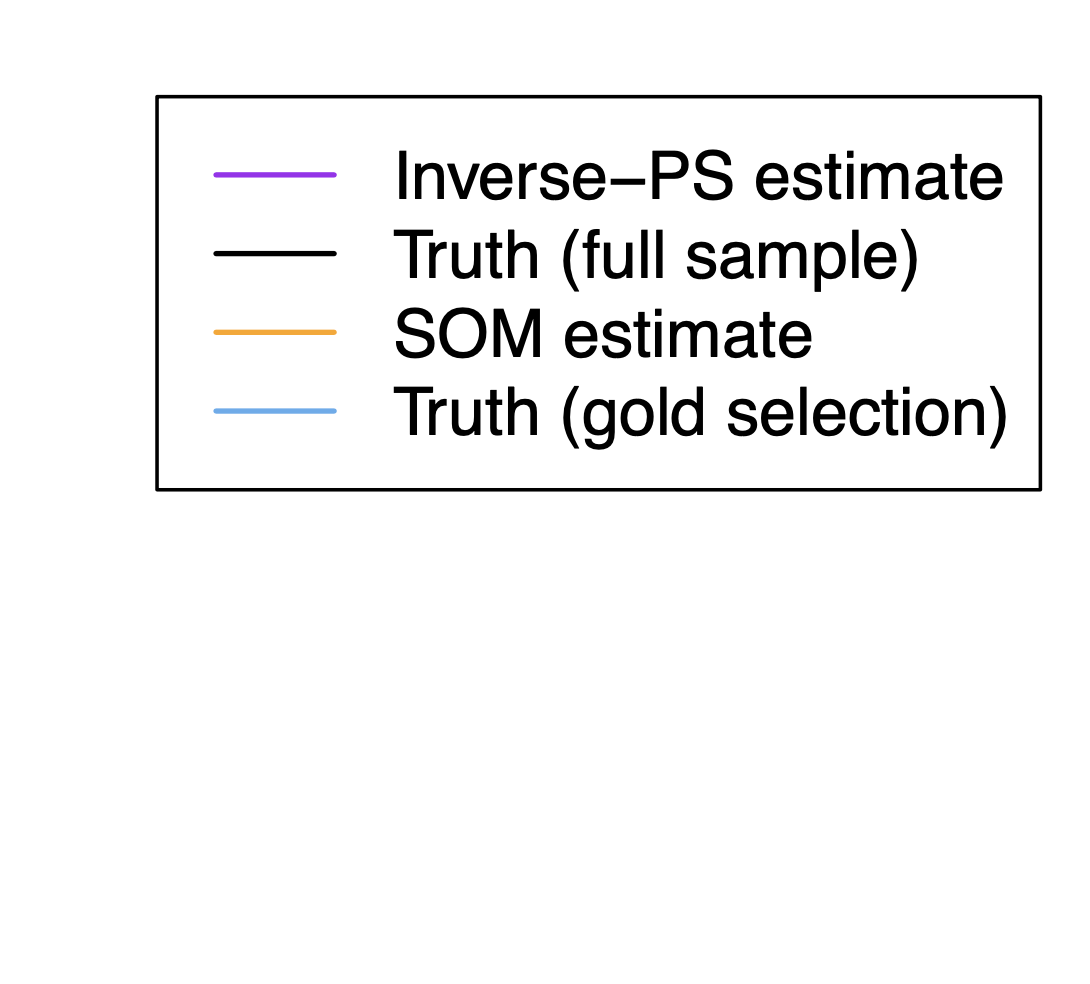}
    \caption{\baselineskip=15pt 
    Redshift population distribution (estimates) per tomographic bin, with tomographic bins obtained as described in Section~\ref{sec:StratLearn_bin_assignment} via \textit{StratLearn}-based binning. The figure illustrates the inverse-PS (purple) and SOM (orange) distribution estimates.
    The underlying true photometric redshift population distributions per tomographic bin (not known in practice) are illustrated in black for the full sample truth, and in light blue for the gold selected true distributions. 
    The averaged (estimated) distributions across the 100 LoS are illustrated per tomographic bin. 
    \label{fig:population_distributions}} 
\end{minipage}
\end{figure*}

In Figure~\ref{fig:population_distributions}, we show the inverse-PS
weighted redshift distributions per tomographic bin (in purple), obtained via the procedure described in Section~\ref{sec:inverse-PS}, and based on the \textit{StratLearn} tomographic bin assignment (following Section~\ref{sec:StratLearn_bin_assignment}).
The true redshift distributions per tomographic bin (not known in practice) are shown in black.
The purple inverse-PS weighted distributions exhibit a similar shape as the black true redshift population distributions 
recovering reasonably well the true photometric population distribution shapes, particularly throughout tomographic bins 1 to 3. 
Figure~\ref{fig:population_distributions} further illustrates the SOM estimated population distributions (in orange), and its underlying true redshift population distributions (in light blue) obtained on the \textit{StratLearn} tomographic binning after applying the gold selection quality cuts \citepalias{wright2020photometric}. We note that Figure~\ref{fig:population_distributions} presents the average (estimated) redshift population distributions across the 100 LoS per each tomographic bin.\footnote{For illustration purposes, a mild Gaussian kernel density smoothing (with bandwidth 0.00294) was applied to the presented distributions in Figure~\ref{fig:population_distributions}. The non-smoothed distributions are illustrated in Figure~\ref{supp_fig:IPS_SOM_populations_notsmoothed} in the Appendix. } 

In Figure~\ref{fig:pp-plot_inverse_PS_SOM}, we assess the quality of the two estimation methods (inverse-PS and SOM) w.r.t. their underlying true distributions via probability-probability plots (pp-plots)\footnote{pp-plots are obtained by plotting two (empirical) cumulative distribution functions (CDF) against each other. The distributions are equal iff the pp-plot falls on the diagonal line from (0,0) to (1,1).}: the figure shows the average pp-plot (across the 100 LoS) for the inverse-PS estimated distributions vs. the true (full) photometric redshift distributions per tomographic bin in purple lines, and the average pp-plot of the SOM estimated distributions vs. the gold selected true distributions in orange dashed lines. 
The vertical bars gives $95\%$ intervals indicating the dispersion of the central 95 pp-plot lines from the 100 LoS. 
%

Both estimates (inverse-PS and SOM) are close to the diagonal line throughout bins 1 to 3, with larger deviations in bins 4 and 5. Notably, the Inverse-PS and SOM pp-plot lines exhibit very similar deviation patterns from the diagonal line; both methods are based on reweighting of the spectroscopic samples (following (\ref{form:joint_weighted_source3})), which explains similarities in their estimates.  
The purple (average) inverse-PS lines are closer to the diagonal line than SOM in tomographic bins 1,3,4 and 5, and almost identical with SOM in bin 2. 
In addition, the vertical $95\%$ intervals are generally smaller for inverse-PS compared to SOM (particularly in bins 1 to 3), indicating less variability in the estimate across the 100 LoS.
Overall, the inverse-PS estimate thus approximates its underlying truth (the full binned photometric distribution) better than the SOM method its underlying (gold-selected) true distribution, with the additional advantage that no quality cuts are required for inverse-PS, leading to $\sim18\%$ more galaxies in the photometric sample available for scientific analysis. For additional visualization of the distribution differences presented in Figure~\ref{fig:pp-plot_inverse_PS_SOM}, 
Figure~\ref{fig:pp-plot_inverse_PS_SOM_subtracted} in the Appendix illustrates a slightly modified version of Figure~\ref{fig:pp-plot_inverse_PS_SOM} by subtracting the x-axis values (the quantiles of the true distributions) from the y-axis values (the quantiles of the estimated distributions) in each tomographic bin.\footnote{While here we are mostly interested in the population estimates obtained for the newly proposed and more accurate \textit{StratLearn}-based tomographic binning, we provide similar assessment of the population distribution estimates obtained for $z_B$-based tomographic binning in Figures~\ref{fig:pp-plot_inverse_PS_SOM_zB} and  \ref{supp_fig:IPS_SOM_populations_smoothed_zB} in the Appendix.}

\begin{figure*}
\centering
\begin{minipage}{.95\textwidth}
  \centering
    \includegraphics[width=0.32\columnwidth]{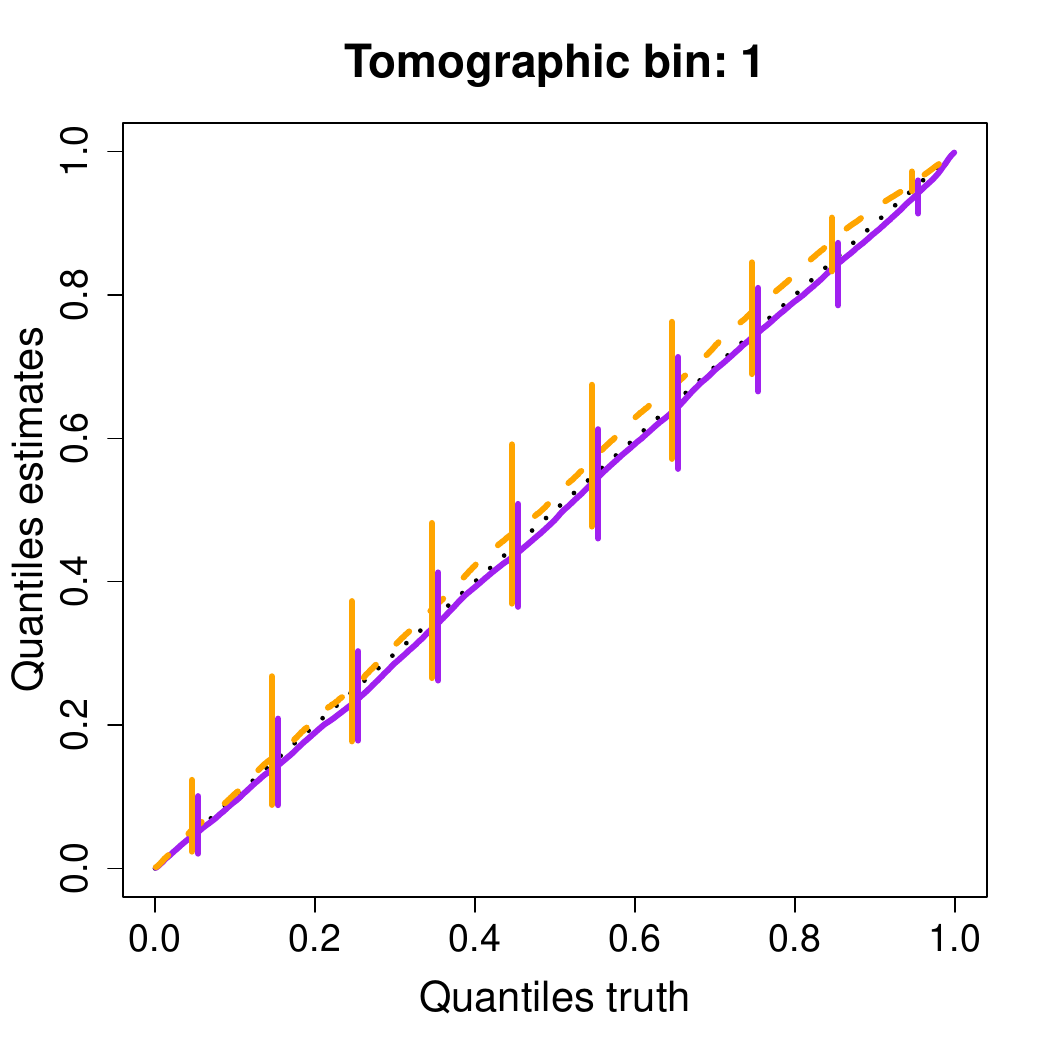}
    \includegraphics[width=0.32\columnwidth]{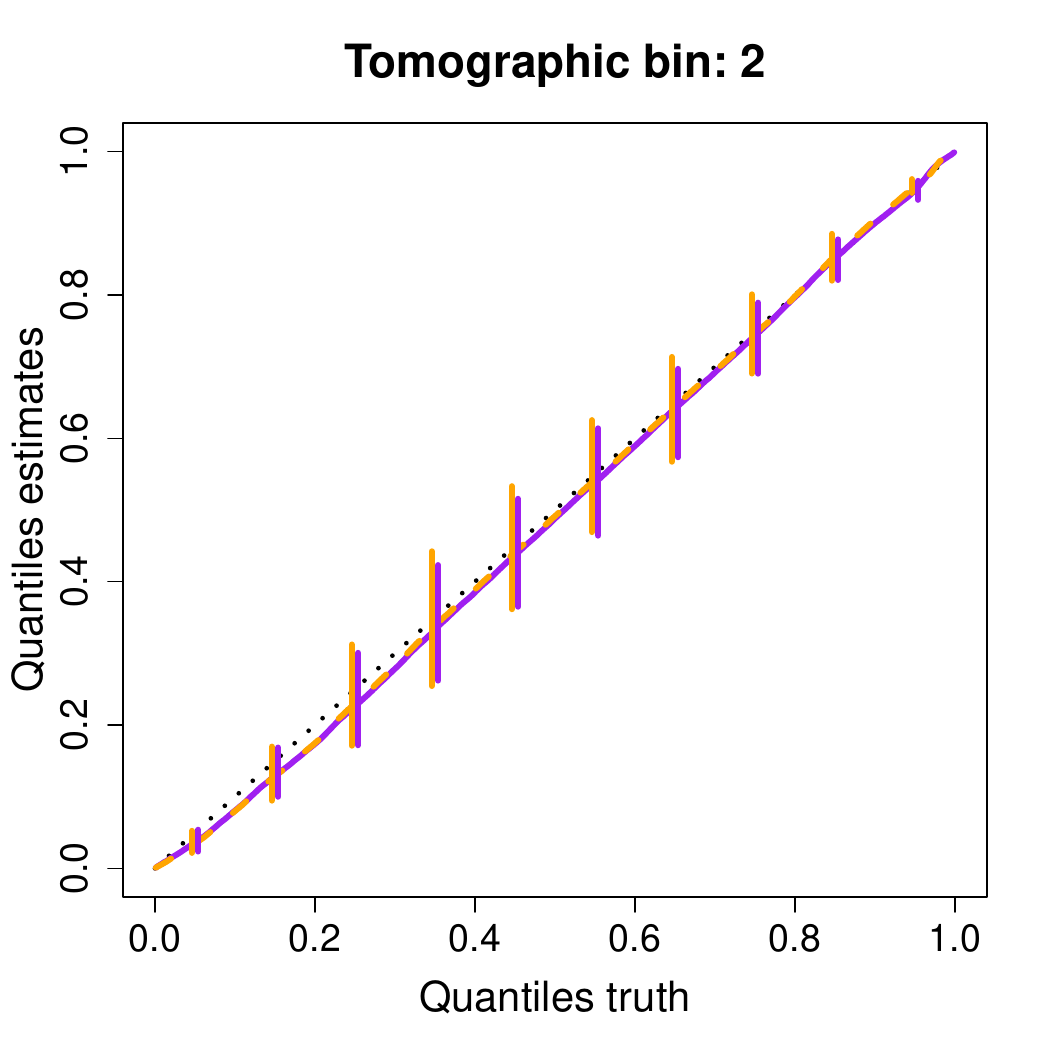}
    \includegraphics[width=0.32\columnwidth]{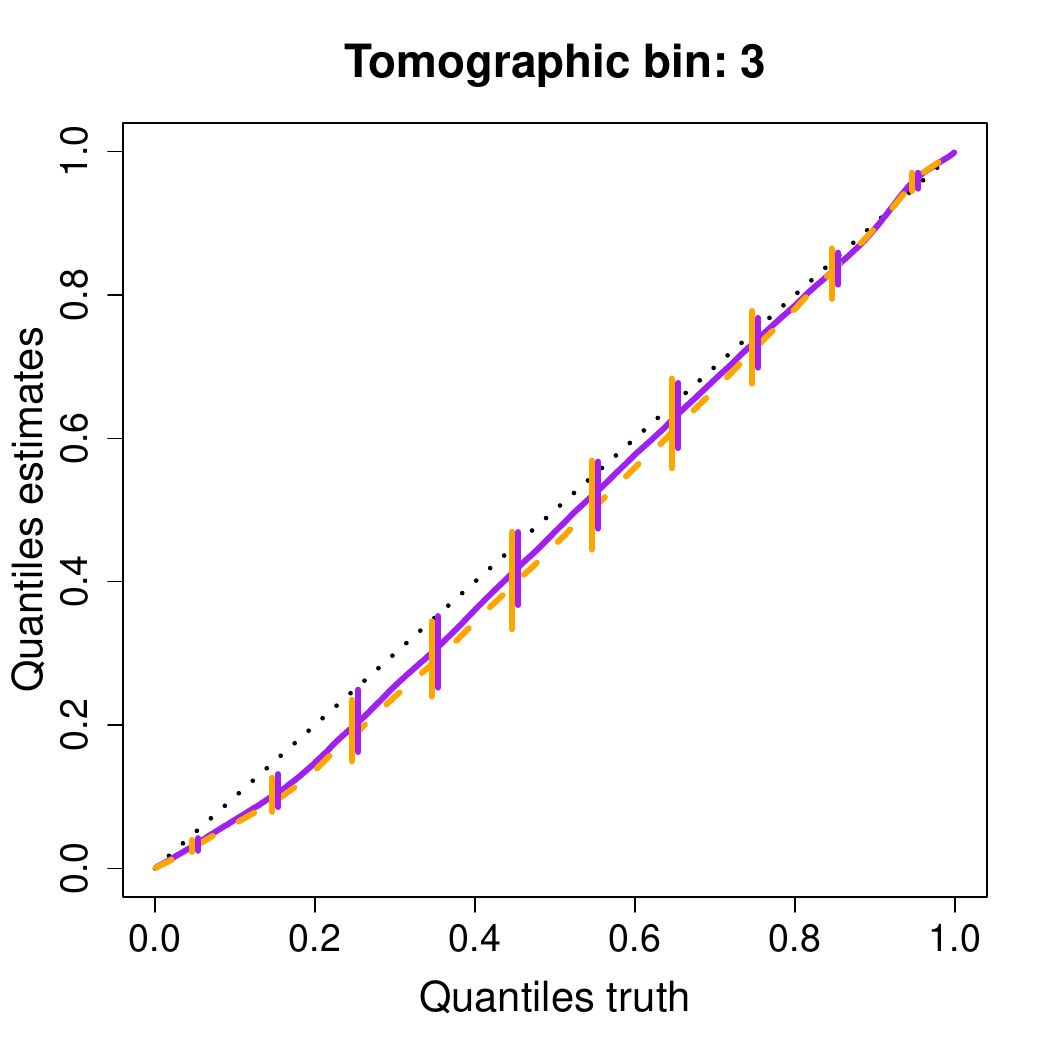}
        \includegraphics[width=0.32\columnwidth]{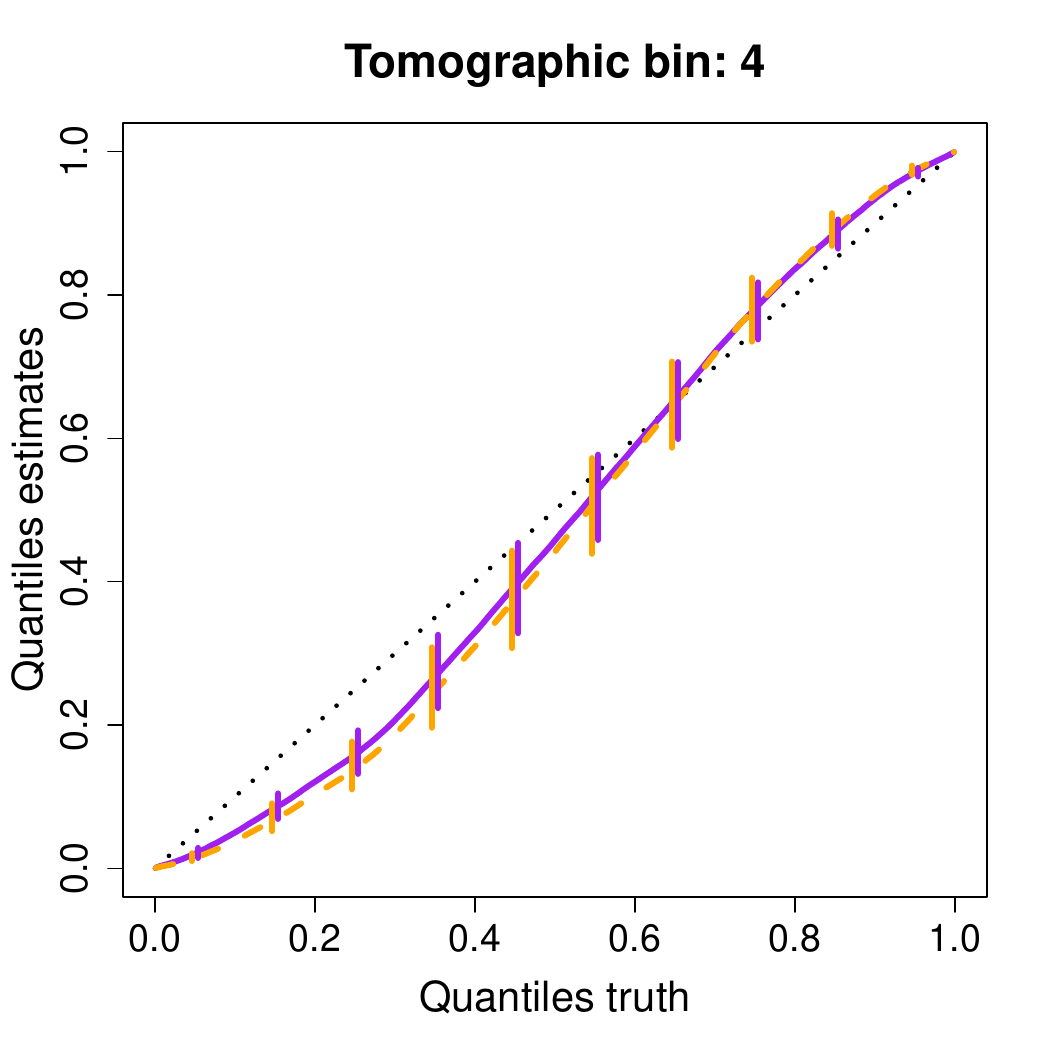}
        \includegraphics[width=0.32\columnwidth]{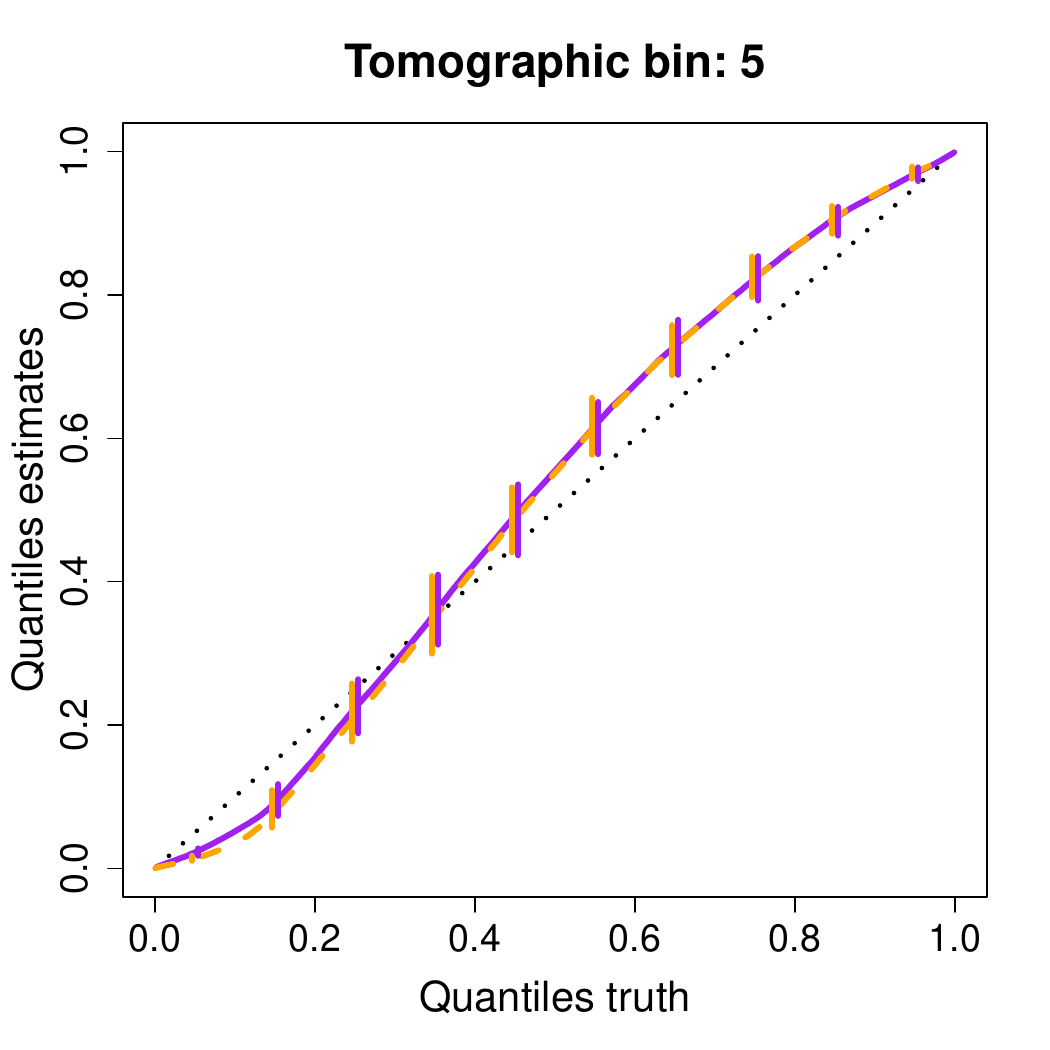}
       \includegraphics[width=0.32\columnwidth]{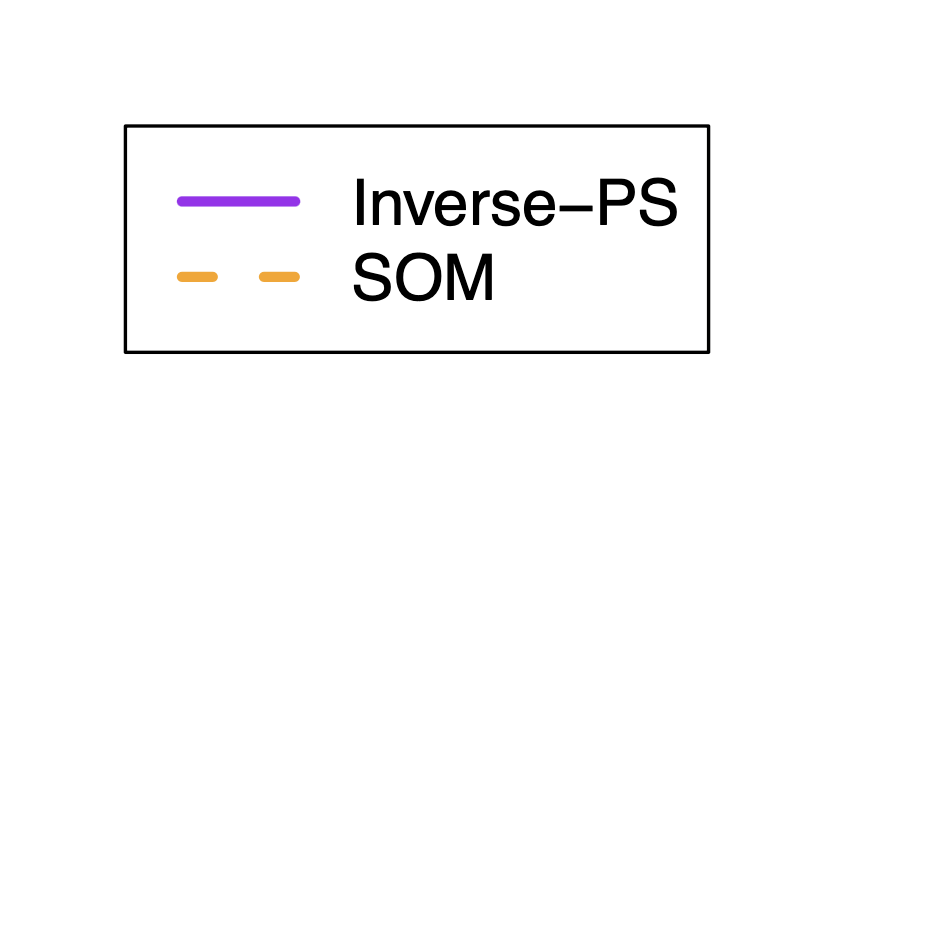}
    \caption{\baselineskip=15pt 
    Probability-probability plots (pp-plot) for the inverse-PS estimated distributions vs. the true (full) photometric redshift distributions 
    in purple lines, and pp-plots of the SOM estimated distributions vs. the gold selected true distributions in orange dashed lines, based on the \textit{StratLearn} tomographic binning (following Section~\ref{sec:StratLearn_bin_assignment}). For each tomographic bin, the averaged pp-plots across the 100 LoS are presented, with vertical bars illustrating $95\%$ intervals indicating the range of the central 95 pp-plot lines from the 100 LoS.  
    \label{fig:pp-plot_inverse_PS_SOM}} 
\end{minipage}
\end{figure*}

\begin{figure}
\centering
\begin{minipage}{.99\columnwidth}
  \centering
    \includegraphics[width=0.95\columnwidth]{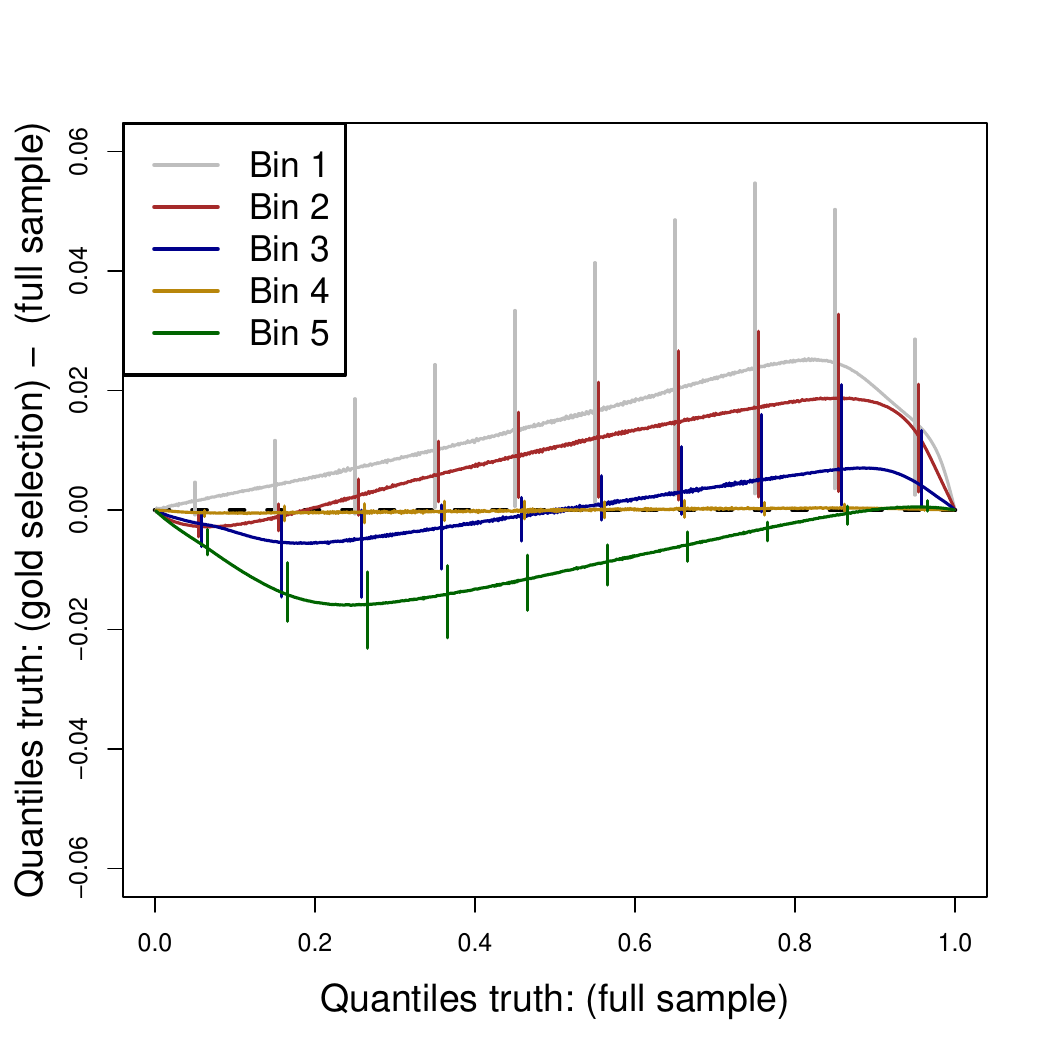}
    \caption{\baselineskip=15pt The figure illustrates (modified) pp-plots, comparing the true redshift distribution of the full sample (without quality cuts) with the true redshift distribution after applying gold selections, per tomographic bin (based on \textit{StratLearn} tomograhic binning, following Section~\ref{sec:StratLearn_bin_assignment}). For better readability, we illustrate ``modified" pp-plots, with the x-axis placing the quantiles of the full sample truth and the y-axis showing the quantiles of the gold selected truth subtracted by the x-axis values (the quantiles of the full sample true redshift distribution). For each tomographic bin, means of the pp-lines across the 100 LoS are illustrated, as well as $95\%$ intervals (vertical bars).  \label{fig:pp-plot_truth-full_vs_truth-gold}} 
\end{minipage}
\end{figure}

In Figure~\ref{fig:pp-plot_truth-full_vs_truth-gold}, we assess the differences between the true full photometric redshift distribution and the true redshift distribution after gold selection, for each of the five tomographic bins. Figure~\ref{fig:pp-plot_truth-full_vs_truth-gold} presents a (modified) pp-plot, illustrating the full true photometric distributions (on the x-axis) vs. the gold selected true distributions (on the y-axis); with the modification that the x-axis values (full true distribution quantiles) are subtracted from the y-axis (gold selection truth quantiles) for better visibility of the distribution differences. 
Figure~\ref{fig:pp-plot_truth-full_vs_truth-gold} 
illustrates that there are some mild changes in the underlying truth when applying the gold selection quality cuts (compared to the full true photometric sample) for bins 1,2,3, and 5. In bin 4, the true photometric distributions (before and after gold selection cuts) are approximately the same.

Finally, as noted in Section~\ref{sec:related_literature}, direct redshift calibration methods are generally prone to high variance, in 
particularly in the presence of a small number of large weights. While we have demonstrated improvement of inverse-PS upon SOM for estimation of the redshift population distribution shapes on the \textit{StratLearn}-based binning, it is true that the inverse-PS estimate can generally be affected by the same large variance instability. We note however that the formulation of the weights via propensity scores enables the use of methods developed in the (causal inference) propensity score literature to improve assessment and estimation of the propensity scores for direct redshift calibration (e.g., \citealt{imai2004causal,austin2015moving,Pirracchio2014,ridgeway2017toolkit,autenrieth2021stacked}), which will be the subject of a dedicated future work.

\section{Discussion}
\label{sec:discussion}

This paper introduced a novel, statistically principled method that improves photometric redshift calibration for weak lensing.
The central plank of our approach is the estimation of individual galaxy photo-$z$ conditional densities within a Bayesian hierarchical model, coupled with the $\textit{StratLearn}$ framework, a recently proposed statistically principled method for learning under non-representative source/training data in the presence of covariate shift \citep{autenrieth2021stratifiedASA}. The computation of galaxy-level conditional density estimates allows us to introduce an alternative tomographic binning strategy to the previously used $z_B$-based binning \citep{benitez2000bayesian}. 
We presented a hierarchical Bayesian framework, (\textit{StratLearn}-Bayes), to model summaries of the conditional density estimates to obtain nearly unbiased photometric redshift population mean estimates within tomographic bins. 

We evaluated our method on a comprehensive and realistic simulation study  (\citetalias{wright2020photometric}; \citealt{van2020testing}) mimicking the KiDS+VIKING-450 dataset \citep{wright2019kids+,hildebrandt2020kids+}
with realistic photometric noise and spectroscopic incompleteness. The results of this study can be summarized in four points: 
\begin{enumerate}
\item The \textit{StratLearn} conditional density-based tomographic binning strategy substantially improves upon the $z_B$ tomographic bin assignment, with an overall binning accuracy of ${\sim}62.2\%$ using \textit{StratLearn}, compared to ${\sim}52.6\%$ using $z_B$.
\item The \textit{StratLearn}-Bayes model leads to the lowest bias in the estimation of tomographic redshift population means. On average across the five tomographic bins, the proposed \textit{StratLearn}-Bayes method leads to an absolute bias of 0.0052, a substantial improvement over the previously best calibration method, SOM with $z_B$ binning \citepalias{wright2020photometric}, of 0.0085 average absolute bias. The strong reduction of bias is accompanied by a slight increase in uncertainty, leading to an average standard deviation of 0.0067, compared to 0.0051.
Using the \textit{StratLearn}-Bayes framework, we find a maximum bias of
 $\Updelta \langle z \rangle =  0.0095 \pm 0.0089$, slightly below the potentially critical bias value of $ \Updelta \langle z \rangle > 0.01 $, 
compared to SOM based on $z_B$ binning, which leads to maximum biases of $0.0135 \pm 0.0052$ and $0.0147 \pm 0.0040$.
\item While the previously best calibration method, SOM based on $z_B$ binning, requires systematic quality cuts to define a gold sample \citepalias{wright2020photometric}, our method does not require any cuts of the photometric sample. Thus, together with the improved tomographic bin assignment, the \textit{StratLearn}-Bayes framework delivers an increase of ${\sim}18\%$ in the galaxies available for the cosmic shear analysis.
\item We demonstrate how propensity scores can be employed via inverse-PS weighting in a direct redshift calibration approach to obtain realistic estimates of the redshift population distribution shapes per tomographic bin. Given the newly proposed \textit{StratLearn} binning, we show that using inverse-PS leads to a better approximation of the true photometric population distributions compared to employing SOM for estimation of the gold selected population distributions (with the additional advantage of not requiring any quality cuts).
\end{enumerate}

Finally, we believe that the improved tomographic binning assignment, the reduction of population mean bias within tomographic bin, and the increase in the number of galaxies available for cosmic shear analysis will have a substantial impact on the eventual scientific results and cosmological parameter inference.  
Analysing the final KiDS data release with our improved calibration method might 
lead to more precise and more accurate 
constraints on cosmological parameter estimates, particularly on $S_8$, the clustering strength of (predominantly dark) matter. We further believe that the proposed method might provide a powerful tool to improve the analysis of present and upcoming cosmic shear analysis.
We will investigate if the expected availability of larger spectroscopic source data sizes might allow further reduction of bias and variability to meet the stringent accuracy requirements of Euclid \citep{laureijs2011euclid} and 
the Legacy Survey of Space and Time 
\citep[LSST;][]{abell2009lsst}.

\section{Acknowledgements:}
We would like to thank Alan Heavens and Andrew Jaffe for valuable discussions. This work was supported by the UK Engineering and Physical Sciences Research Council [grant number EP/W522673/1]. 
David van Dyk acknowledges partial support from the UK Engineering and Physical Sciences Research Council [EP/W015080/1]; 
Roberto Trotta's work was partially supported by STFC in the UK [ST/P000762/1,ST/T000791/1]; 
and David Stenning acknowledges the support of the Natural Sciences and Engineering Research Council of Canada (NSERC) [RGPIN-2021-03985]. 
Van Dyk, Stenning and Autenrieth acknowledge support from the Marie-Skodowska-Curie RISE [H2020-MSCA-RISE-2019-873089] Grant provided by the European Commission. RT acknowledges co-funding from Next Generation EU, in the context of the National Recovery and Resilience Plan, Investment PE1 – Project FAIR ``Future Artificial Intelligence Research''. This resource was co-financed by the Next Generation EU [DM 1555 del 11.10.22]. RT is partially supported by the Fondazione ICSC, Spoke 3 ``Astrophysics and Cosmos Observations'', Piano Nazionale di Ripresa e Resilienza Project ID CN00000013 ``Italian Research Center on High-Performance Computing, Big Data and Quantum Computing'' funded by MUR Missione 4 Componente 2 Investimento 1.4: Potenziamento strutture di ricerca e creazione di ``campioni nazionali di R\&S (M4C2-19 )'' - Next Generation EU (NGEU).
BJ acknowledges support by STFC Consolidated Grant ST/V000780/1. 
AHW is supported by an European Research Council Consolidator Grant (No. 770935), as well as by the Deutsches Zentrum für Luft- und Raumfahrt (DLR), made possible by the Bundesministerium für Wirtschaft und Klimaschutz, and acknowledges funding from the German Science Foundation DFG, via the Collaborative Research Center SFB1491 "Cosmic Interacting Matters - From Source to Signal".
This work is based on observations made with ESO Telescopes at the La Silla Paranal Observatory under programme IDs 100.A-0613, 102.A-0047, 179.A-2004, 177.A-3016, 177.A-3017, 177.A-3018, 298.A-5015. The MICE simulations have been developed at the MareNostrum supercomputer (BSC-CNS) thanks to grants AECT-2006-2-0011 through AECT-2015-1-0013. Data products have been stored at the Port d’Informació Científica (PIC), and distributed through the CosmoHub webportal (cosmohub.pic.es). 
Finally, this research was enabled in part by support provided by the Digital Research Alliance of Canada (alliancecan.ca) and the BC DRI Group.
\textit{The authors report there are no competing interests to declare.}

\section*{Data Availability}
Upon acceptance, the simulated data underlying this article as well as the documented code files will be made public.



\bibliographystyle{mnras}
\bibliography{bibliography} 



\appendix

\include{Supplement/supplement}


\bsp	
\label{lastpage}
\end{document}

%% file: Supplement/supplement.tex

{\huge\bf Appendix}

\section{Additional Details for Conditional Density Estimation}\label{sec_supp:conditional_densities}
In this section, we provide derivations of the {\it generalized} risk under the $L^2-$loss given in (\ref{formula:fzxrisk_source}). Following \cite{izbicki2017photo}, we start with the risk based on the general $L^2-$loss via
\begin{align}\label{form_supp:L2_risk}
  R_S(\hat{f}) = \iint(\hat{f}(z|x) - f(z|x))^2 d P_S(x)dz,
\end{align}
with $\hat{f}(z|x)$ being the full conditional density estimate of redshift $z$ given the covariates at point $x$, $f(z|x)$ being the true conditional density of $z$ given $x$, and $P_S(x)$ being the distribution of the source covariates. In extended form, (\ref{form_supp:L2_risk}) can be written as 
\begin{align}\label{form_supp:source_risk_extended}
    R_S(\hat{f}) =  \iint \hat{f}^2(z|x) d P_S(x)dz &- 2 \iint\hat{f}(z|x)f(z|x) d P_S(x)dz \nonumber \\
    &+\underbrace{\iint f^2(z|x) d P_S(x)dz}_\text{= constant $C$}, 
\end{align}
which up to the constant $C$ is equal to
\begin{align}\label{form_supp:source_risk_upconstant}
R_S(\hat{f}) =  \iint \hat{f}^2(z|x) d P_S(x)dz - 2 \iint \hat{f}(z|x) d P_S(x,z). 
\end{align}
From (\ref{form_supp:source_risk_extended}) to (\ref{form_supp:source_risk_upconstant}), the equality $dP_L(x,z) = f(z|x) dP_L(x)dz$ (via Radon-Nikodym derivative)  
is employed. Given the labelled source samples $(x_S,z_S)$, we can get an estimate of (\ref{form_supp:source_risk_upconstant}) via
\begin{align} \label{formula_supp:fzxrisk_source}
    \hat{R}_S(\hat{f}) = & \frac{1}{n_S} \sum_{k=1}^{n_S} \int \hat{f}^2 (z|x_S^{(k)}) dz  -  2 \frac{1}{n_S} \sum_{k=1}^{n_S}  \hat{f} (z_S^{(k)}|x_S^{(k)}),
\end{align} 
as presented in (\ref{formula:fzxrisk_source}).

\section{Additional Model Details}\label{supp_sec:add_model_details}

\subsection{Posterior Derivations}
This section provides the theoretical justification of the posterior derivations in Section~\ref{sec:Bayesian_model}. 

Deriving (\ref{form:joint_marginal}) from (\ref{form:joint_posterior_gaussian}) is obtained by integrating over the product of normal densities in (\ref{form:joint_posterior_gaussian}), which can analytically be done via completing the squares. We note that flipping the $z_i$ and $\hat{\zeta}_i$ in (\ref{form:joint_posterior_gaussian}) constitutes the standard normal-normal hierarchical model with Gaussian measurement errors on latent $z_i$ with Gaussian population. The analytical derivation of this model is a standard result in Bayesian statistics \citep[see, e.g.,][page 117]{gelman1995bayesian}, demonstrating that
\begin{align}
    \int_{\mathbb{R}}
    N(\hat{\zeta}_i| z_i , \hat{\tau}_i^2)  N(z_i | \mu, \sigma^2) dz_i = N(\hat{\zeta}_i | \mu, \hat{\tau}_i^2 + \sigma^2).
\end{align}
Mathematically, the densities   $N(\hat{\zeta}_i| z_i , \hat{\tau}_i^2) $ and     $N(z_i| \hat{\zeta_i} , \hat{\tau}_i^2) $ are identical, due to the symmetry of the normal distribution. It thus directly follows that
\begin{align}
    \int_{\mathbb{R}}
    N(z_i| \hat{\zeta}_i, \hat{\tau}_i^2)  N(z_i | \mu, \sigma^2) dz_i = N(\hat{\zeta}_i | \mu, \hat{\tau}_i^2 + \sigma^2),
\end{align}
which concludes (\ref{form:joint_marginal}) from (\ref{form:joint_posterior_gaussian}). 

With a uniform conditional prior density $p(\mu_b|\sigma_b)$, the Gaussian conditional posterior of $\mu_b$ given $\sigma_b, \hat{X}_b$ in (\ref{form:post_mu_cond_sigma}) then follows directly as another standard result \citep[see, e.g.,][page 117]{gelman1995bayesian}. 
More precisely,  (\ref{form:joint_marginal}) is a product of Gaussian densities, which yields a Gaussian density (the log-posterior is quadratic in $\mu_b$). The parameters of the Gaussian conditional posterior in (\ref{form:post_mu_cond_sigma}) are obtained by considering the $\hat{\zeta}_i$ as independent estimates of $\mu_b$ with variances $(\hat{\tau}_i^2 + \sigma_b^2 )$.  

\subsection{Posterior Distribution of $\sigma$}
In our hierarchical Bayesian model, the marginal posterior distribution of the population variance $\sigma_b$ can be obtained via
\begin{align}\label{supp_form:posterior_sigma}
    p(\sigma_b |  \hat{X}_{\bold{{n_{Tb}}}}) \propto p(\sigma_b) V_{\mu}^{1/2} \prod_{j = 1}^J (\tau_j^2 + \sigma_b^2)^{-1/2} \exp{\left( -\frac{ (\hat{\zeta}_j - \tilde{\mu}_b)^2}{2(\hat{\tau}_j^2 + \sigma_b^2)} \right)}
\end{align}
with $\tilde{\mu}_b$ and $V_{\mu_b}$ as defined in (\ref{form:post_mu_cond_sigma_params}). Choosing a uniform prior on $\sigma_b$, $p(\sigma_b) \propto 1$, makes (\ref{supp_form:posterior_sigma}) a proper posterior density \citep[see, e.g.,][page 117]{gelman1995bayesian}.

\subsection{Justification of Stacked Population Variance Estimate}\label{section:append_stacking_justification}

In this section, we justify the stacked estimator of the marginal redshift population distribution $p_b(z)$ in (\ref{form:stack_marginal_est}). Precisely, we can express $p_b(z)$ via
\begin{align}\label{form:stack_marginal_est_1}
    p_b(z) = \int p_b(z|x) p_b(x) dx,
\end{align}
with $x$ being photometric magnitudes/colors,  
$p_b(x)$ being the distribution of the covariates (magnitudes/colors) of bin $b$, and $p_b(z|x)$ being the conditional distribution of redshift $z$ given covariates $x$ of bin $b$. 
By assuming the set of photometric magnitudes/colors are finite, and assuming that the conditional distribution of $z$ given $x$ is the same for all bins (i.e., $p_b(z|x) = p(z|x)$), we obtain
\begin{align}\label{form:stack_marginal_est_2}
    p_b(z) = \sum_i p(z|x = x_i) p_b(x = x_i).
\end{align}
We have an estimate of the conditional densities $p(z_i |x_i) \approx f(z_i |x_i)$. Since $z_i \overset{\text{iid}}{\sim}  p(z)$, it holds that $p(z| x = x_i) = p(z_i |x = x_i) \approx \hat{f}(z_i |x_i)$. 
Further, we can approximate $p_b(x = x_i)$ by counting occurrences of $x_i$ in the sample of observed magnitudes/colors within tomographic bin $b$.
Alternatively, averaging over all estimated conditional densities $\hat{f}(z_i |x_i)$ of galaxies in bin $b$ directly incorporates these occurrence frequencies,   
leading to the estimator in (\ref{form:stack_marginal_est}).

\bigskip

\section{Additional Figures}\label{sec:additional_figures}
\begin{figure}
\centering
\begin{subfigure}[b]{0.49\columnwidth}
   \includegraphics[width=0.89\columnwidth]{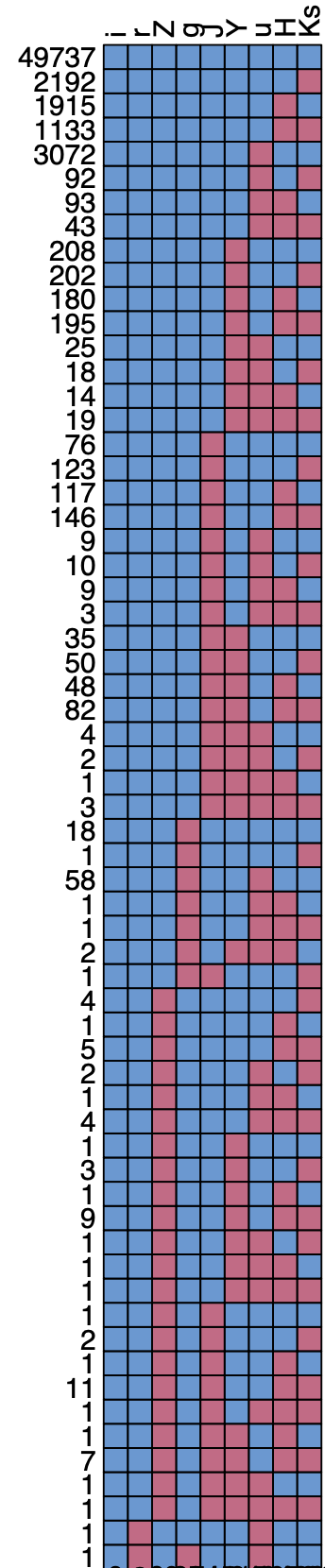}
   \caption{Magnitudes.}
   \label{figure:missing_magnitudes} 
\end{subfigure}
\begin{subfigure}[b]{0.49\columnwidth}
    \vspace{0.3cm}
   \includegraphics[width=0.89\columnwidth]{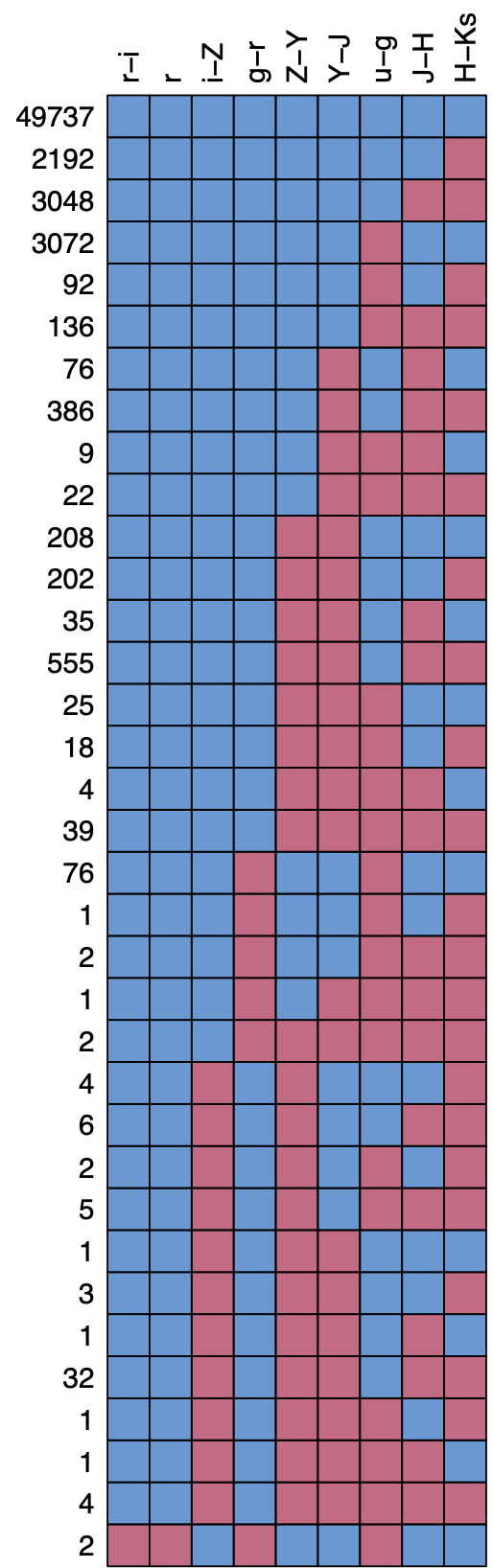}
   \caption{Colors.}
   \label{figure:missing_colors}
\end{subfigure}
\caption{``Missing data" pattern for (a) magnitudes, and (b) colors, of a random subsample of 60K galaxies from the photometric survey. \label{figure:missing_magnitudes_and_colors}}
\end{figure}

\begin{figure}
\centering
\begin{minipage}{.85\columnwidth}
  \centering
 \includegraphics[width=0.99\columnwidth]{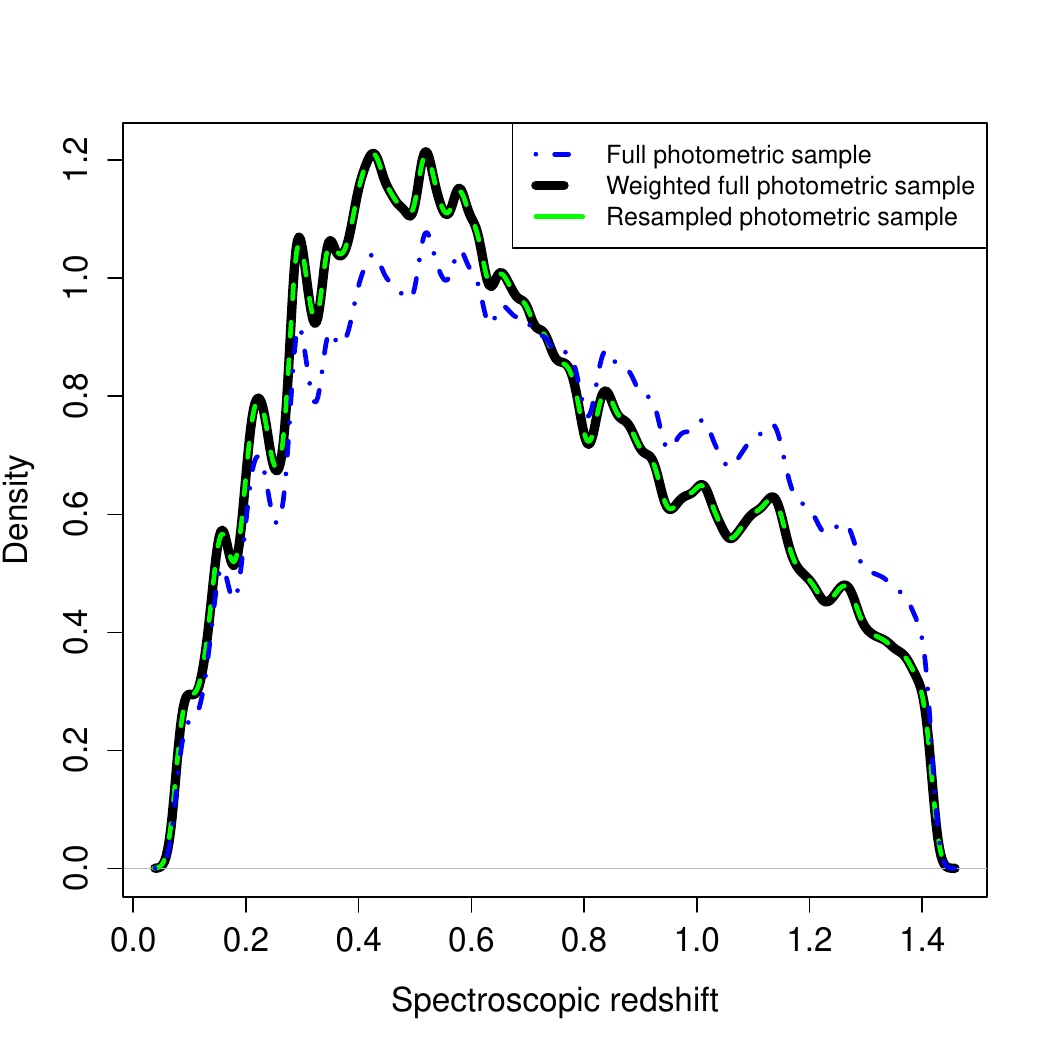}
 \caption{\baselineskip=15pt Spectroscopic (true) redshift distributions of the photometric survey (not known in practice), with and without incorporation of shear-measurements weights. The blue dashed line shows the redshift distribution of the full photometric sample $D_T^{u}$ (before resampling). The black line shows the redshift density of the full photometric density weighted by the shear-measurment weights $\hat{w}$, and the green line illustrates the redshift density of the resampled sample $D_T$. The black and green line perfectly match, with a mean difference of ${\sim}10^{-4}$, illustrating that there is negligible difference of targeting the resampled distribution (as in this study) and targeting the weighted distribution (as in \citetalias{wright2020photometric}). 
 \label{fig:resample_vs_reweight}}
\end{minipage}
\end{figure}

\begin{figure}
\centering
\begin{minipage}{.78\columnwidth}
  \centering
 \scalebox{.9}{  \includegraphics[width=0.99\columnwidth]{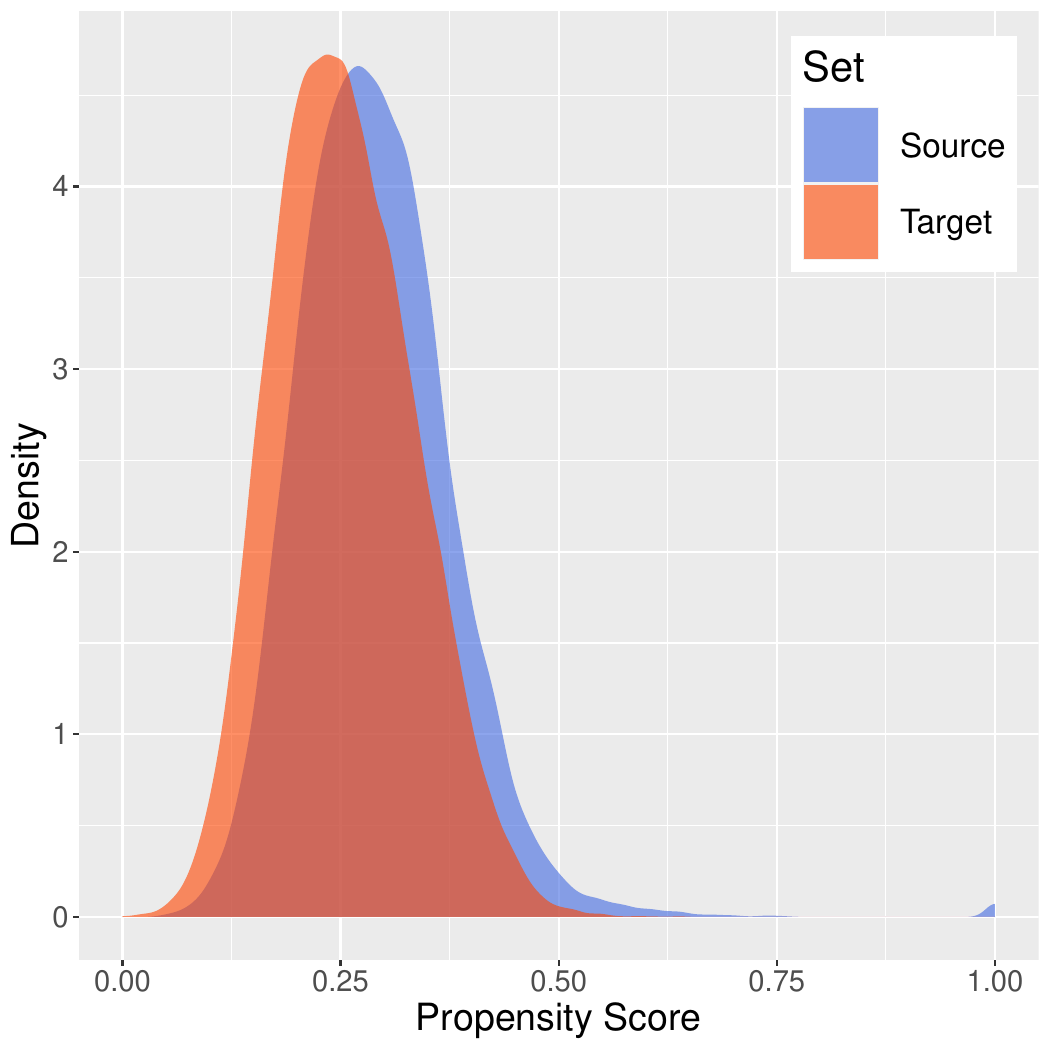}}
 \caption{\baselineskip=15pt  Propensity score distributions of source and target data in Section~\ref{sec:numerical_demonstrations}. \label{figure:PS_distributions}}
\end{minipage}
\end{figure}

\begin{figure*}
\centering
\begin{minipage}[t]{.99\textwidth}
  \centering
    \includegraphics[width=0.19\columnwidth]{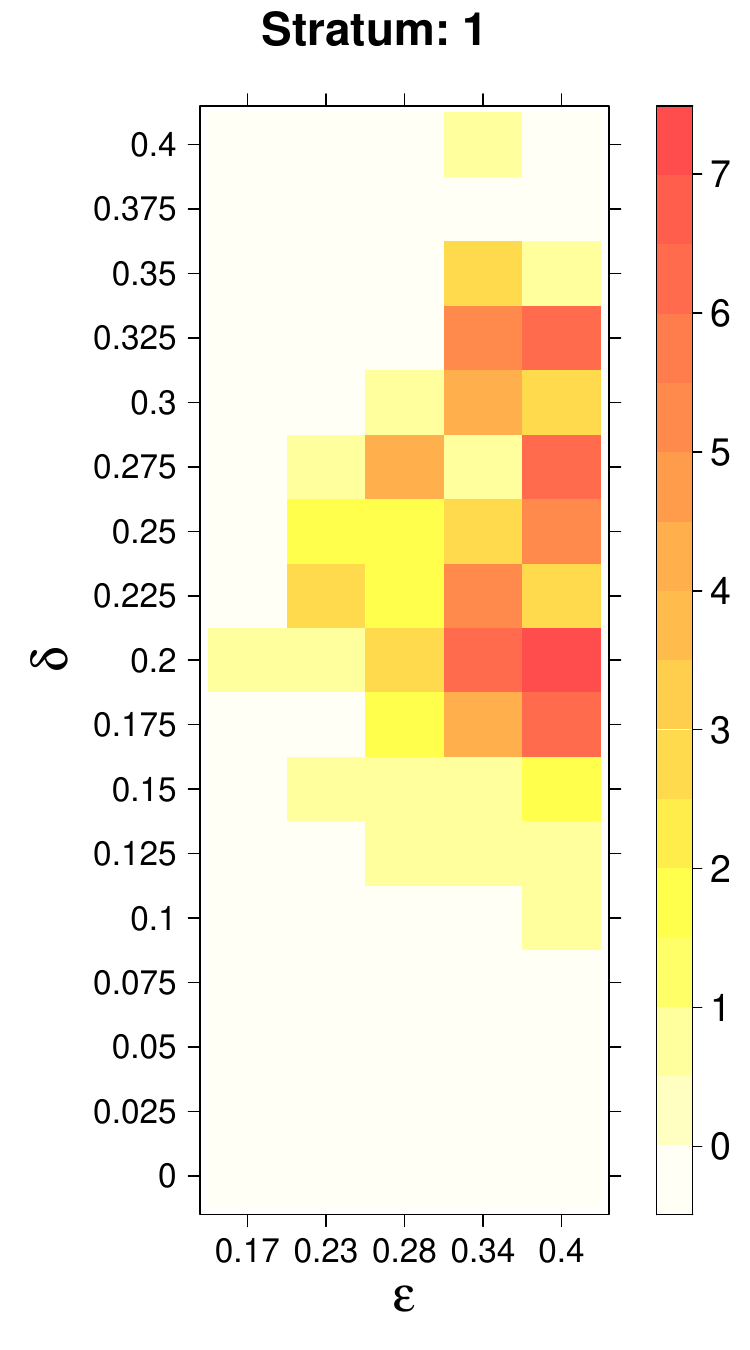}
    \includegraphics[width=0.19\columnwidth]{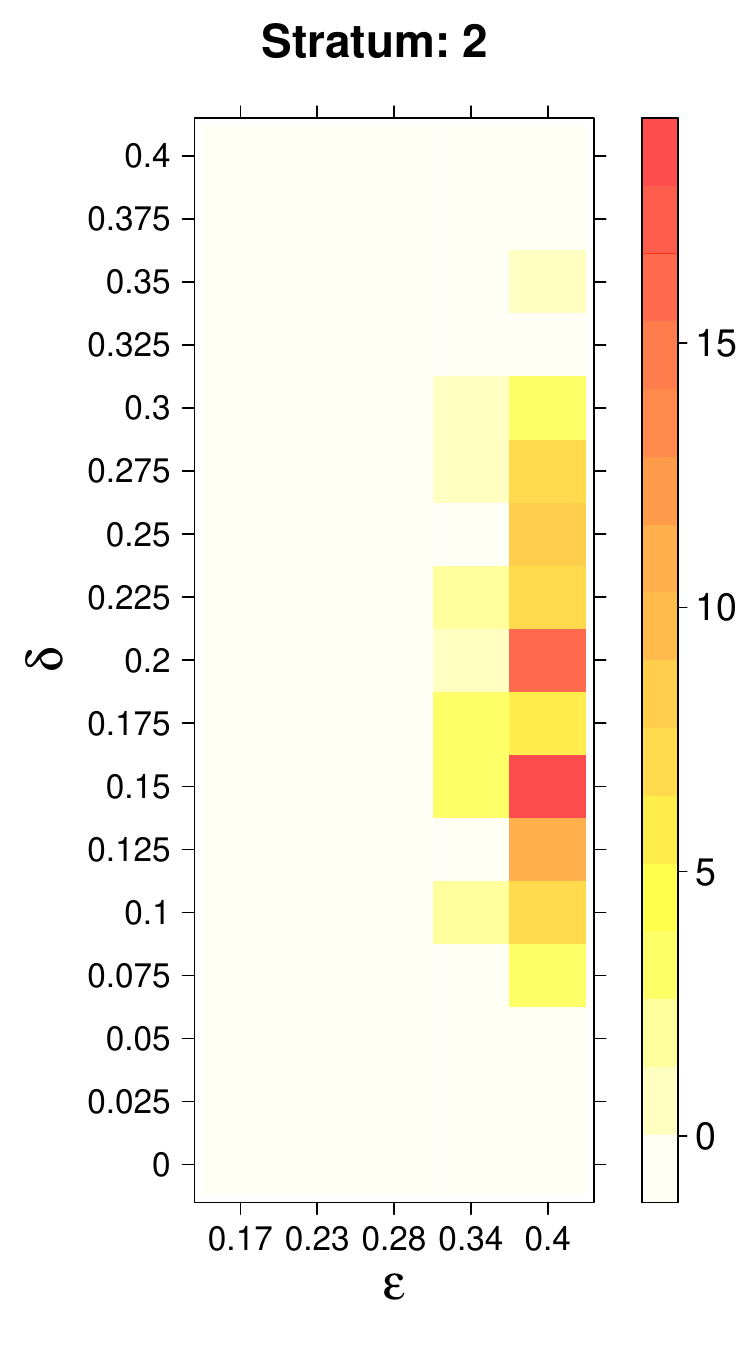}
    \includegraphics[width=0.19\columnwidth]{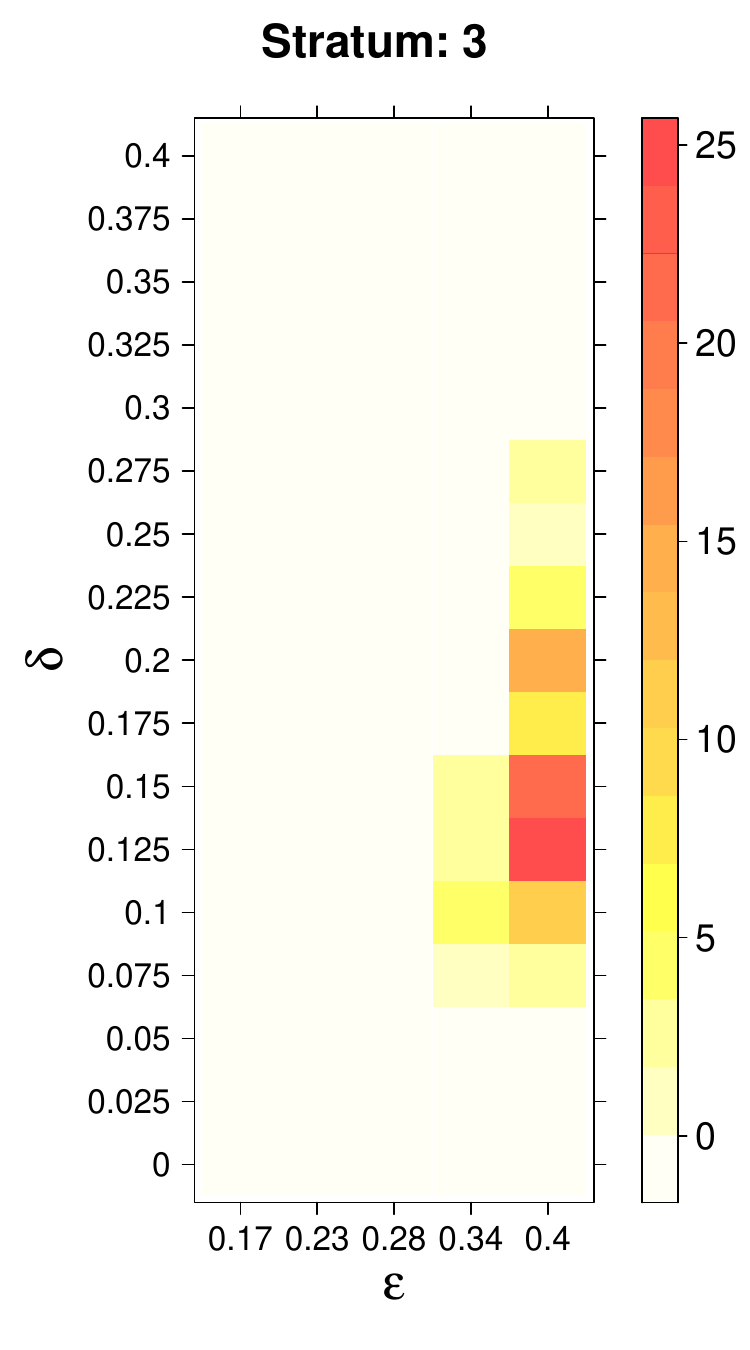}
    \includegraphics[width=0.19\columnwidth]{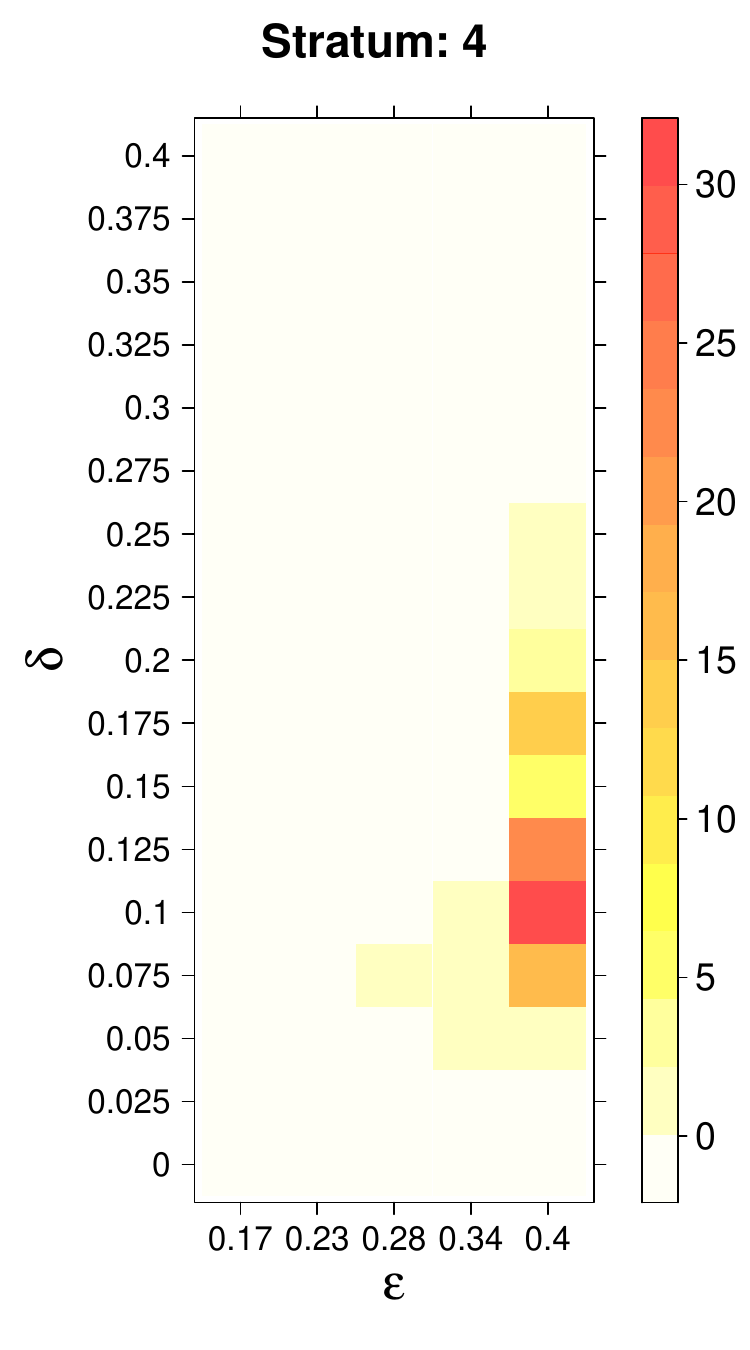}
    \includegraphics[width=0.19\columnwidth]{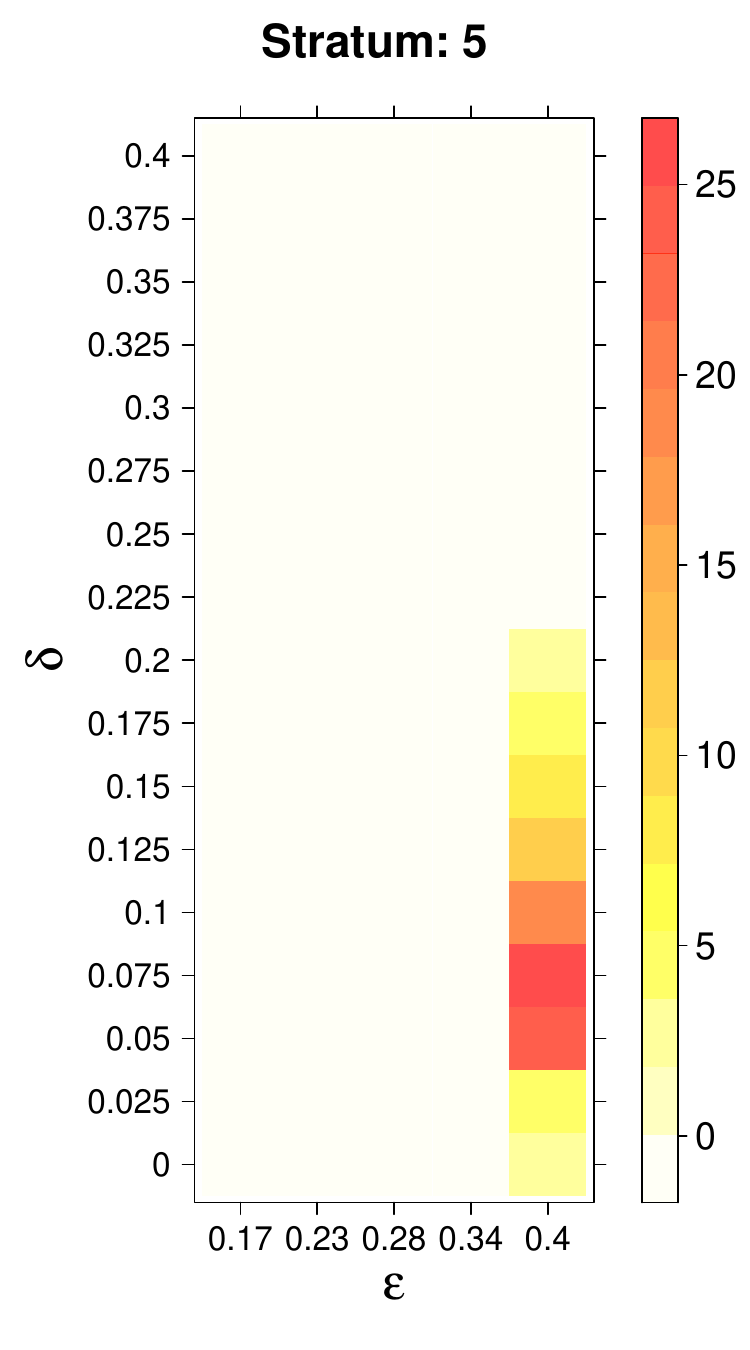}
    \caption{\baselineskip=15pt Illustration of the optimized hyperparameters for the Series conditional density estimator (described in Section~\ref{sec:StratLearn}). The heatmaps illustrate the prevalence of the various hyperparameter combinations across the 100 LoS for each stratum, respectively. Hyperparameters were optimized as described in Sections~\ref{sec:StratLearn} and \ref{sec:data}. We note that for the $\epsilon$ hyperparameter two additional grid values (0.05 and 0.11) were available, but never selected as the optimal hyperparameter, and thus not illustrated here for better readability. In addition, we note that our preliminary investigation showed that values for $\epsilon$ which were greater than 0.4 did not (or only marginally) lead to risk improvements; we thus used 0.4 as the maximum value for the $\epsilon$ grid. \label{figure:hyperparam_Series}}  
\end{minipage}
\end{figure*}

\begin{figure*}
\centering
\begin{minipage}[t]{.99\textwidth}
  \centering
    \includegraphics[width=0.19\columnwidth]{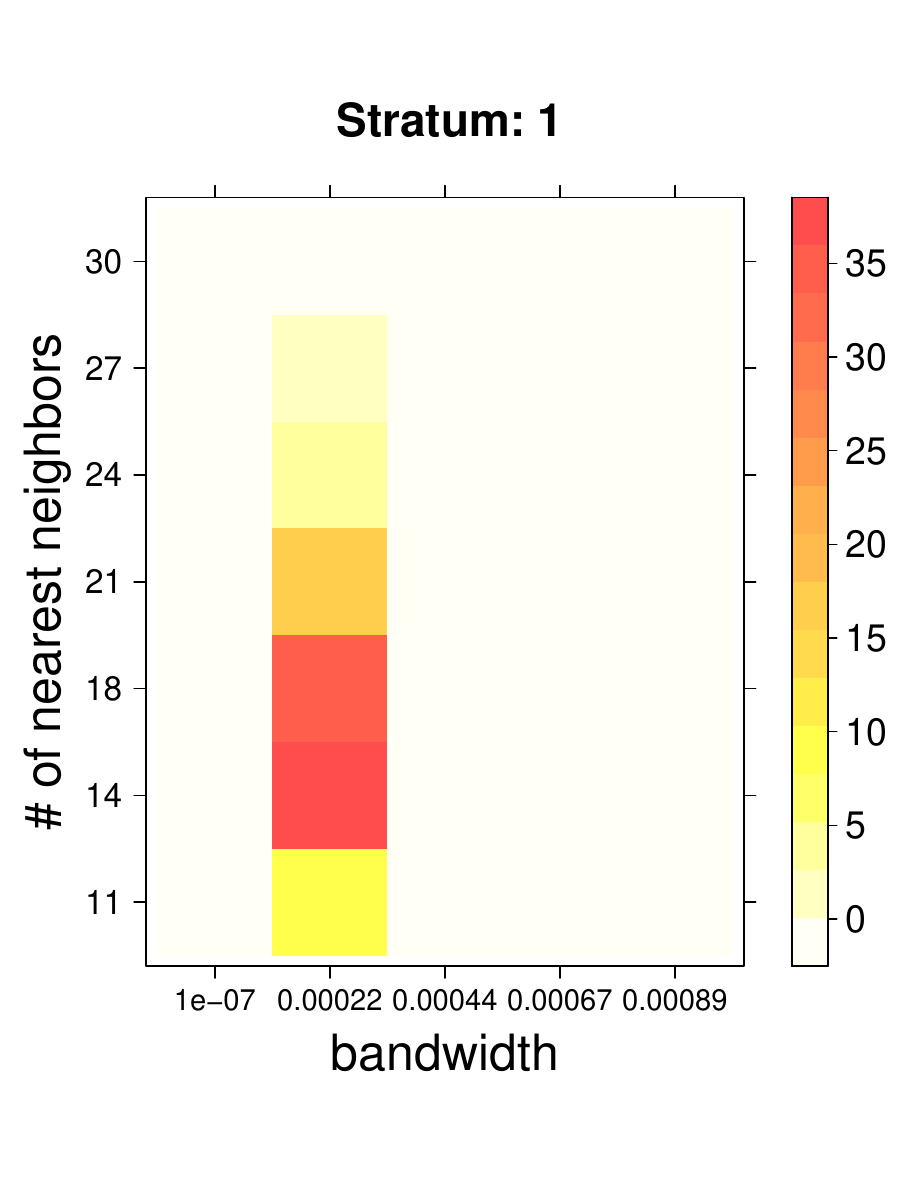}
    \includegraphics[width=0.19\columnwidth]{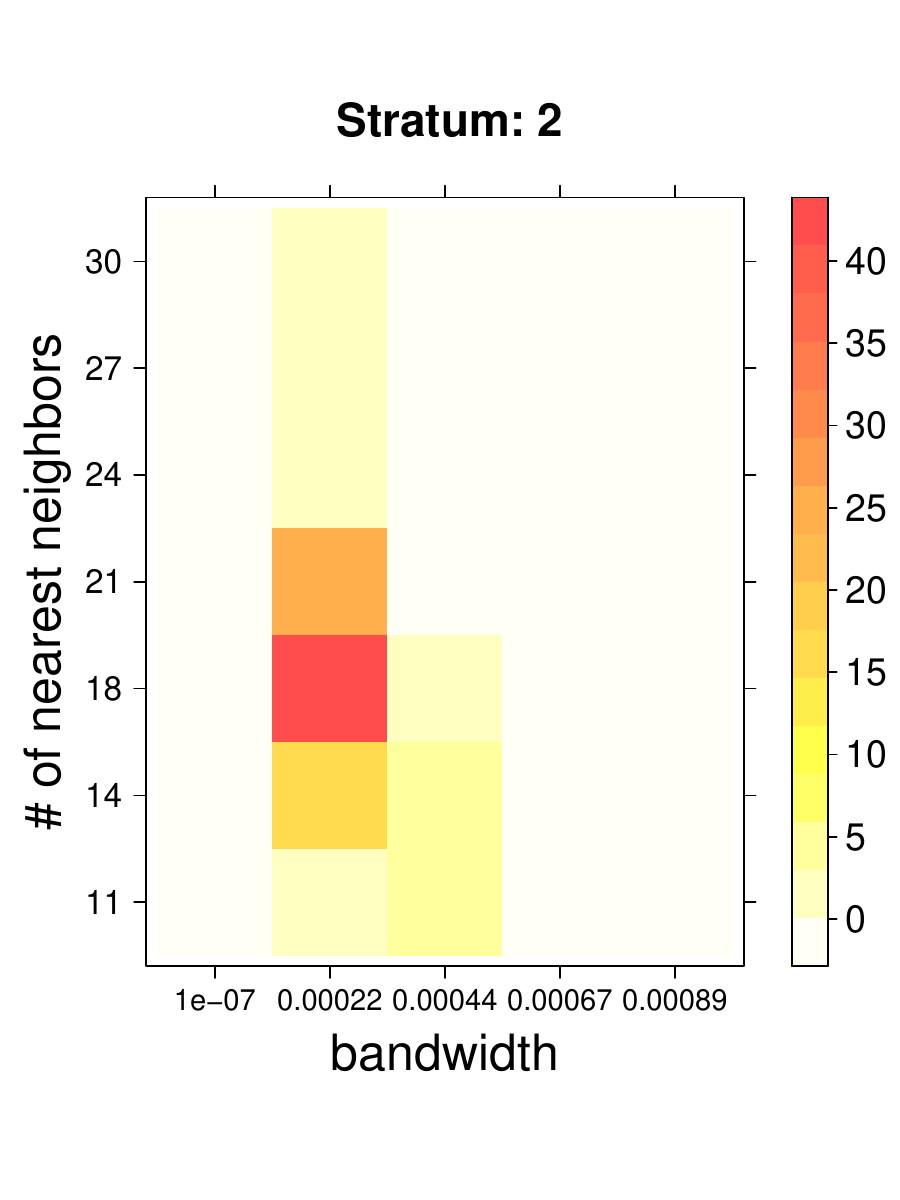}
    \includegraphics[width=0.19\columnwidth]{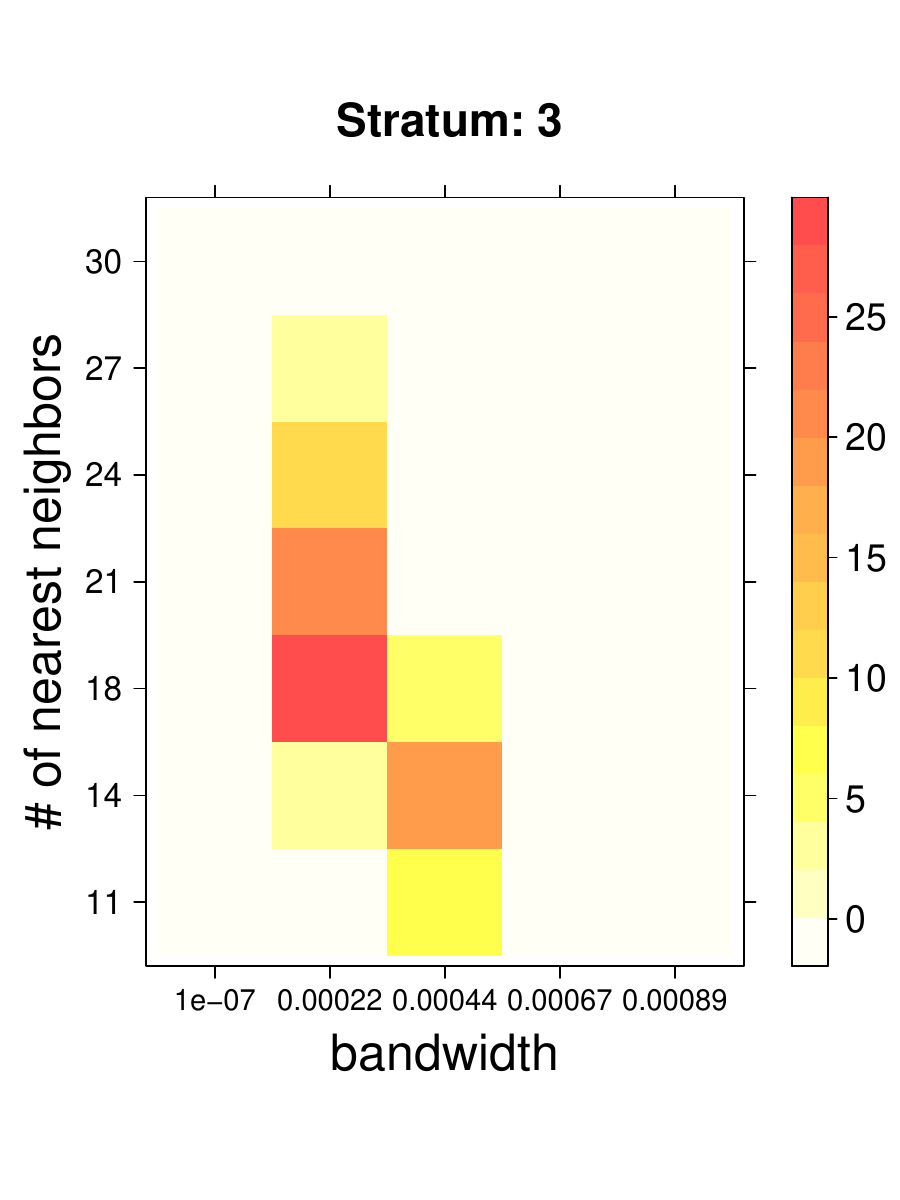}
        \includegraphics[width=0.19\columnwidth]{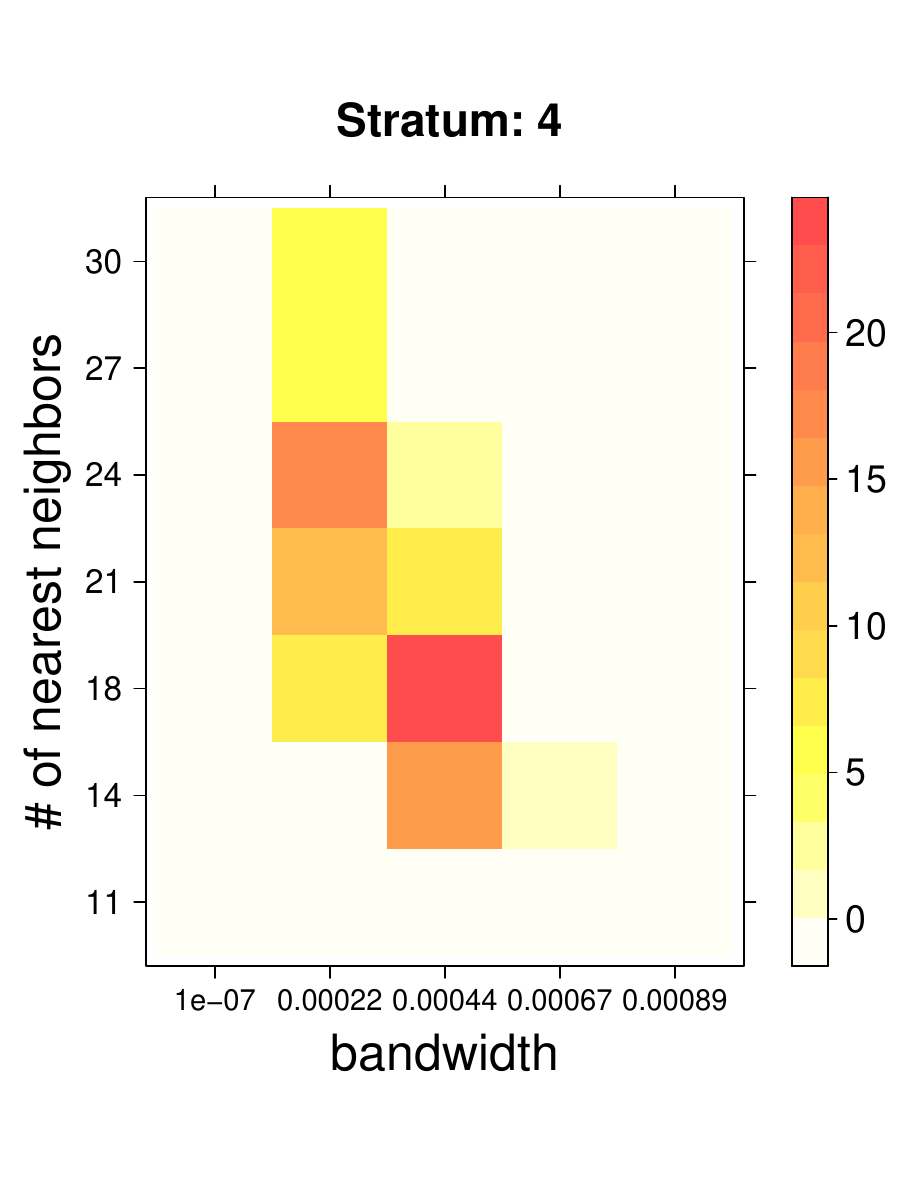}
        \includegraphics[width=0.19\columnwidth]{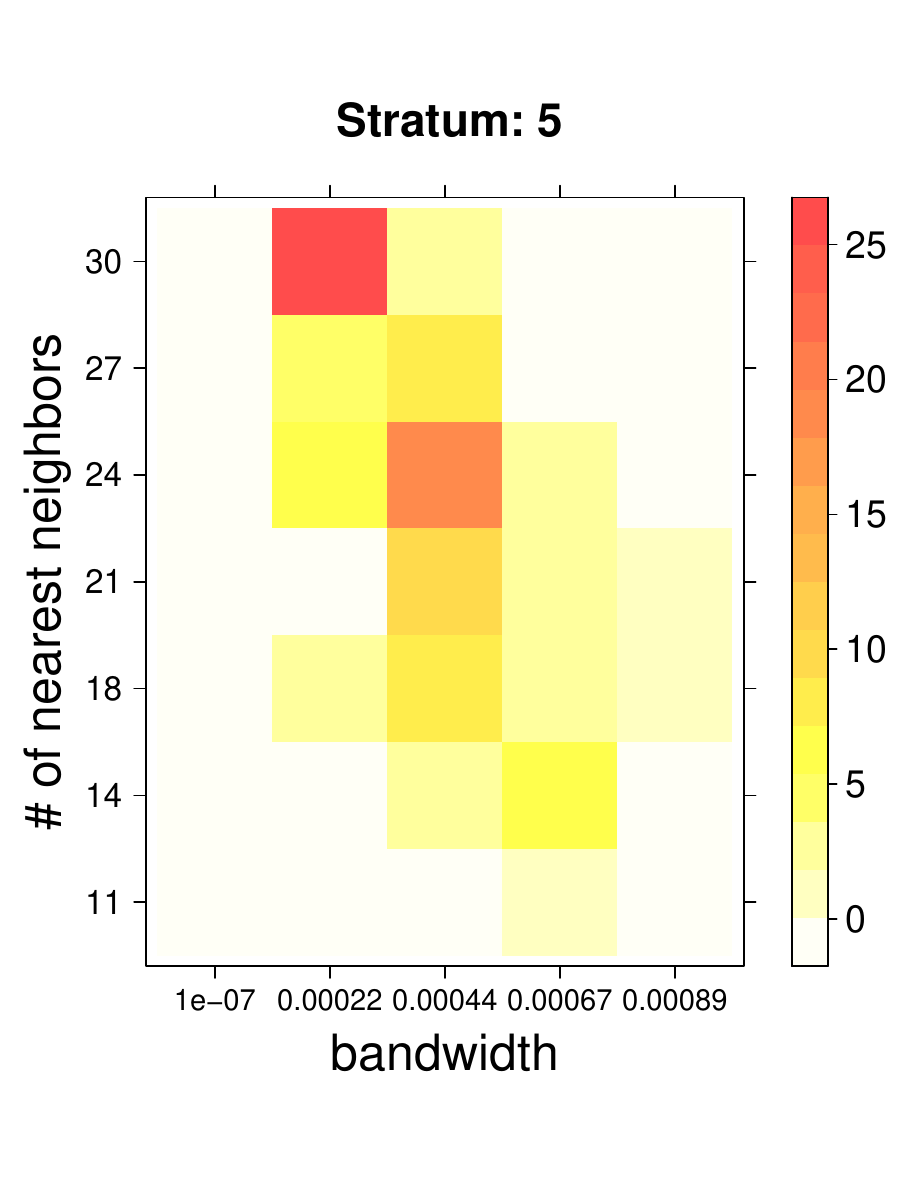}
    \caption{\baselineskip=15pt Illustration of the optimized hyperparameters for the Ker-NN conditional density estimator (described in Section~\ref{sec:StratLearn}). The heatmaps illustrate the prevalence of the various hyperparameter combinations across the 100 LoS for each stratum, respectively. Hyperparameters were optimized as described in Sections~\ref{sec:StratLearn} and \ref{sec:data}. 
    For the `bandwidth' hyperparameter five additional grid values (0.0011 0.0013 0.0015 0.0018 0.002) were available, but never selected as the optimal hyperparameter. For the `$\#$ of nearest neighbors parameter' parameter three additional grid values (2,5,8) were available but never selected as the optimal hyperparameter. These grid values are not illustrated in the heatmaps for better readability. 
    \label{figure:hyperparam_KerNN} } 
\end{minipage}
\end{figure*}

\begin{figure}
\centering
\begin{minipage}{.99\columnwidth}
  \centering
 \includegraphics[width=0.95\columnwidth]{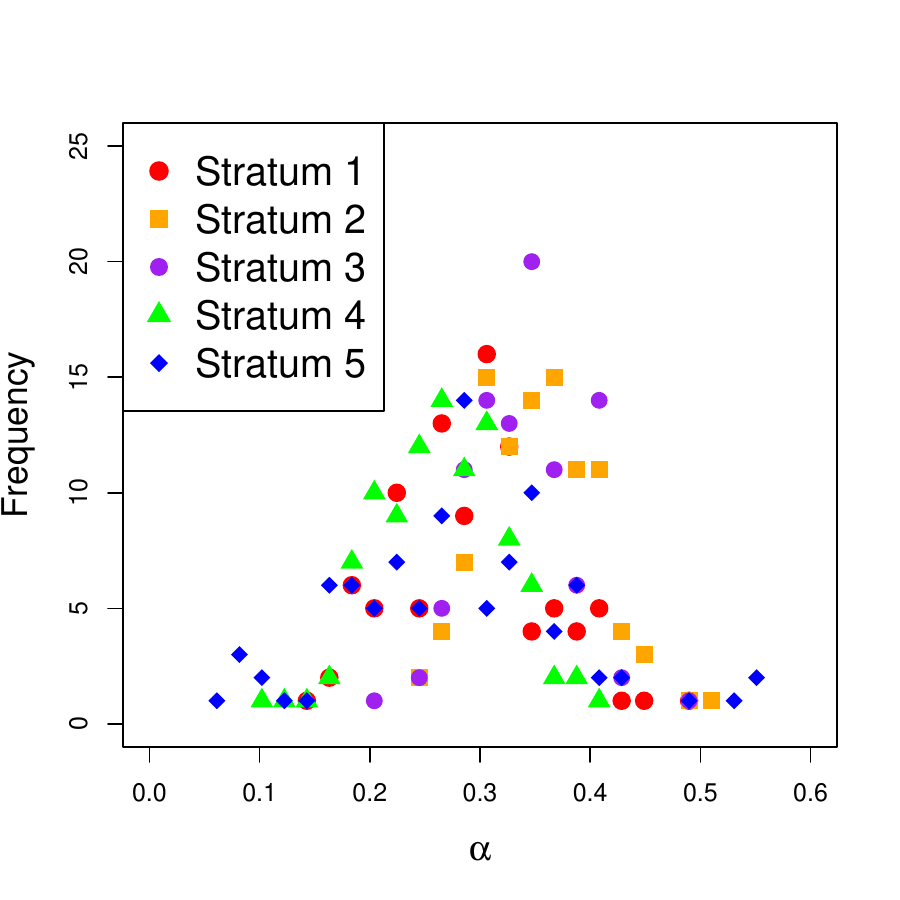} 
\end{minipage}
\caption{\baselineskip=15pt  Frequency of the optimized $\alpha$ hyperparameter for the Comb conditional density estimator (described in Section~\ref{sec:StratLearn}) across the 100 LoS, for the five strata respectively. The $\alpha$ hyperparameter was optimized as described in Sections~\ref{sec:StratLearn} and \ref{sec:data}. 
\label{figure:hyperparam_Comb}} 
\end{figure}

\begin{figure}
\centering
\begin{minipage}{.99\columnwidth}
  \centering
 \includegraphics[width=0.95\columnwidth]{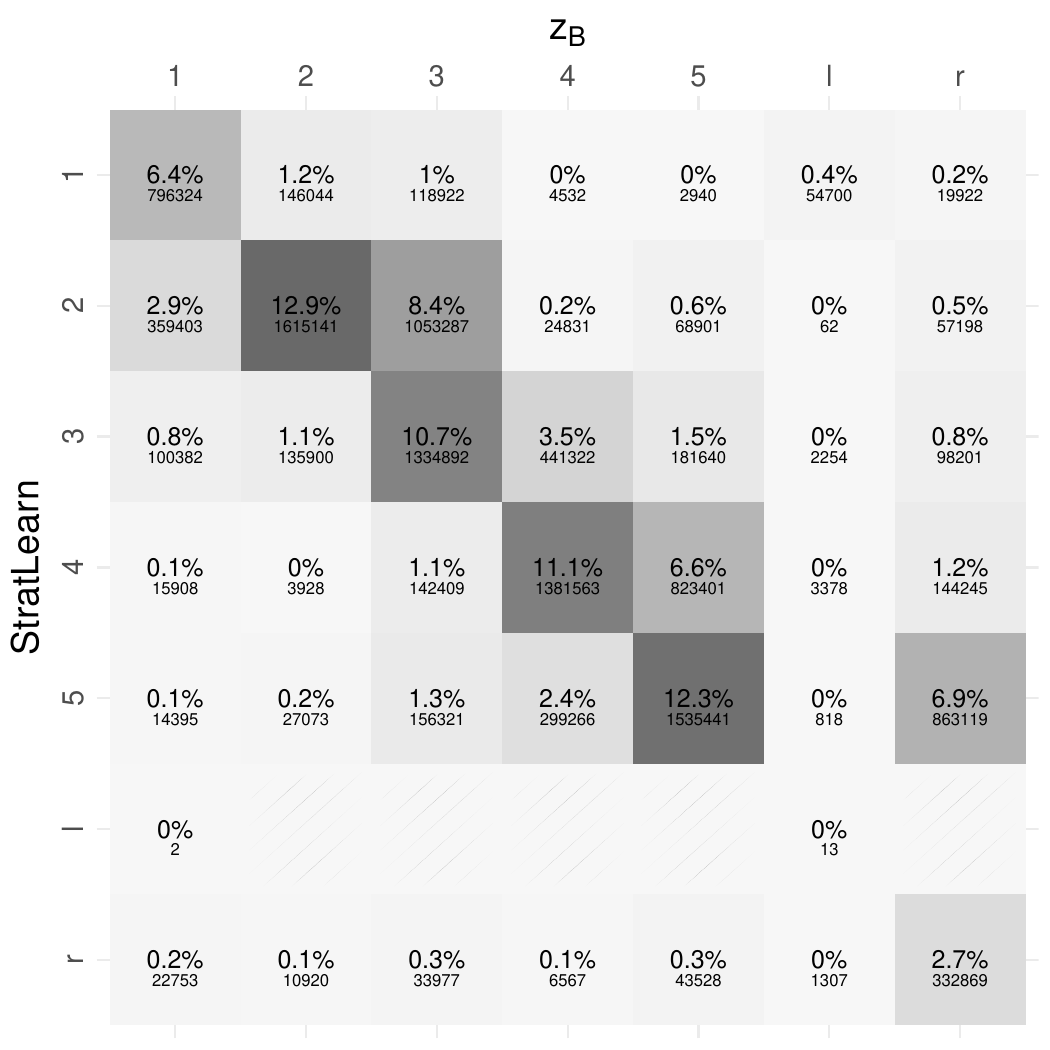} 
\end{minipage}
\caption{\baselineskip=15pt Changes in bin assignment using \textit{StratLearn} vs $z_B$ binning, averaged across the 100 LoS. 
    \label{figure:conf_matrix_StratLearn_vs_ZB}} 
\end{figure}

\begin{figure}
\centering
\begin{minipage}{.99\columnwidth}
  \centering
    \includegraphics[width=0.99\columnwidth]{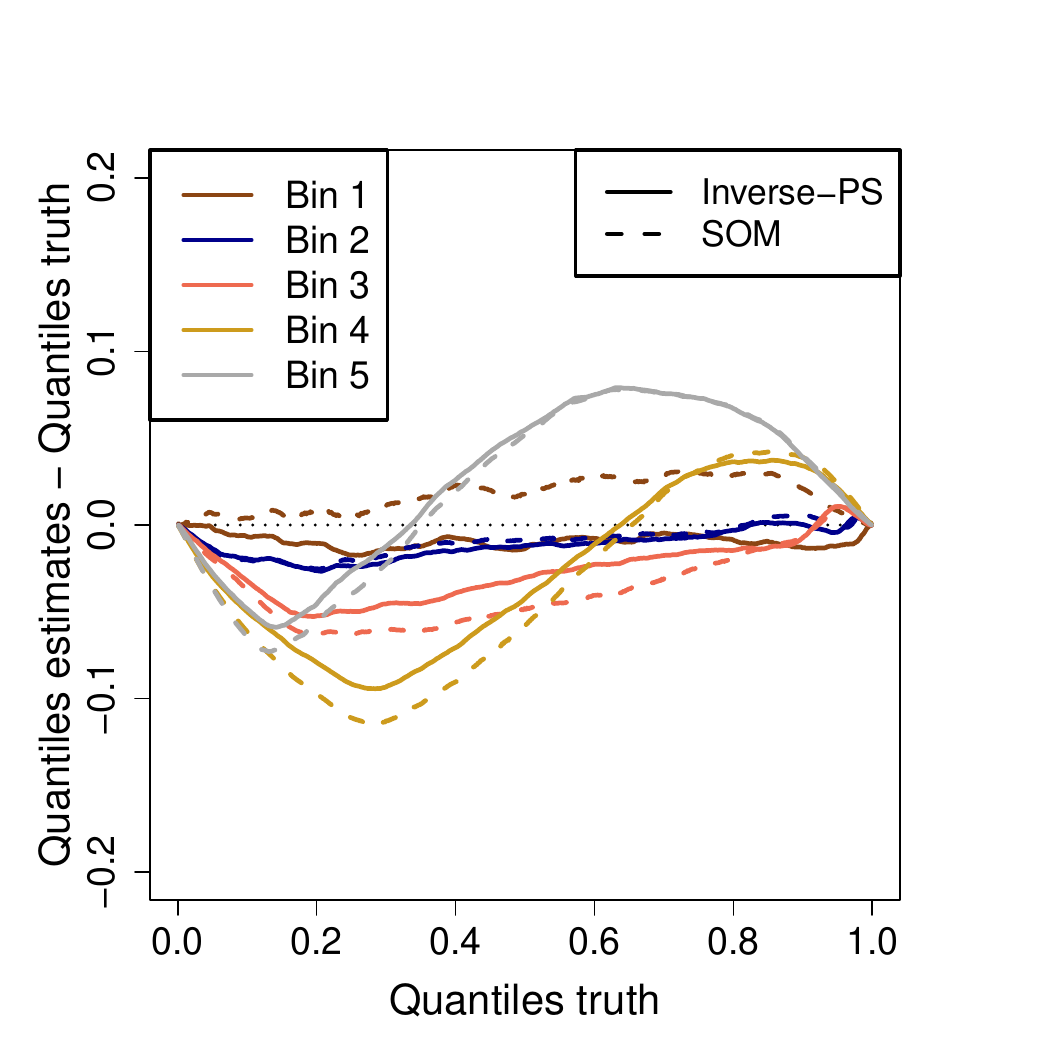}
    \caption{\baselineskip=15pt 
    Modification of the pp-plots illustrated in Figure~\ref{fig:pp-plot_inverse_PS_SOM} to visualize differences between the estimated distributions (Inverse-PS and SOM) with their underlying ground truth (full photometric truth and gold selected truth); modified by subtracting the x-axis values (the quantiles of the true distributions) from the y-axis values (the quantiles of the estimated distributions) in each tomographic bin illustrated in Figure~\ref{fig:pp-plot_inverse_PS_SOM}. Solid lines illustrate the inverse-PS results, and dashed lines illustrate the SOM results. The $95\%$ intervals (vertical bars) are omitted for clarity. 
    \label{fig:pp-plot_inverse_PS_SOM_subtracted}} 
\end{minipage}
\end{figure}

\begin{figure*}
\centering
\begin{minipage}{.95\textwidth}
  \centering
    \includegraphics[width=0.32\columnwidth]{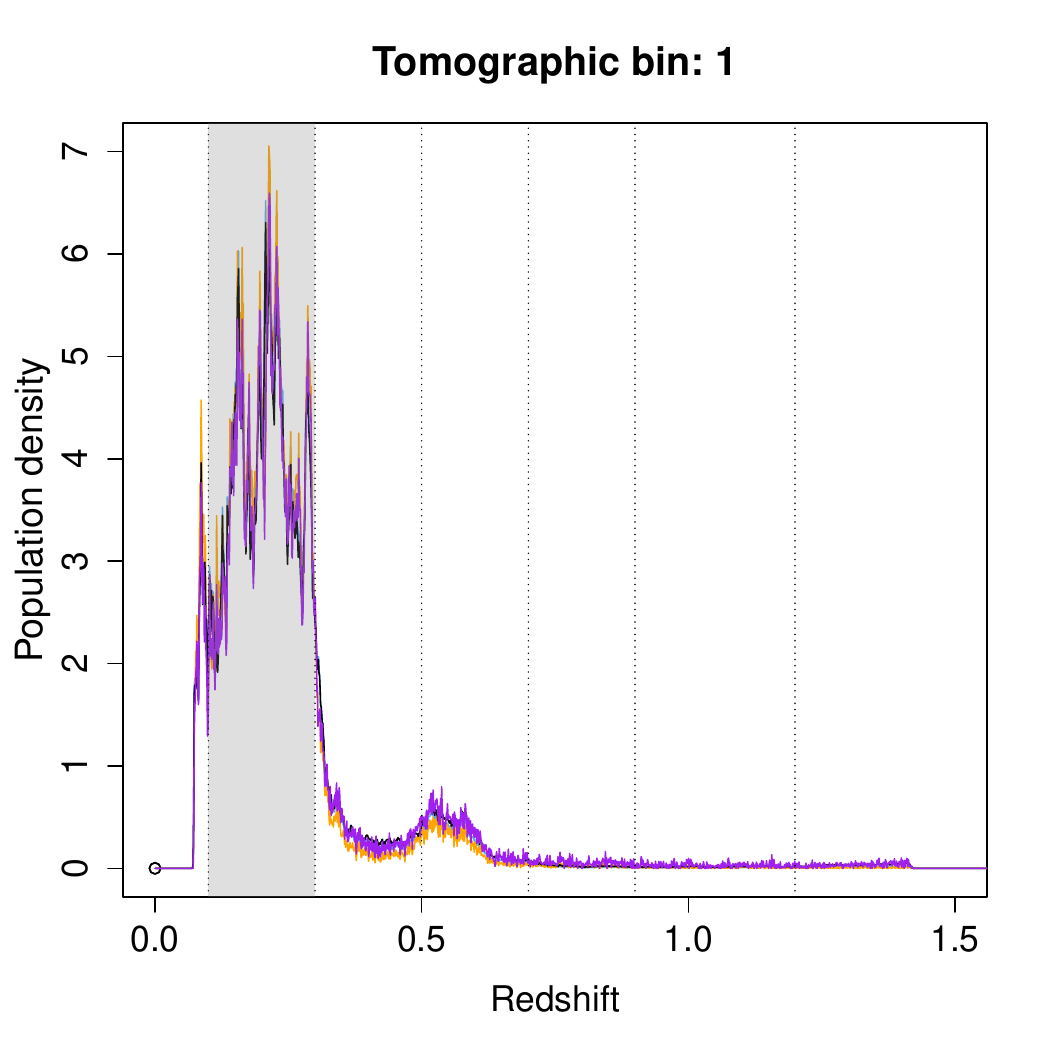}
    \includegraphics[width=0.32\columnwidth]{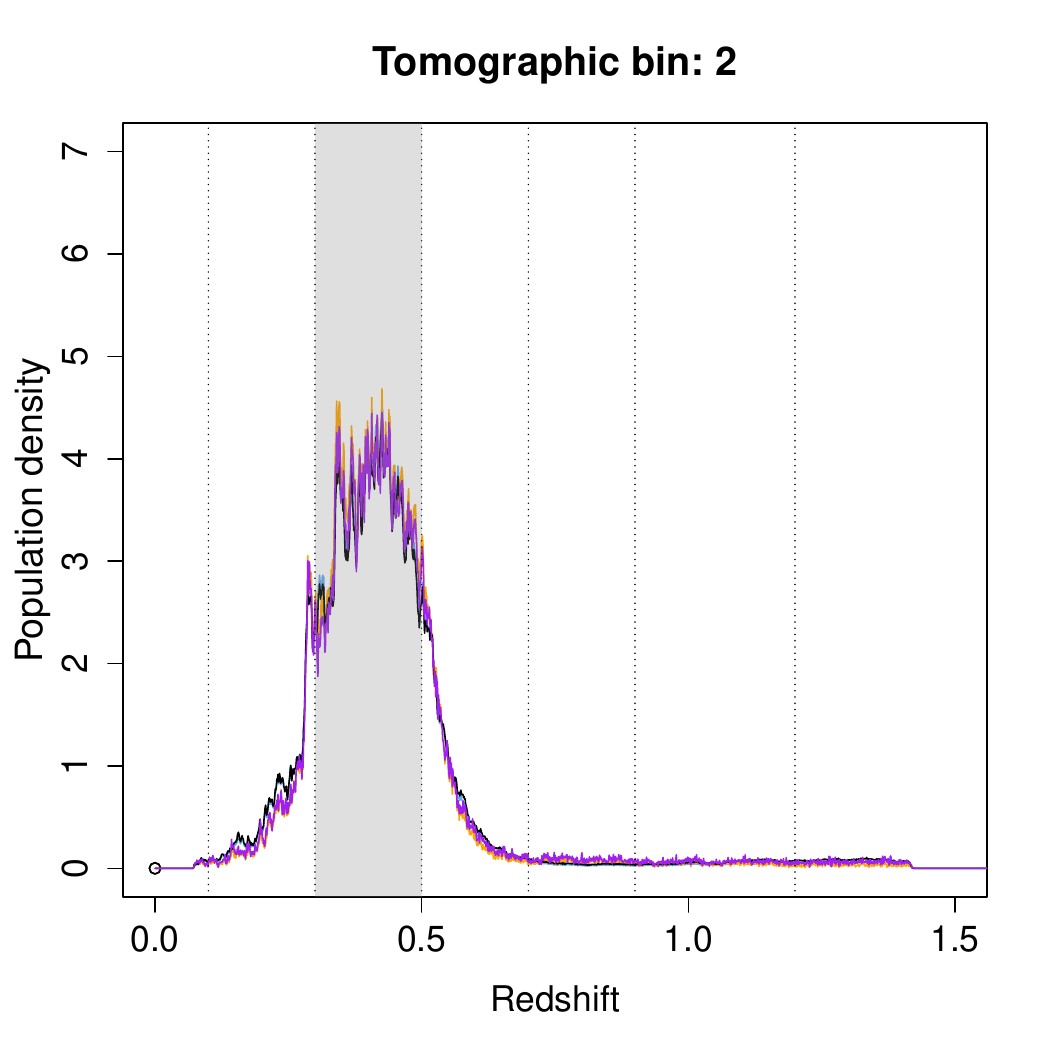}
    \includegraphics[width=0.32\columnwidth]{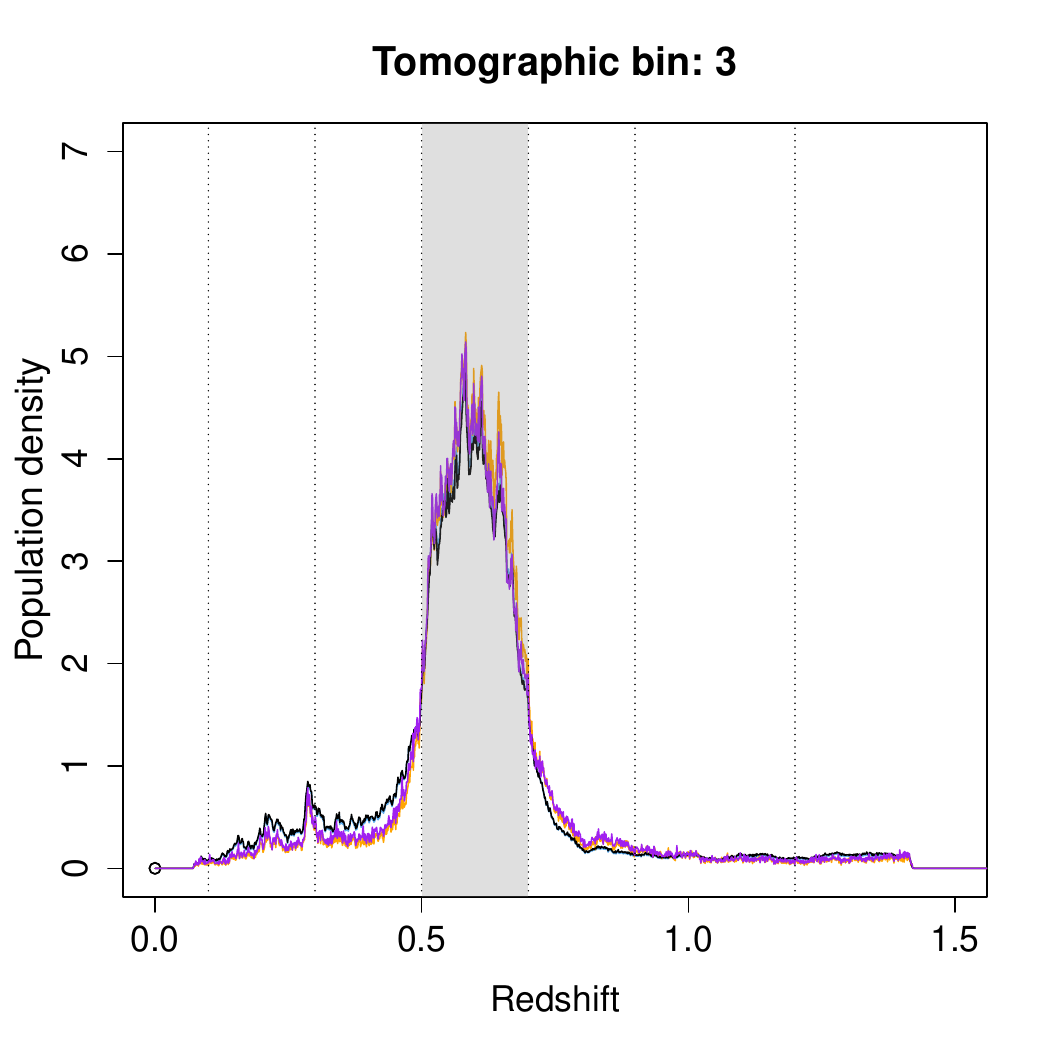}
        \includegraphics[width=0.32\columnwidth]{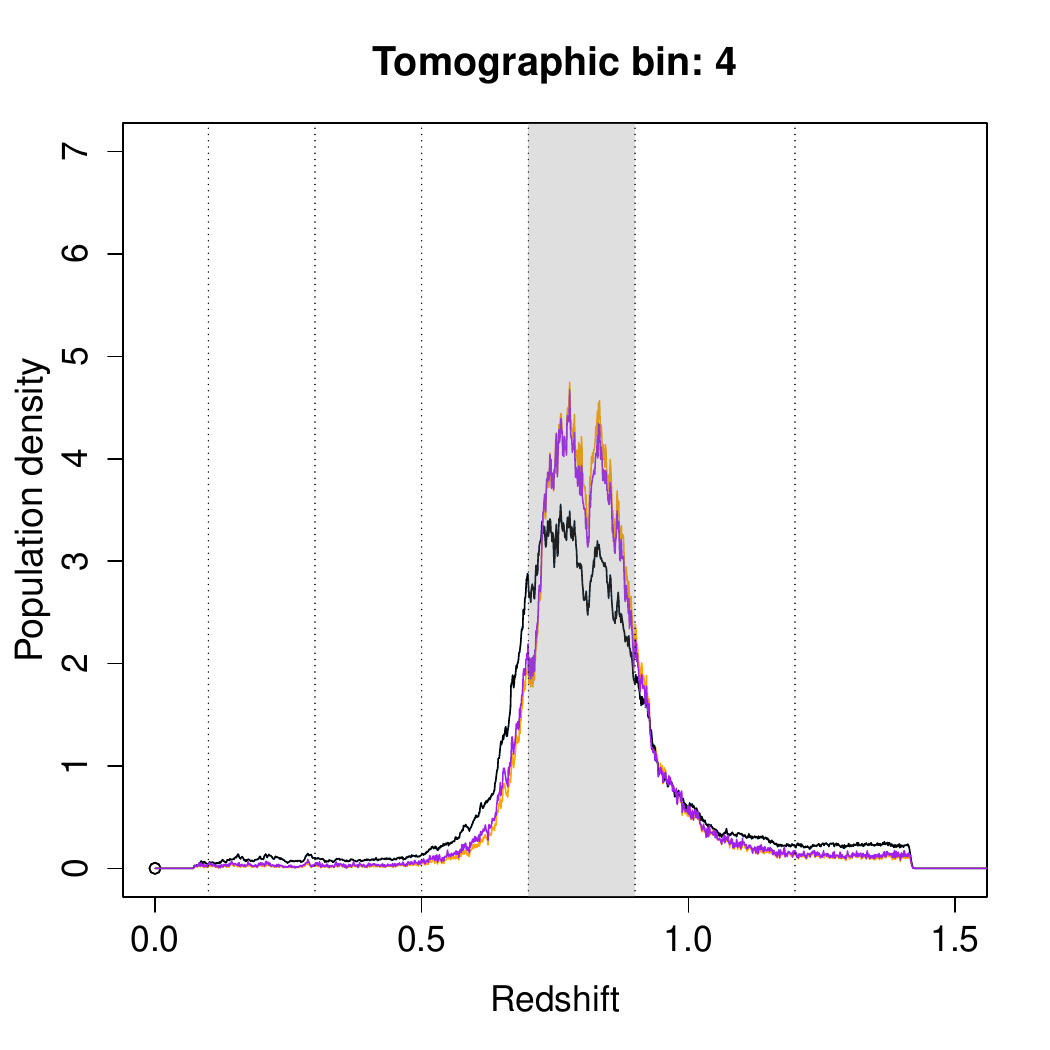}
        \includegraphics[width=0.32\columnwidth]{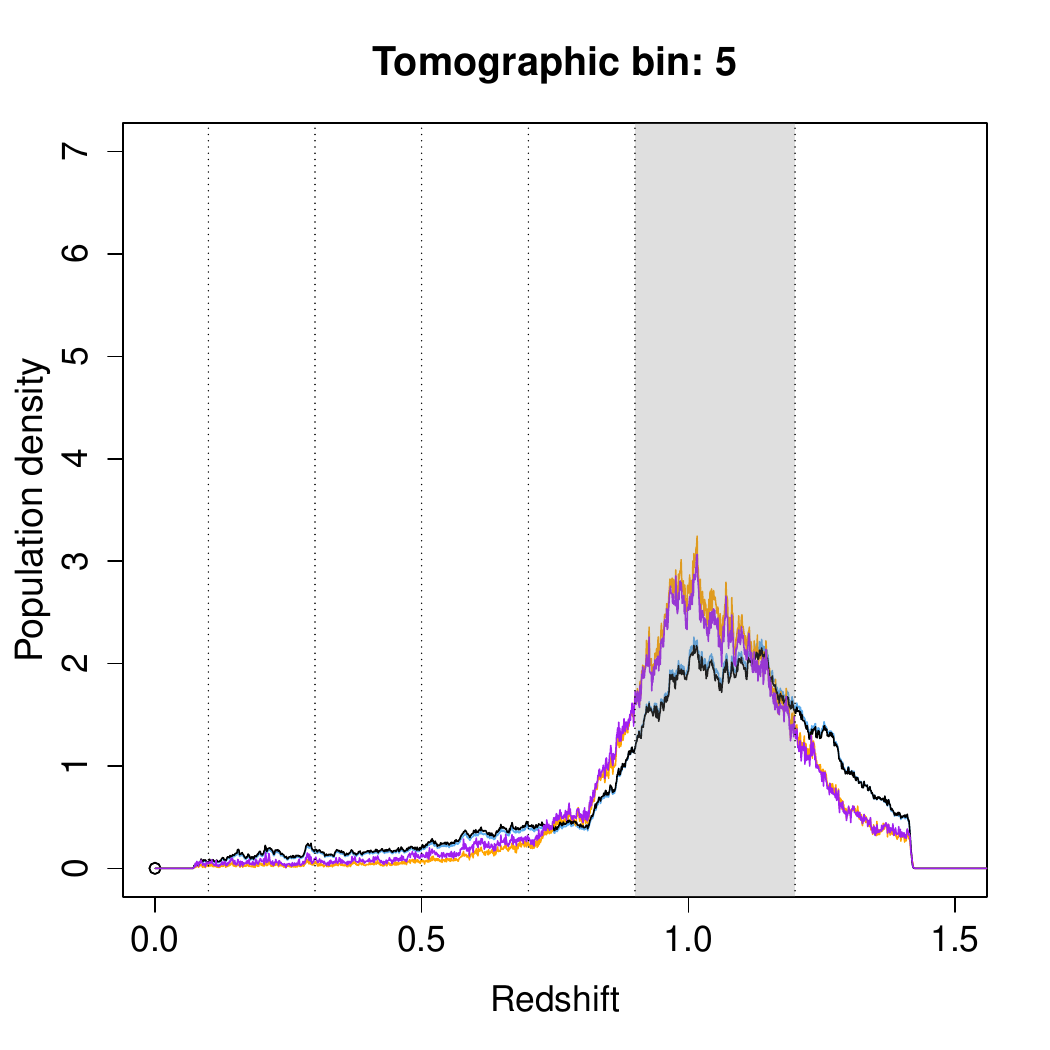}
        \includegraphics[width=0.32\columnwidth]{Figures/Population_densities/Legend_Population_SOM_IPS_sb.png}
    \caption{\baselineskip=15pt 
    The same as Figure~\ref{fig:population_distributions}, but without additional Gaussian Kernel smoothing of the population distributions. More precisely, the figure illustrates the redshift population distribution (estimates) per tomographic bin, with tomographic bins obtained as described in Section~\ref{sec:StratLearn_bin_assignment} via \textit{StratLearn}-based binning. The figure illustrates the inverse-PS (purple) and SOM (orange) distribution estimates.
    The underlying true photometric redshift population distributions per tomographic bin (not known in practice) are illustrated in black for the full sample truth, and in light blue for the gold selected true distributions. 
    The averaged (estimated) distributions across the 100 LoS are illustrated per tomographic bin. 
    \label{supp_fig:IPS_SOM_populations_notsmoothed}} 
\end{minipage}
\end{figure*}

\begin{figure*}
\centering
\begin{minipage}{.95\textwidth}
  \centering
    \includegraphics[width=0.32\columnwidth]{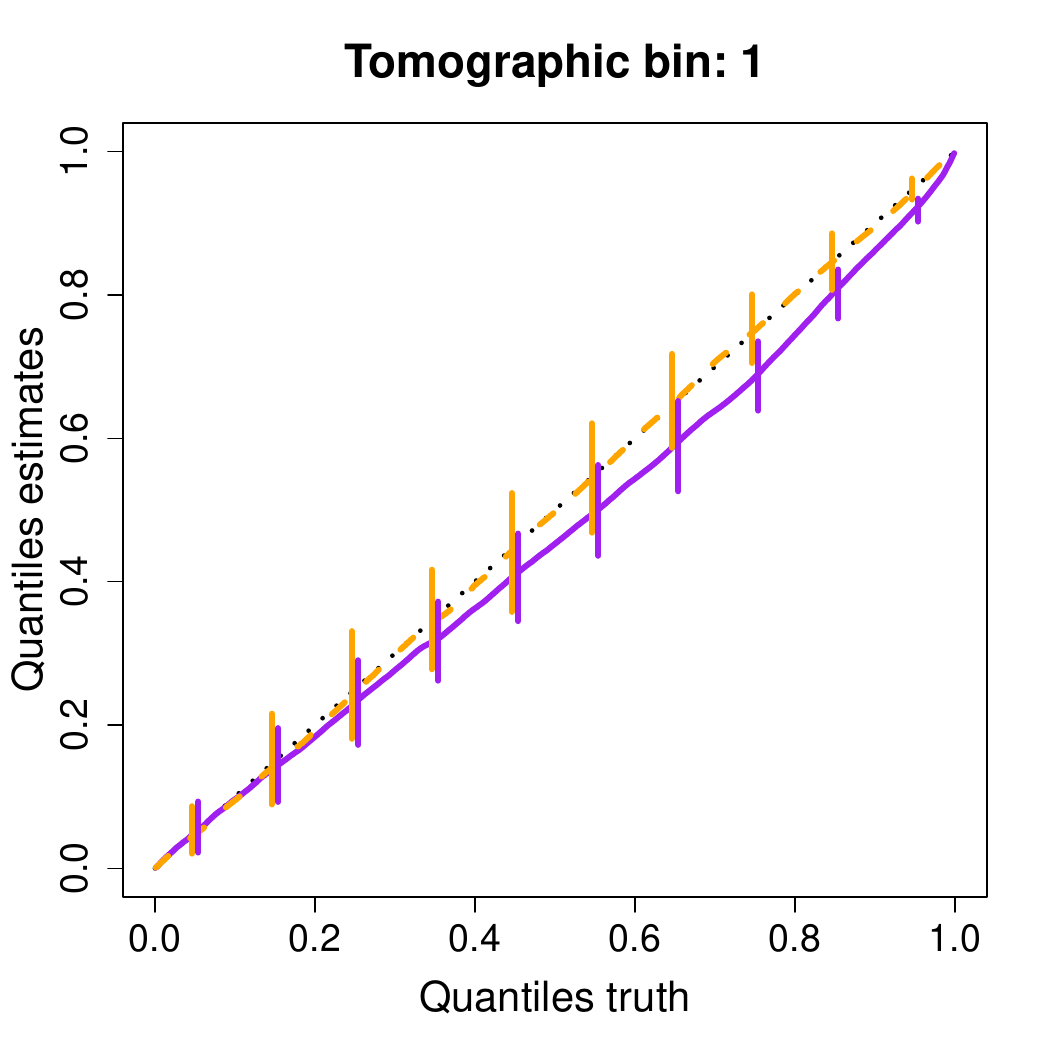}
    \includegraphics[width=0.32\columnwidth]{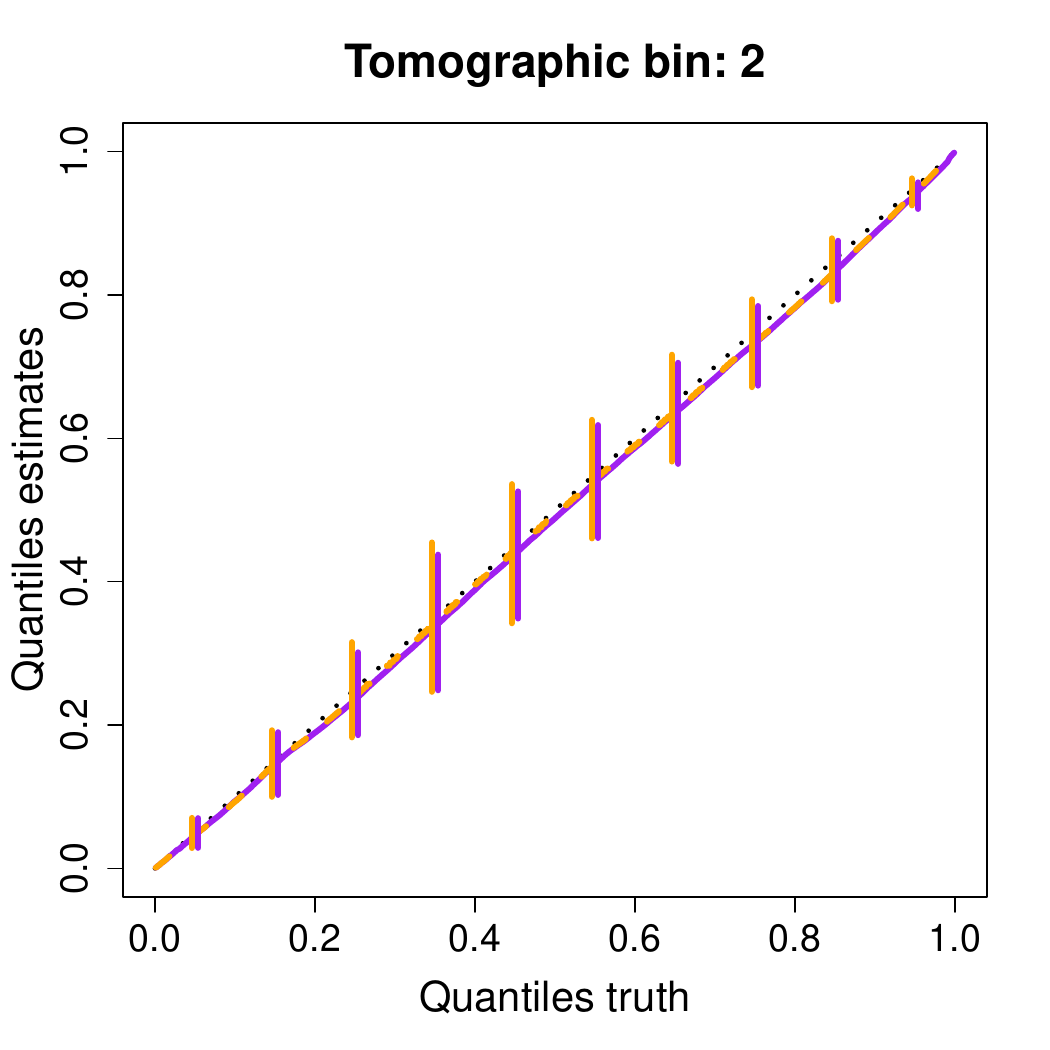}
    \includegraphics[width=0.32\columnwidth]{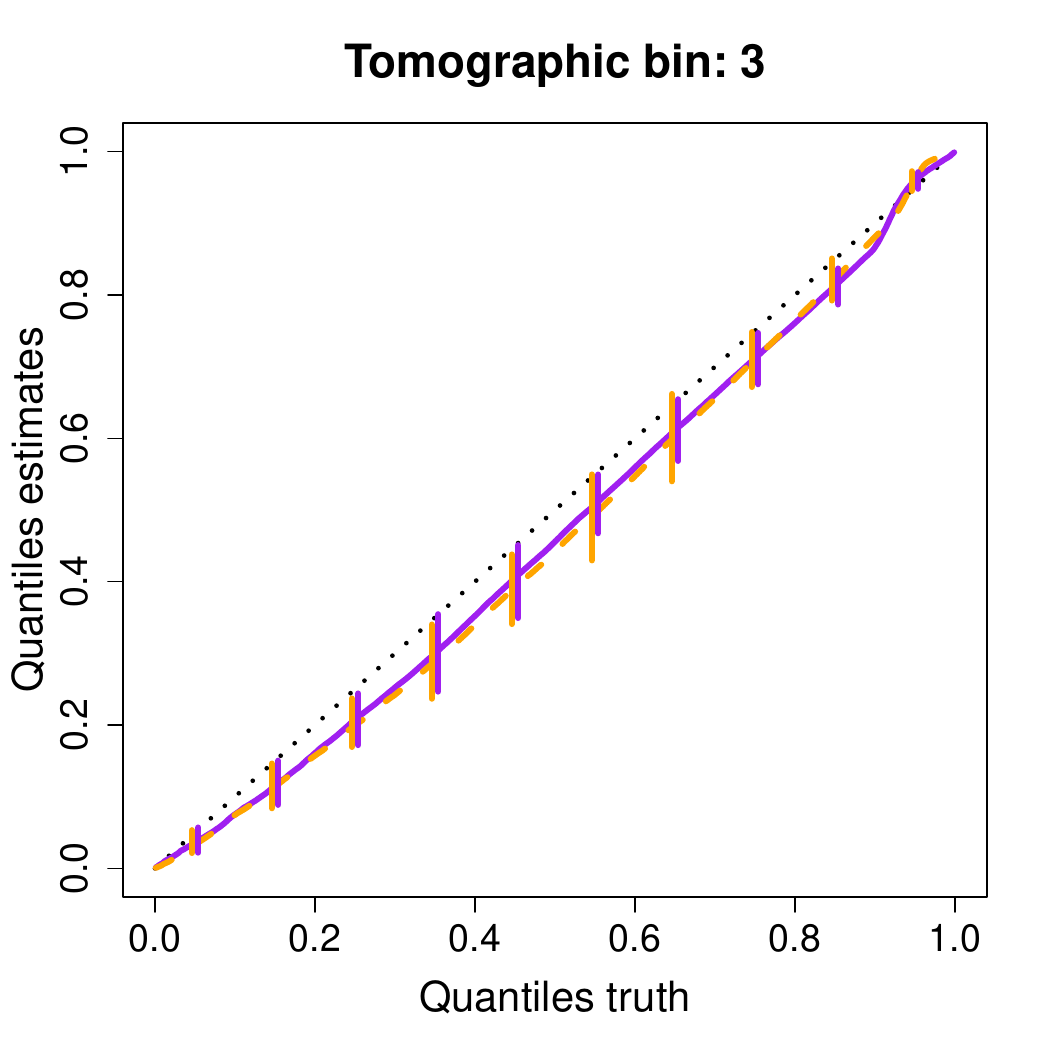}
        \includegraphics[width=0.32\columnwidth]{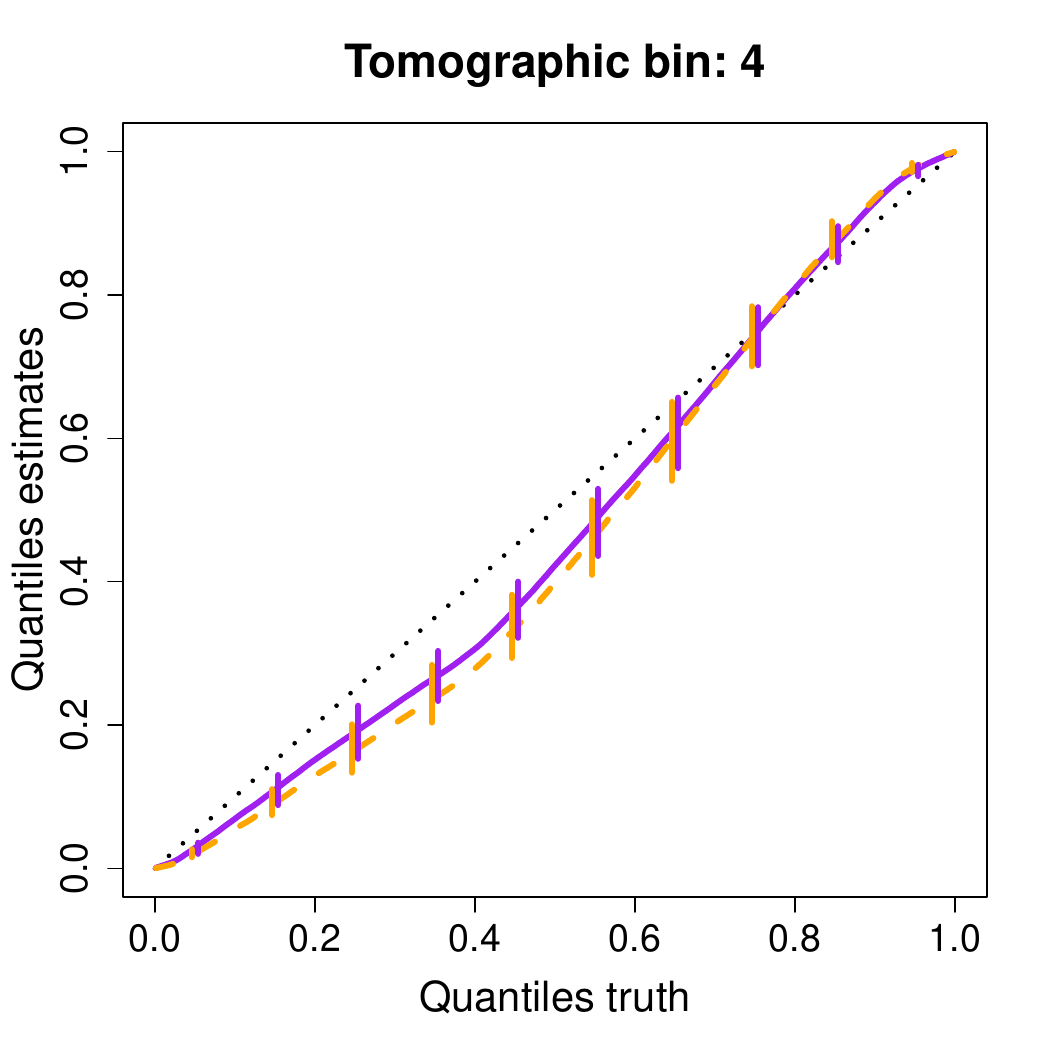}
        \includegraphics[width=0.32\columnwidth]{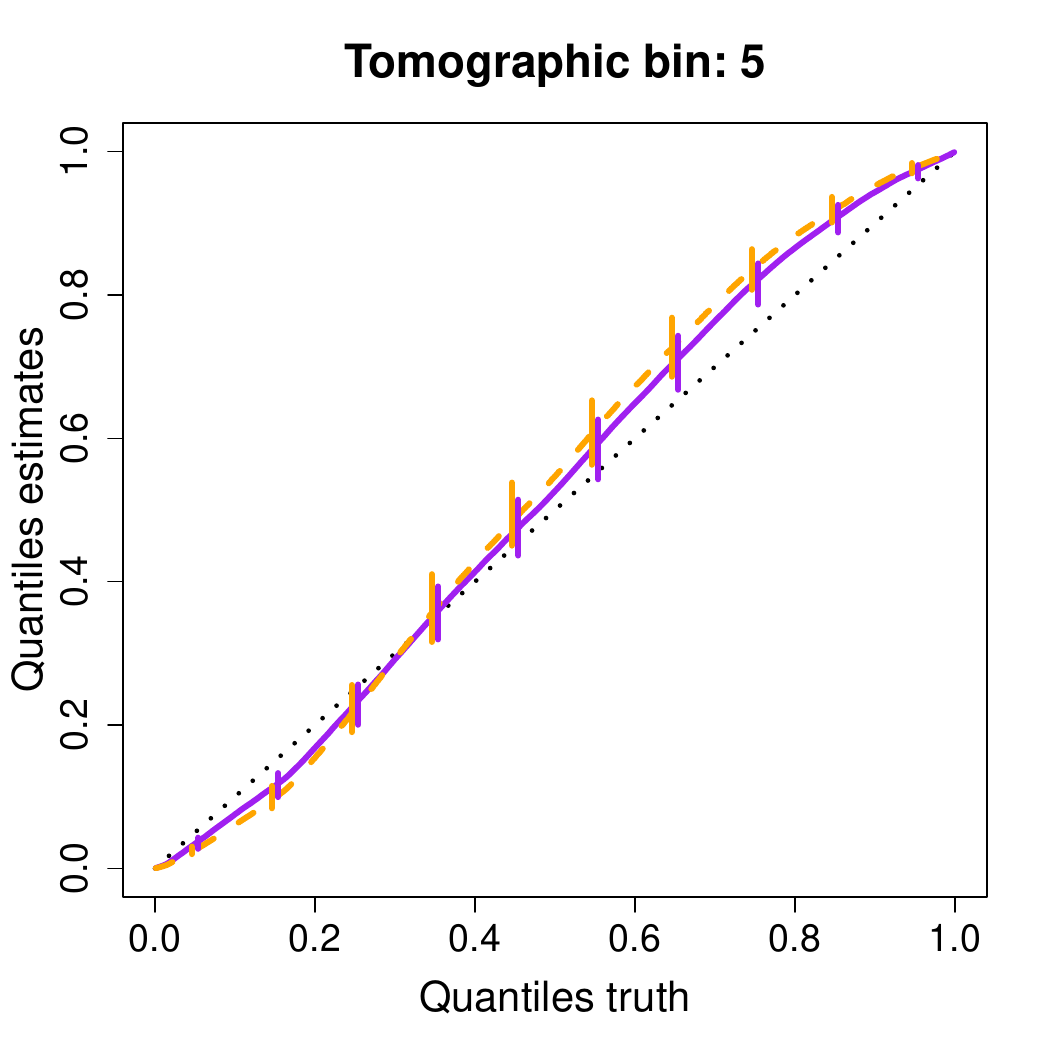}
       \includegraphics[width=0.32\columnwidth]{Figures/pp-plots/Legend_ppplots.png}
    \caption{\baselineskip=15pt  The same as in Figure~\ref{fig:pp-plot_inverse_PS_SOM}, but on $z_B$-based binning instead of \textit{StratLearn}-based binning. More precisely, the figure presents pp-plots  for the inverse-PS estimated distributions vs. the true (full) photometric redshift distributions 
    in purple lines, and pp-plots of the SOM estimated distributions vs. the gold selected true distributions in orange dashed lines, based on the $z_B$ tomographic binning (following Section~\ref{sec:StratLearn_bin_assignment}). For each tomographic bin, the averaged pp-plots across the 100 LoS are presented, with vertical bars illustrating $95\%$ intervals indicating the range of the central 95 pp-plot lines from the 100 LoS. In bin 1, the SOM pp-plot line is closer to the $45^{\circ}$ line,  
    while in tomographic bins 3 to 5 the inverse-PS pp-plot line is slightly closer to the $45^{\circ}$ line, with almost identical performance in tomographic bin 2. Given the $z_B$-based binning none of the estimators (inverse-PS or SOM) is thus  consistently closer to its underlying ground-truth (throughout the tomographic bins). 
    \label{fig:pp-plot_inverse_PS_SOM_zB}} 
\end{minipage}
\end{figure*}

\begin{figure*}
\centering
\begin{minipage}{.95\textwidth}
  \centering
    \includegraphics[width=0.32\columnwidth]{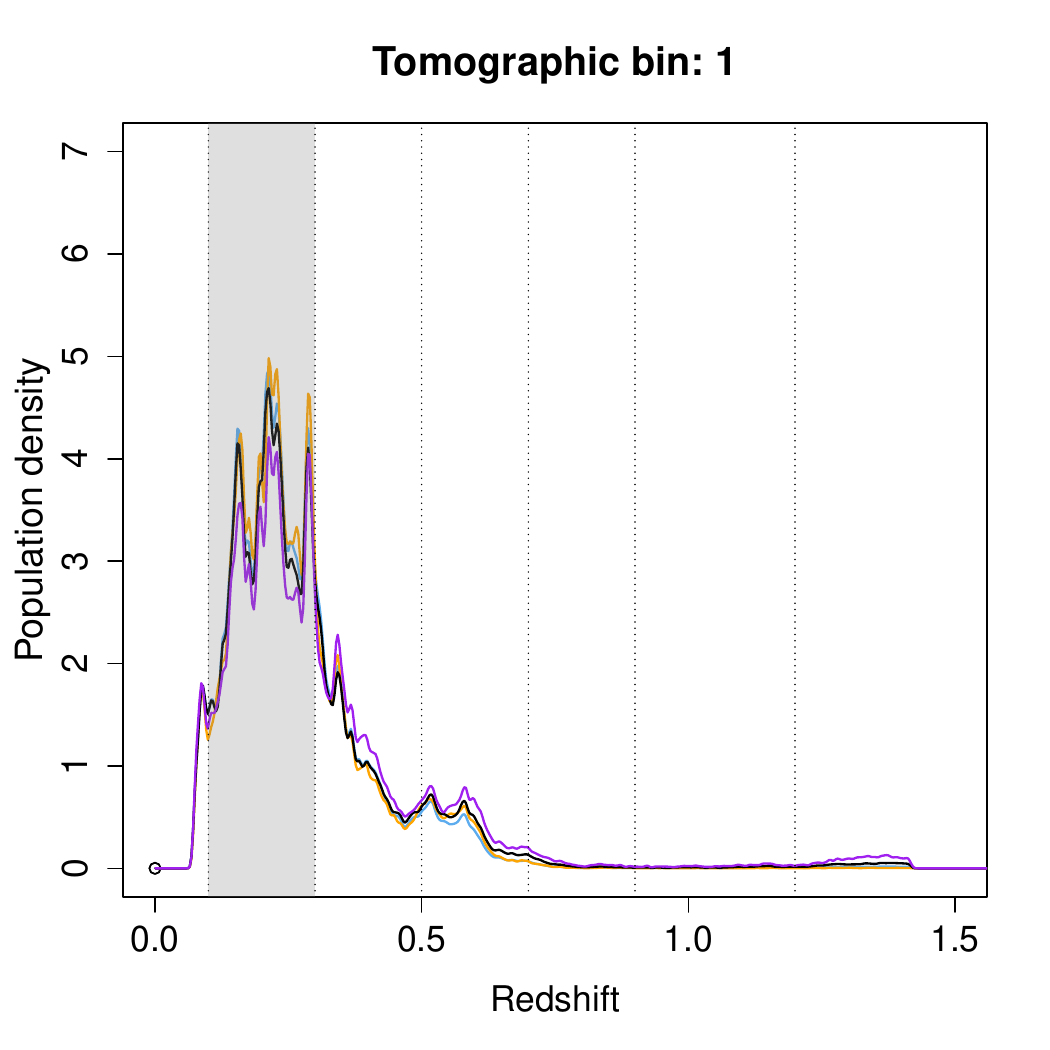}
    \includegraphics[width=0.32\columnwidth]{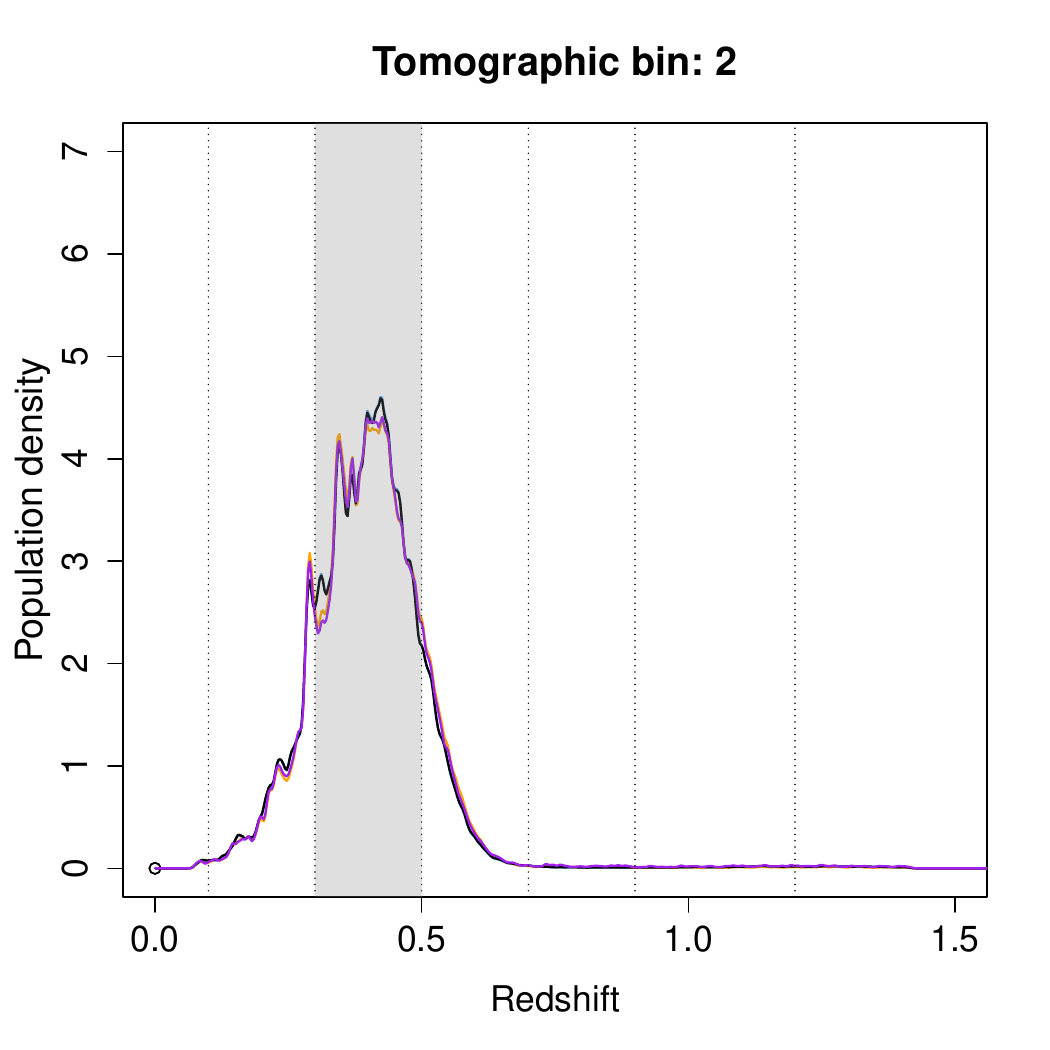}
    \includegraphics[width=0.32\columnwidth]{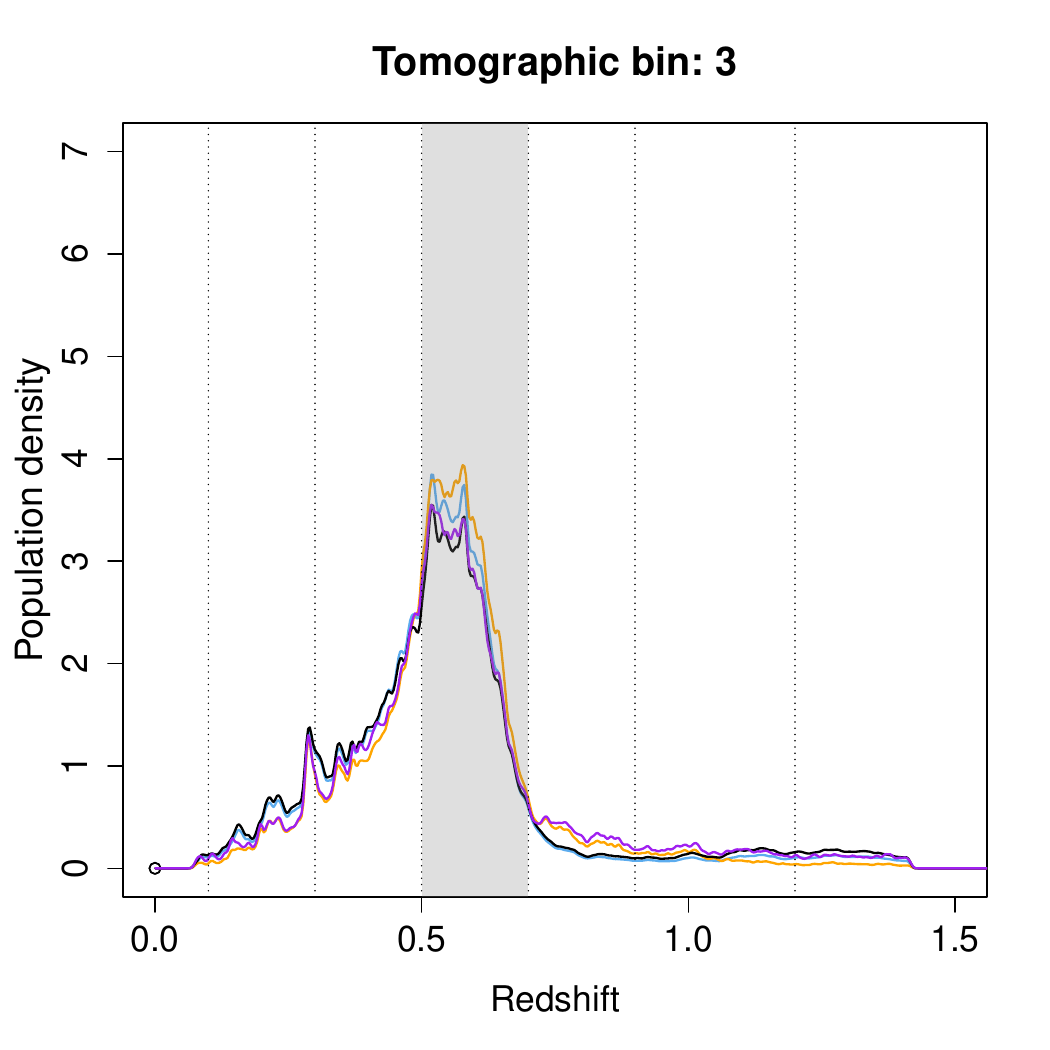}
        \includegraphics[width=0.32\columnwidth]{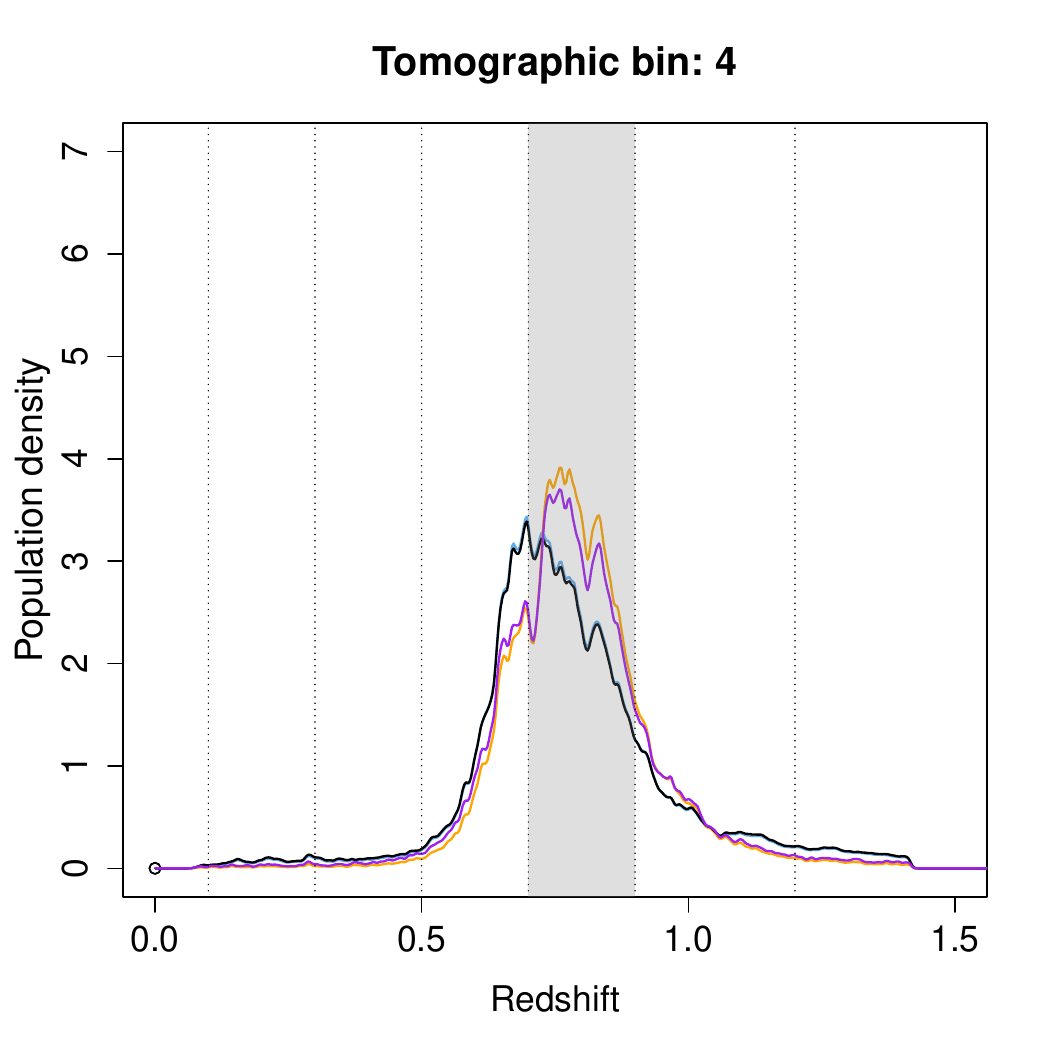}
        \includegraphics[width=0.32\columnwidth]{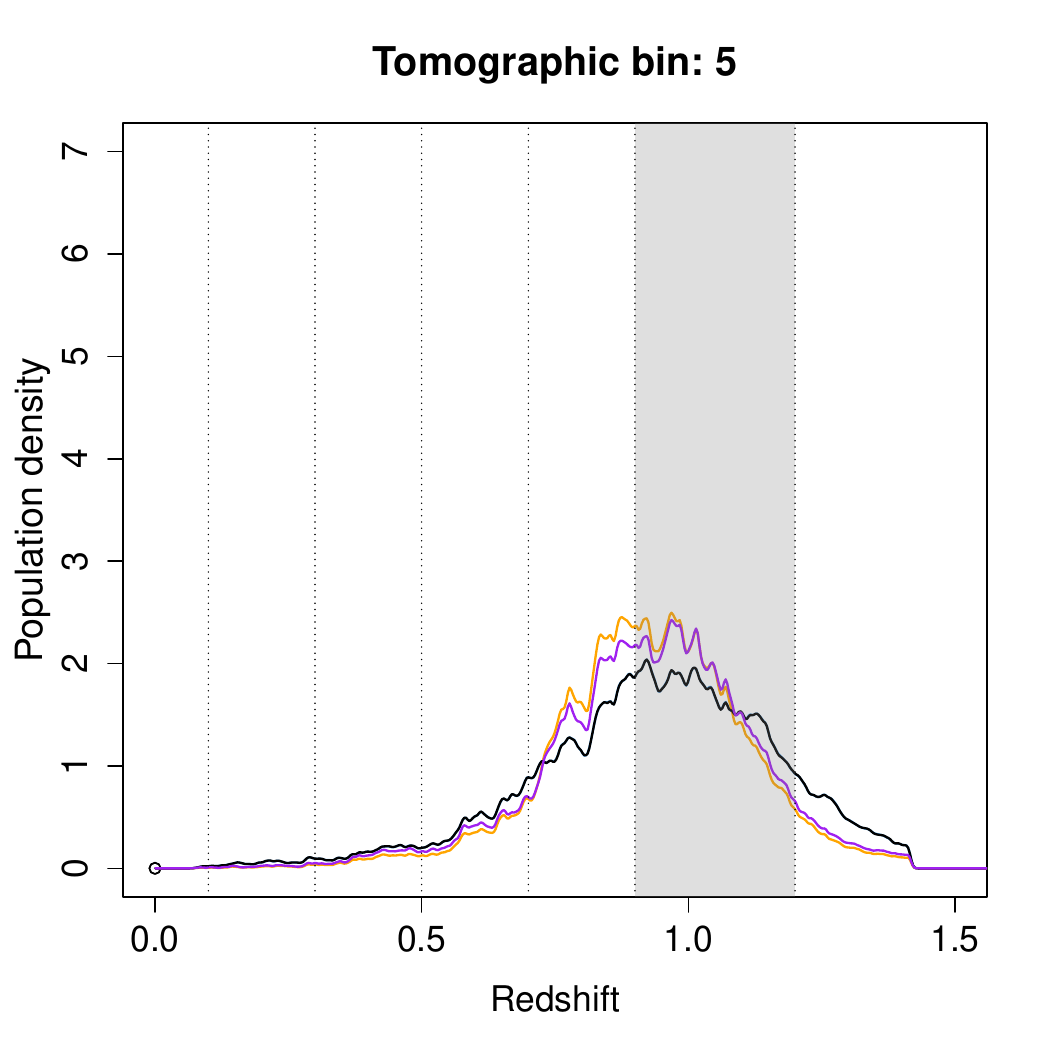}
        \includegraphics[width=0.32\columnwidth]{Figures/Population_densities/Legend_Population_SOM_IPS_sb.png}
    \caption{\baselineskip=15pt     The same as in Figure~\ref{fig:population_distributions}, but on $z_B$-based binning instead of \textit{StratLearn}-based binning. More precisely, the figure illustrates the redshift population distribution (estimates) per tomographic bin, with $z_B$-based tomographic binning. The figure illustrates the inverse-PS (purple) and SOM (orange) distribution estimates.
    The underlying true photometric redshift population distributions per tomographic bin (not known in practice) are illustrated in black for the full sample truth, and in light blue for the gold selected true distributions. 
    The averaged (estimated) distributions across the 100 LoS are illustrated per tomographic bin. 
    \label{supp_fig:IPS_SOM_populations_smoothed_zB}} 
\end{minipage}
\end{figure*}

This section presents additional figures, as previously referred to in the main paper. More precisely, the figures provide additional data/simulation study details, such as Figures~\ref{figure:missing_magnitudes_and_colors}, \ref{fig:resample_vs_reweight}; and additional numerical results, Figures~\ref{figure:PS_distributions}, \ref{figure:hyperparam_Series}, \ref{figure:hyperparam_KerNN}, 
\ref{figure:hyperparam_Comb}, \ref{figure:conf_matrix_StratLearn_vs_ZB}, \ref{fig:pp-plot_inverse_PS_SOM_subtracted}, \ref{supp_fig:IPS_SOM_populations_notsmoothed}, \ref{fig:pp-plot_inverse_PS_SOM_zB}, and \ref{supp_fig:IPS_SOM_populations_smoothed_zB}.

\bigskip

